\documentclass[rmp,twocolumn,author-year,notitlepage]{revtex4-2}

\usepackage{amsmath}
\usepackage{amssymb}
\usepackage[unicode=true,colorlinks=true,citecolor=blue,urlcolor=blue]{hyperref}

\usepackage{amsmath}
\usepackage{amssymb}
\usepackage[unicode=true,colorlinks=true,citecolor=blue,urlcolor=blue]{hyperref}

\usepackage{natbib}
\usepackage{graphicx}

\makeatletter
\def\@bibdataout@aps{%
\immediate\write\@bibdataout{%
@CONTROL{%
apsrev41Control%
\longbibliography@sw{%
    ,author="08",editor="1",pages="1",title="0",year="1"%
    }{%
    ,author="08",editor="1",pages="1",title="",year="1"%
    }%
 }%
}%
\if@filesw \immediate \write \@auxout {\string \citation {apsrev41Control}}\fi 
}
\makeatother

\usepackage{bm}

\newcommand{\blue}[1]{{\color{blue}{#1}}}

\newcommand{\nix}[1]{}

\renewcommand {\Im}{\mathop\mathrm{Im}\nolimits}
\renewcommand {\Re}{\mathop\mathrm{Re}\nolimits}

\renewcommand {\phi}{{\varphi}}
\newcommand {\rmi}{{\rm i}}
\newcommand {\rmd}{{\rm d}}
\newcommand {\sign}{\mathop{\mathrm{sign}}\nolimits}
\newcommand {\e}{{\rm e}}
\newcommand {\ve}[1]{\bm{#1}}
\newcommand {\eps}{\varepsilon}

\newcommand {\rot}{\mathop\mathrm{rot}\nolimits}
\newcommand{\GO}{\gamma_{\rm 1D}}
\newcommand{\GOR}{\gamma^\rightarrow}
\newcommand{\GOL}{\gamma^\leftarrow}
\newcommand{\GORL}{\gamma^{\rightarrow/\leftarrow}}
\begin{document}

\title{Waveguide quantum electrodynamics: collective radiance and photon-photon correlations
 }
 \author{Alexandra S. Sheremet}
\email{a.sheremet@metalab.ifmo.ru}
\affiliation{Russian Quantum Center, Skolkovo, 143025 Moscow, Russia}
\affiliation{Department of Physics and Engineering, ITMO University, Saint-Petersburg, 197101, Russia}
\author{Mihail I. Petrov}
 \author{Ivan V. Iorsh}
\affiliation{Department of Physics and Engineering, ITMO University, Saint-Petersburg, 197101, Russia}
 \author{Alexander V. Poshakinskiy}
 \affiliation{Ioffe Institute, St. Petersburg 194021, Russia}
  \author{Alexander N. Poddubny}
\email{a.poddubny@fastmail.com}
\affiliation{Rehovot 763051, Israel}

\begin{abstract}
This review describes the emerging field of waveguide quantum
electrodynamics (WQED) concerned with the interaction of photons
propagating in a waveguide with localized quantum emitters. The
collective emitter-photon interactions can lead to both enhanced and
suppressed coupling compared to the case of independent emitters. Here,
we focus on guided photons and ordered arrays, leading to
super- and sub-radiant states, bound photon states and quantum
correlations with promising quantum information applications. We
highlight recent groundbreaking experiments performed with different
quantum platforms, including cold atoms, superconducting qubits,
semiconductor quantum dots, quantum solid-state defects, and we
provide a comprehensive introduction to theoretical techniques to
study the interactions and dynamics of these emitters and the photons
in the waveguide.
\end{abstract}
\date{\today}

\maketitle
\tableofcontents


\section{Introduction}
Arrays of atoms coupled to photons present a paradigmatic system for quantum optics since at least the fundamental discovery of Dicke superradiance \cite{Dicke1954}. More recently, the rapid development of quantum technologies \cite{Chang2014,Polzik2010} has led to emergence of novel experimental platforms of emitters, coupled to propagating photons in a waveguide. These systems can be based on natural or artificial atoms, such as superconducting qubits, or solid state quantum dots and defects, and can employ different types of optical and microwave waveguides \cite{KimbleRMP2018, lodahl2015interfacing,Sile2016,Roy2017}. 
The resulting emerging field of research is termed 
waveguide quantum electrodynamics (WQED) that offers novel opportunities both for fundamental physics and for quantum information processing.

Since photons are confined within the waveguide,  atom-photon interactions become much stronger than in free space, similar to the case of cavity QED. The  WQED is also closely linked to circuit QED that studies networks of coupled superconducting qubits  interacting with microwave photons~\cite{blais2020circuit,Gu2017,Carusotto2020}. The distinguishing feature of WQED is the coupling of quantum emitters just to a single or several propagating photon modes. This inherently one-dimensional  geometry is beneficial for cascaded processing of photons, enabling efficient generation and detection of quantum states of light~\cite{Prasad2020}. 
From the fundamental side, WQED systems can be viewed as  artificial media with strong optical nonlinearities at a single-photon
level~\cite{Chang2014}. The combination of strong atom-photon interaction with the light-mediated coupling between atoms at large distances makes WQED setups also quite unusual from the points of view of condensed matter physics. They can act as quantum simulators of many-body effects ranging from superfluid-Mott insulator transitions \cite{Shi2018} to topological states of matter \cite{kim2020quantum} and many-body localization \cite{fayard2021manybody}. 

There already exists a number of excellent topical reviews related to the waveguide quantum electrodynamics \cite{Roy2017,KimbleRMP2018,Trschmann2019}, quantum optics with atoms and fibers \cite{Sile2016},
quantum light-matter interfaces \cite{Polzik2010,lodahl2015interfacing}, quantum simulations and many-body physics with light~\cite{Noh2016}. The state-of-the-art structures for a single photon processing are discussed in the review~\cite{uppu2021singlephoton}. The goal of the current review is twofold. First, we discuss in detail several recent representative experiments in the WQED setups and beyond, including demonstrations of tunable photon bunching and antibunching from atomic arrays \cite{Prasad2020}, generation and detection of collective entangled atom-photon states \cite{Corzo2019} and subradiant atom-made mirrors \cite{Rui2020}. We also compare  different state-of-the-art experimental WQED platforms. Second, we provide a comprehensive theoretical background on the cooperative emission effects and photon-photon interactions for one-dimensional ordered atomic arrays, starting from the basics and proceeding to the advanced theoretical techniques. Due to certain similarities between the Dicke problem of quantum optics and the Kondo problem \cite{Leggett1987}, many techniques have been adopted from the condensed matter physics (the Bethe ansatz) and the quantum field theory (functional integral approach). While the Dicke problem  for two-level atoms located in exactly the same point can be solved by  the Bethe ansatz~\cite{Yudson1984,Yudson1985}, the  generalized case of non-zero interatomic spacing is still far from being  completely understood despite significant recent theoretical progress. It manifests a plethora of phenomena familiar from other fields, such as formation of bound photon states, fermionization of photons, interaction-induced topological states and quantum Hall phases. Thus, we hope that this review might be useful to both experimentalists and theorists already working on WQED or coming to WQED from other fields of physics.

We start the  main part of the review in Sec.~\ref{sec:current} with comparison of various experimental platforms for the waveguide quantum electrodynamics, differing by the choice of natural or artificial atoms and different waveguide realizations.
We try to put them in an universal perspective based on the typical numbers of emitters and the light-matter coupling strength and then provide an outlook by discussing emerging platforms in Sec.~\ref{sec:outlook}. Section~\ref{sec:effects} presents a detailed theoretical consideration  of polaritonic quantum states in the waveguide with periodic  emitter arrays. We discuss collective super- and subradiant states  and the resulting correlations between photons scattered on such structures. Several experimental demonstrations for different platforms  are discussed in Sec.~\ref{sec:applications}, where we put a special emphasis on  the demonstrations of superradiance (Sec.~\ref{sec:superradiance} ) and subradiance (Sec.~\ref{sec:ustinov}), generation of quantum states of light (Sec.~\ref{sec:generation}), quantum memory applications (Sec.~\ref{sec:memory} ) and physics of Bragg-spaced atomic arrays (Sec.~\ref{sec:Bragg}).
Final Sec.~\ref{sec:summary} presents summary and outlook. In order to make the main text more accessible, we reserve most of the theoretical details for Appendices~\ref{sec:input_output}--\ref{app:lattice-sum}.

\section{Waveguide QED systems}\label{sec:platforms}

\subsection{Tuning light-matter coupling in atomic arrays}\label{sec:principles}
In this Section, we consider various WQED platforms that  have different advantages depending on  the waveguide realization and the type of quantum emitters.  For example, both cold atom arrays and solid-state emitters can be used to generate and detect quantum light. The former are also especially beneficial for quantum memory due to their high coherence. Superconducting qubit structures, operating in the microwave spectral range, have tremendous tunability that can be exploited to process quantum states. Before proceeding to the specifics, it is instructive to first  discuss  general advantages of the waveguide coupling  for a specific case of cold atom ensemble as compared to atoms in a free space.

The main idea behind the WQED is to controllably enhance or suppress the light-matter interaction of a $N$-atom ensemble with a given propagating photon mode. As such, the two most important parameters are   the  number of atoms $N$ and the coupling efficiency $\beta$. We define the latter  as the ratio of the radiative decay rate of an individual emitter into the waveguide mode $\Gamma_{\rm 1D}$ to its total decay rate $\Gamma_{\text{tot}} = \Gamma_{\rm 1D}+\Gamma_{\rm nonrad}+\Gamma_{\rm ng}$ \cite{Arcari2014,Scarpelli2019}
\begin{equation}\label{eq:beta}
\beta=\frac{\Gamma_{\rm 1D}}{\Gamma_{\rm 1D}+\Gamma_{\rm ng}+\Gamma_{\rm nonrad}}\:.
\end{equation}
Here,  $\Gamma_{\rm ng}$ is a radiative decay rate into all other electromagnetic modes (nonguided modes), and $\Gamma_{\rm nonrad}$ is a homogeneous nonradiative decay rate.  It is instructive to discuss how  the values of $\Gamma_{\rm 1D}$ and $\Gamma_{\rm ng}$ can be tailored to increase $\beta$. 

\begin{figure}[!t]
 \centering\includegraphics[width=0.48\textwidth]{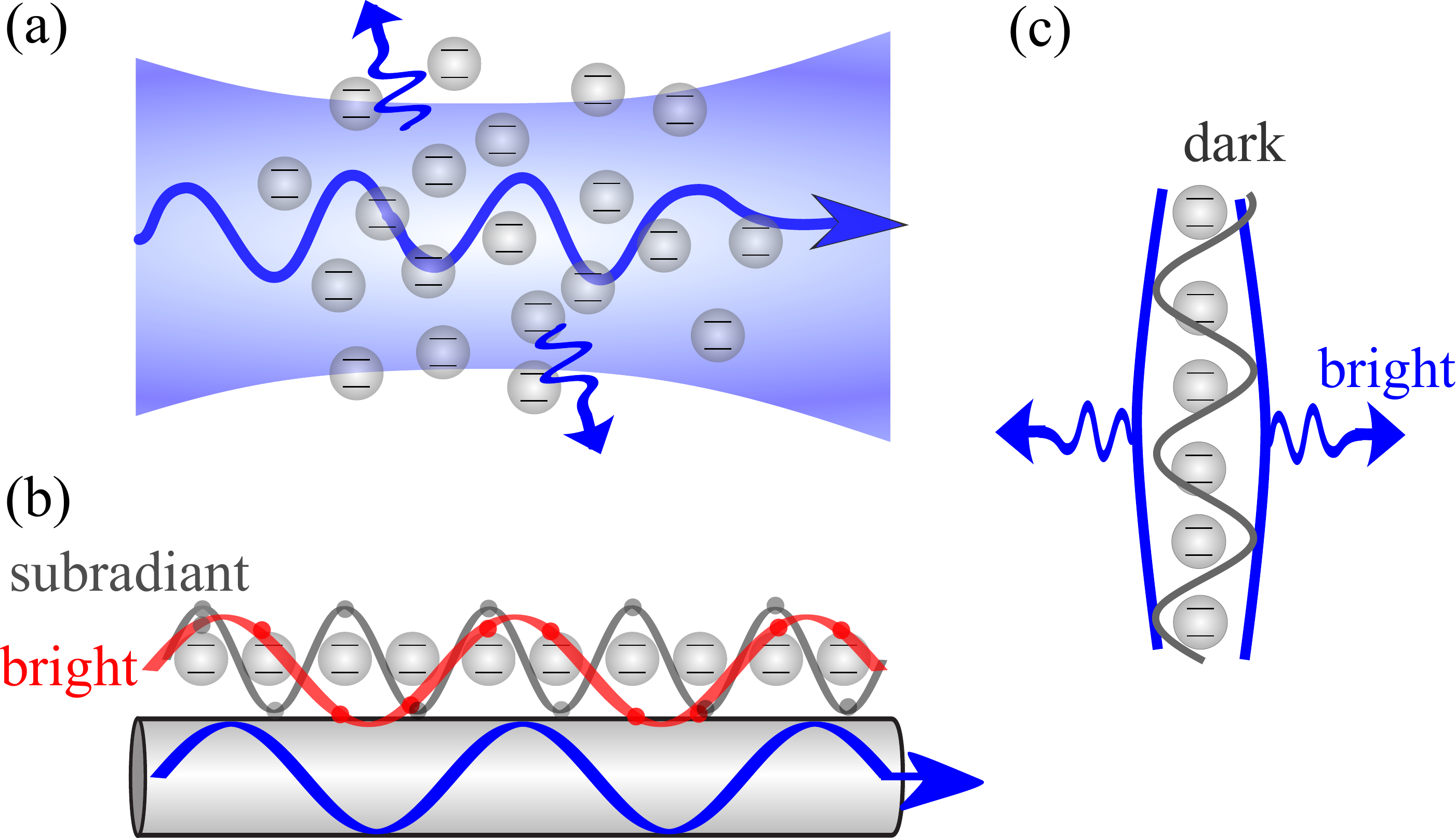}
 \caption{Schematic illustration of light-matter interaction in various atomic ensembles. (a) Light beam propagation and scattering in a disordered dilute atomic array in free space. (b) Interaction of an ordered atomic array with the guided photon mode. Subradiant and bright collective array excitations that are out-of-phase and in-phase with the photon mode (blue wave) are shown. (c) Interaction of ordered atomic array with a photon mode propagating through a waveguide in a transverse direction. Similarly to (b), bright and dark collective excitations are shown. }
 \label{fig:beta}
\end{figure}

Let us start with the beam propagation in a dilute disordered atomic array in free space. In this case, the efficiency of light interaction with an atom can be estimated as $\beta=\sigma_{0}/A$, where  $\sigma_0$ is the light scattering cross section and $A$ is the effective beam area. For an ideal two-level atom at the electric dipole resonance one has $\sigma_0 = 3\lambda_0^2/2\pi$ . The area of $A$ is limited from below by the diffraction limit or the sample area and it is typically much larger than $\lambda_0^2$ \cite{Polzik2010}. For instance, in a Cs atom sample with a diameter of 60~$\mu$m, considered in \cite{Polzik2008}, the  effective $\beta$-factor was low, on the order of $10^{-4}$. The crude estimation of the $\beta$-factor $\beta=\sigma_{0}/A$ clearly indicates that it can be enhanced by reducing the effective area of photon mode $A$, which can be done by confining photons to the waveguide, see Fig.~\ref{fig:beta}(b). In realistic atomic arrays near a fiber waveguide with a radius ~$400$ nm, this factor is on the order of $\beta\sim 10^{-2}$, which is by two orders of magnitude larger than in free space.
\begin{table*}[t]
\begin{tabular}{|c|c|c|cc|c|}
\hline
System  &Enhanced & Long-ranged & \multicolumn{2}{|c|}{Collective eigenmodes} &Details \\

& interaction &  coupling  & Superradiant&  Subradiant   & \\ \hline	

two atoms in free space & $-$ & $-$ & $\checkmark$ & $\checkmark$ & Sec.~\ref{sec:superradiance} \\

two atoms near a waveguide & \checkmark & \checkmark & $\checkmark$ & $\checkmark$ & Sec.~\ref{sec:superradiance} \\

dense array near a waveguide  &\checkmark & $-$ & $-$ & $-$ & Sec.~\ref{sec:single_photon},\ref{sec:generation},\ref{sec:ustinov} \\

Bragg-spaced array near the waveguide &\checkmark  & \checkmark & \checkmark & \checkmark & Sec.~\ref{sec:Bragg} \\

array in a cavity & \checkmark & $-$ & \checkmark & \checkmark & \\

Ordered 2D array in free space & \checkmark & \checkmark & \checkmark & \checkmark & Sec.~\ref{sec:2d} \\
\hline 
\end{tabular}
\caption{Light-matter coupling phenomena in  various quantum optical systems }\label{table:buzz}
\end{table*}

One more important parameter that should be minimized to increase the $\beta$-factor is the decay rate of the nonguided modes to the transverse direction $\Gamma_{\rm ng}$, see Fig.~\ref{fig:beta}(b). In typical disordered fiber-coupled arrays, this rate is on the order of $\Gamma_0$ \cite{LeKien2005}, but it can be optimized in the ordered arrays. Namely, only the collective modes that are inside the light cone, i.e. have the wavevector $k$ along the array smaller than $\omega_0/c$, can emit into the free space. The modes with $|k|>\omega_0/c$ are evanescent in the direction transverse to the array. Provided that, the array spacing is smaller than the light wavelength, most of modes will be guided ones.

Another crucial figure of merit for light interaction strength with the whole array is the resonant optical depth given by ${\rm OD} = -\ln T(\omega_0) \approx 2N\Gamma_{1 \rm{D}}/\Gamma_0$, where $T(\omega_0)$ is the transmission coefficient at the resonance frequency. We should mention that the expression ${\rm OD} = 2N\Gamma_{1{\rm D}}/\Gamma_0$ is valid only for non-Braagg arrays, see \cite{AsenjoGarcia2017atom}. In free space, if the atoms are far apart, the OD is emerging from forward propagation and it can be expressed as ${\rm OD} = 2n\sigma_0L$, where $n = N/A$ is density of atoms in the ensemble of the length $L$. Increasing OD is important for many quantum information applications, for example, for quantum memory \cite{Gorshkov2007}. However, achieving high OD is quite challenging, and $\rm OD\sim 10$ for an atomic ensemble 
 in free space required  $\sim 10^5\div 10^6$ atoms~\cite{Polzik2008}. From Eq.~\eqref{eq:beta}, one can see that the optical depth also can be expressed as ${\rm OD} = 2N\beta$. As it was mentioned above, in realistic atomic arrays near a fiber waveguide $\beta$ factor is on the order of $\beta\sim 10^{-2}$, which means that the optical density ${\rm OD} \sim 10$ can be already reached for $N\sim 1000$ of atoms, which is smaller by two-three orders of magnitude than in free space~\cite{Corzo2019}. 

\begin{figure*}[t]
\centering\includegraphics[width=\textwidth]{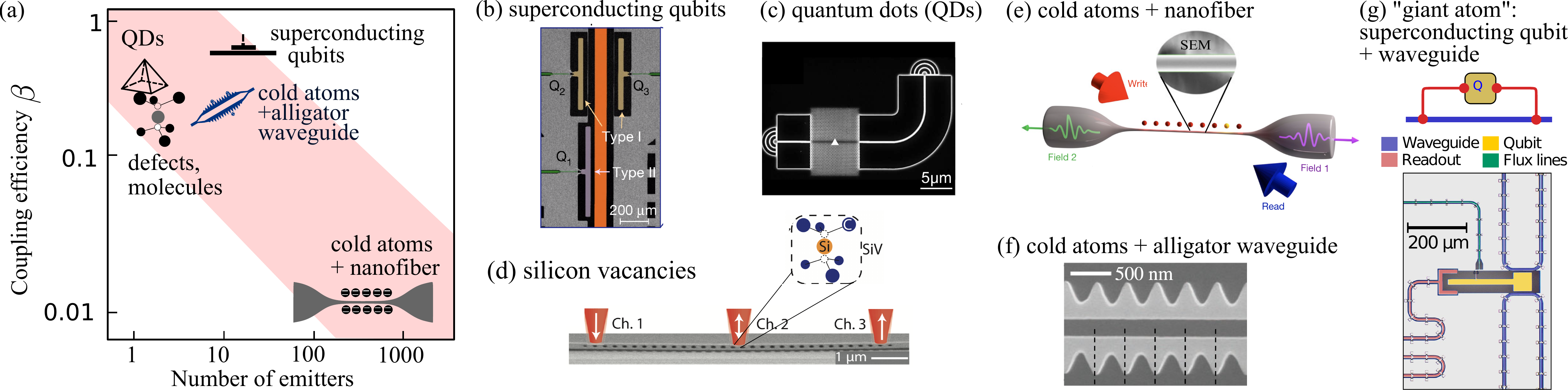}
\caption{(a) Comparison of waveguide coupling efficiency $\beta$ and number of emitters $N$ for different platforms of waveguide quantum electrodynamics. (b--e) schematic illustrations  of different platforms reproduced from Refs.: (b)~\cite{Mirhosseini2019},(c)~\cite{Foster2019},(d)~\cite{Sipahigil2016}, (e)~\cite{Corzo2019}, (f)~\cite{goban2015superradiance}, (g)~\cite{Kannan2020}. More information is given in Table~\ref{table:WQED}.}
\label{fig:platforms}
\end{figure*}

At the same time the coupling between the emitters, mediated by the waveguide photons, is enhanced and becomes long-ranged. If the position of emitters along the waveguide can be controlled, as in the case of solid-state quantum emitters and superconducting qubits, the formation of collective atomic excitations becomes important and their interaction with light can be further optimized. The basic idea is that only the array excitations that are in-phase with the guided wave couple to light efficiently. For the rest modes, the interaction is suppressed and they become subradiant as illustrated in Fig.~\ref{fig:beta}(b). 
This allows one to increase the lifetime of a stored quantum state which is beneficial for quantum information and in particular quantum memory applications. Also, subradiant states have enhanced sensitivity to external fields and suppressed decoherence, making themselves important in metrology.
We should note, that the expression ${\rm OD} = 2N\beta$ treats in fact the atomic array as a homogeneous medium, all the role of inhomogeneity is reduced to the scattering losses. The formation of  collective atomic excitations, that can have decay rates different from the free-space atomic decay rate $\Gamma_0$, is also not captured by the concept of OD. However, in disordered dilute arrays this  effect is  relatively unimportant~\cite{Chang2021refr}.

Since the concepts of waveguide-enhanced coupling, long-ranged photon-mediated interactions, collective superradiant and subradiant excitations are extensively used in the rest of the review,  in Table~\ref{table:buzz}  we have tried to summarize how these phenomena are manifested in various systems. Specifically, collective excitations can form already for two atoms in free space~ \cite{Devoe1996,Kaiser2016}, but photon-mediated coupling between two atoms decays with distance.
On the other hand, if such a pair of atoms is  placed near a waveguide, the coupling between them is enhanced and becomes long-ranged. The situation with a dense atomic array near the waveguide is more subtle. 
If the waveguide-enhanced  $\beta$-factor  remains much smaller than unity, the interaction of light with the atoms can be still considered independently, as in (1D) free space, and characterized by the concept of OD. For  large OD the transmission of photons between atoms on the opposite sides of the array is suppressed, so that the role of long-ranged interaction and collective modes is less important. 
In another words, most of the phenomena can be still described by treating the array as an effective homogeneous resonant optical medium. Thus, while the long-ranged coupling is inherent to this problem, it is not directly manifested in a dense array. 
A more detailed discussion is given  below in  Sec.~\ref{sec:single_photon}.
For large $\beta$-factor  collective subradiant modes can form and can be experimentally observed if the $\beta$-factor is high, see details in Sec.~\ref{sec:ustinov}. 
The situation drastically changes for a Bragg-spaced array, where collective superradiant mode is formed due to the waveguide-mediated interaction between all the atoms. This is somewhat similar to the cavity-QED setup, where collective states can also form, but the interaction is not long-ranged.
An ordered 2D array in free space  presents one more very flexible setup. Depending  on the lattice period one can tune the strength of atom-photon interactions for various collective modes of this array. This  will be discussed in more detail 
in Sec.~\ref{sec:2d}.

Having stated these basic principles of light-matter coupling engineering, we will now present a more specific overview of the different WQED platforms, shown in Fig.~\ref{fig:platforms}. Typical value of  parameters $N$, $\beta$ and others are listed in Table~\ref{table:WQED} and we now discuss them in more detail.

\subsection{Established platforms}\label{sec:current}

We start from artificial atoms, semiconductor quantum dots (QDs)~\cite{versteegh2014observation,Thyrrestrup2018,Foster2019,Jeannic2021}, that operate in the near infrared or visible spectral range, see Fig.~\ref{fig:platforms}(a, c). The main advantage of the QD platform is the fact that the dots are incorporated in the bulk of the photonic structure, which results in the relatively high coupling factors $\beta$ up to 99$\%$~\cite{Scarpelli2019}. 
Precise control over the position of the quantum dot within the photonic structure allows for the flexible tuning of the local field properties at the dot position which facilitates effective Purcell factor engineering~\cite{lodahl2004controlling, liu2018high}. For example, Ref.~\cite{Foster2019} has reported increase of the radiative decay rate for a quantum dot coupled to a photonic crystal waveguide by the Purcell factor of $5$. Another interesting direction in quantum optics is the chiral light-matter interaction. This  can be realized by tuning the polarization properties of the local electromagnetic field near the quantum dot \cite{Sllner2015,coles2016chirality,Lodahl2017}  and applying static magnetic field \blue{or inducing spin polarization}~\cite{javadi2018spin}. This will be discussed in more detail in Sec.~\ref{sec:chiral}. 
Quantum dots have numerous decoherence mechanisms typical for a solid state system, including charge and spin fluctuations and phonon mediated decoherence. These factors lead to fluctuations of the resonant frequency, and even when not affecting the strength of the coupling to a given photon mode  directly, they limit the coherence of the system and the  indistinguishability of emitted photons. In a very crude approximation the decoherence can be incorporated in  the extra  nonradiative decay term $\Gamma_{\rm nonrad}$ in the denominator of  the $\beta$-factor in Eq.~\eqref{eq:beta}~\cite{Arcari2014}. Thus, the $\beta$-factor \eqref{eq:beta}  depends not only on the electromagnetic properties of the environment, but also on the QD material properties and temperature. The record value $\beta\sim 99\%$ is reached at cryogenic temperatures when $\Gamma_{\rm nonrad}$ is quenched~\cite{Scarpelli2019}. 
While the nonradiative decay and decoherence can be relatively effectively suppressed by a combination of specific techniques~\cite{kuhlmann2015transform,dreessen2018suppressing}, the main challenge for the scaling of the self-organized quantum dot platform is the inhomogeneous broadening. WQED structures with a large number of QDs do not seem feasible because the strong inhomogeneous broadening  typically greatly exceeds the radiative linewidth.

\begin{table*}[bt]
\begin{tabular}{|c|ccccc|}
\hline  & Number of  &Transition & Free-space radiative  & Coupling  &\\
 Material system & resonant emitters, & energy, & linewidth (FWHM), & efficiency, &\\
& $N$ & $\hbar\omega_0$ & $\Gamma_{0}/2\pi$ & $\beta$&\\
\hline
 Cs atoms + nanofiber & $1\div 10^3$ & $1.5~\rm eV$ &  $5.2~${\rm MHz} &  $10^{-2}$ &\\
Rb atoms + nanofiber & $1\div 6$ &{$1.6~\rm eV$}&  $6.1~${\rm MHz} &  $10^{-1}$ &\\
Cs atoms + alligator waveguide  & $1\div 3$ &{$1.4~\rm eV$}&  $4.6~${\rm MHz} & $0.5$ &\\
Superconducting transmon qubits & $10$ & $0.03~{
\rm meV}$ (7~{\rm GHz}) & $10\div 100~${\rm MHz} & $0.999$ &\\
Quantum dots & 1 & $1.4~\rm {eV}$ & $0.2~\rm GHz$ & $0.99$  &\\
{Si vacancies in diamonds} & 2 & $1.7~\rm {eV}$ & $100~\rm MHz$ &$\sim 0.5$ &\\
Organic molecules & 1 & $1.6~$eV & $30~\rm MHz$ & $0.2$ &  \\
\hline \end{tabular}
\caption{Parameters of different state-of-the-art  platforms  waveguide quantum electrodynamics.  The indicated numerical values are approximate and have been taken from Refs.
\cite{Corzo2016},\cite{Solano2017},\cite{goban2015superradiance},\cite{Mirhosseini2019},\cite{Foster2019}, \cite{Sipahigil2016},\cite{faez2014coherent}, respectively. }\label{table:WQED}
\end{table*}

An alternative solid state platform is presented by solid state defects, such as silicon vacancies 
~\cite{Sipahigil2016} or germanium vacancies \cite{bhaskar2017quantum}, see Fig. \ref{fig:platforms} (d). In this case, defects can be selectively placed in diamond waveguides using the focused ion beam implantation. This results in enhancement of light-matter interaction, and Purcell factor $\Gamma_{\rm 1D}/\Gamma_0\sim 2\div 3$ as reported in Ref.~\cite{Sipahigil2016}. Inhomogeneous broadening seems to be less of an issue than for the quantum dot system: generation of entangled state of two excited qubits has already been demonstrated~\cite{Sipahigil2016}. {However, the overall coupling efficiency $\beta$ is lower than for quantum dots. We have estimated $\beta \sim 0.5$ for Ref.~\cite{Sipahigil2016} using the experimentally reported cooperativity value $C\sim 1$ that is related to the $\beta$-factor as $C=\beta/(1-\beta)$~\cite{Arcari2014}}.
The reasons for lower $\beta$-factor may involve complex energy structure of an individual vacancy with many optical transitions of close energies as well as the interaction with the phonon environment and nonradiative decay processes~\cite{Becker2017}.

Another interesting system is offered by organic molecules such as dibenzoterrylene (DBT) \cite{faez2014coherent} or terrylene~\cite{Skoff2018}, coupled to a waveguide. Similarly to quantum dots, molecule arrays exhibit strong inhomogeneous broadening, limiting scalability of the system. About $5000$ spectral lines, corresponding to different molecules, have been revealed in the experiment described in Ref.~\cite{faez2014coherent}, where the DBT molecules were put in naphthalene, filling a nanocapillary waveguide.  Individual lines could be resolved spectrally that demonstrate relatively high coupling efficiency and strong antibunching. {Apparently, the homogeneous nonradiative decay $\Gamma_{\rm nonrad}$ is not an issue for molecules. The $\beta$-factor in Ref.~\cite{faez2014coherent} has been determined solely by electromagnetic properties, the competition of the emission into the waveguide and the emission into free space. The maximum value $\beta=0.18$ has been reached for emitters positioned in the center of the fiber and the optimal fiber core radius was equal to $300$~nm.}

We now turn to superconducting transmon qubits, operating at microwave frequencies, and shown in Fig.~\ref{fig:platforms} (b).
In a simplified description, such  qubit presents a high-quality transmission-line resonator with a Josephson junction providing strong nonlinearity on a single-photon level~\cite{Koch2007,Jung2014}. The typical resonance frequency of the qubits is on the order of 5$\div$10~GHz. Thus, in order to suppress thermal noise, they need operate at the low temperatures, on the order of 10~mK.
The first experimental demonstrations of a single-photon scattering, Mollow triplet formation, generation of quantum states of microwave photons have been made  more than a decade ago~\cite{Astafiev2010,Hoi2011,Hoi2012}.
Now, superconducting qubits have become the leading architecture for quantum information processing in circuit QED~\cite{blais2020circuit}. Their main advantage is the  possibility of individual control of every qubit. This makes them also quite suitable for quantum information processing and quantum simulations in the waveguide QED, where the waveguide coupling efficiencies $\beta$ can exceed  $99.9\%$ \cite{Mirhosseini2019}. At present, most of the experimental WQED studies with superconducting qubit arrays have been focused on the  single-excited states ~\cite{Mirhosseini2019,kim2020quantum,brehm2020waveguide}.   
The reason is that it is hard to selectively access higher-excited quantum states by using just a single waveguide mode.
Moreover, large amount of higher-excited states are strongly subradiant \cite{Molmer2019,Poshakinskiy2020} and weakly coupled to the waveguide photons. In order to excite them selectively, one could drive the qubits from the side of the waveguide. This requires more complicated samples, but is technologically possible. Double-excited subradiant states in the four-qubit array have been recently observed in such way~\cite{zanner2021coherent}.

However, potential challenges, limiting the performance of state-of-the-art circuits include individual defects such as charged two-levels systems residing in the tunnel barrier of the Josephson junction or weakly coupling defects on the surfaces and interfaces of circuit electrodes~\cite{Barends2013,Burnett2019,Bilmes2020}. As a result, the  maximum coherence time of qubits is still on the order of hundreds of microseconds~\cite{Rigetti2012,Bilmes2020}.

Another waveguide quantum electrodynamics platform is presented by arrays of laser-cooled atoms of cesium ~\cite{vetsch2010optical,Goban2012,Corzo2016,Polzik2016} or rubidium \cite{Solano2017}, trapped in the vicinity of an optical nanofiber~\cite{Sile2016}, see Fig.~ \ref{fig:platforms}(e). 
The main idea is that the evanescent field surrounding the fiber creates a trapping potential for atoms near the fiber wall. The resulting values of the $\beta$-factor and the number of trapped atoms are very sensitive to the specific trap design. One of the designs includes two pairs of counter-propagating beams in the fiber, one attractive red-detuned and another repulsive blue-detuned, operating at the specific  wavelengths~\cite{LeKien2004}. The whole system is overlapped with a magneto-optical trap. When compared with the superconducting qubits platforms or with the solid-state structures, the  waveguide coupling efficiency is relatively low, $\beta\sim 1\%$. 
It is controlled by the competition of the emission into free space $\Gamma_{\rm ng}$ and into the waveguide $\Gamma_{\rm 1D}$,  $\beta=\Gamma_{\rm 1D}/(\Gamma_{\rm 1D}+\Gamma_{\rm ng})$, while the homogeneous broadening $\Gamma_{\rm nonrad}$ is negligible. Access to individual atoms near the fiber is challenging, which rules out many applications for quantum information processing.
On the positive side, the traps can host thousands of atoms. The waveguide-mediated interactions between atoms are much stronger than in free space and involve all atoms in the array while coherence remains high. The inhomogeneous broadening is weak, on the order of the free-space atom linewidth $\Gamma_0$~\cite{Corzo2016}.
This makes the fiber-coupled arrays beneficial for quantum memory applications~\cite{Corzo2019} and for generation and detection of quantum light \cite{Prasad2020}. Another possibility is opened when the external magnetic field is applied that induces chiral one-way interactions between the atoms. In this case, one can develop non-reciprocal devices and deterministic light-matter interfaces which can be useful for quantum communications~\cite{Lodahl2017}.
 
The optical trapping scheme can be tailored to decrease the distance from atoms to the nanofiber which leads to larger coupling efficiency  $\beta=0.13$~\cite{Solano2017}. However, the number of $^{87}$Rb atoms studied in this experiment has been considerably  smaller, just up to $N=6$. The atom positions have been random so that different collective super- or sub-radiant states were observed for subsequent experimental realizations. A more detailed calculation of the $\beta$-factor for realistic multilevel atoms, coupled to the fiber waveguide, is presented in Sec.~\ref{sec:real}. It shows that the maximum value reached for atoms at the waveguide surface is $\beta\sim 0.3$.

The $\beta$-factor can be increased even further up to $\beta\sim 50\%$ by replacing a nanofiber with an alligator photonic crystal waveguide illustrated in Fig.~\ref{fig:platforms}(f). The distinct near-field maxima between the "alligator scales", very close to the waveguide surface, enable efficient trapping of atoms with high coupling efficiency.  However, the delivery of atoms to submicron-size optical traps of alligator waveguide is even more challenging and requires careful engineering of the trapping beams~\cite{Beguin2020}. The number of trapped atoms realized in practice is small, for example, Refs.~\cite{goban2014atom,goban2015superradiance} reported the atomic number of $N \sim 3$ on average. In the parameter space of Fig.~\ref{fig:platforms}(a), the alligator waveguide platform seems to be closer to solid-state quantum emitters. The possibility to create photonic band gaps in the alligator waveguide also opens more possibilities to tailor light-matter interactions and create atom-photon bound states, see a more detailed discussion in the following subsection.

One more interesting waveguide QED platform is based on giant atoms, i.e. atoms that are coupled to a waveguide at multiple points which can be spaced by a wavelength distance or more, see \ref{fig:platforms} (g). The main advantage of such a system is that multiple coupling points of giant atoms give rise to interference effects that are not present in quantum optics with point-like atoms. These interference effects can lead to a coherent exchange interaction between atoms mediated by a waveguide, and it can result in suppression of relaxation of one or more atoms into the waveguide \cite{Kockum2014, Kockum2018, Kockum2020prr}. Such systems can be implemented both with superconducting qubits coupled either to microwave transmission lines~ \cite{Kannan2020} or surface acoustic waves~\cite{gustafsson2014propagating}, see Fig.~\ref{fig:platforms}(g), and cold atoms \cite{Kockum2021}.
{Specifically, 
the spontaneous decay rate for a superconducting qubit shown in Fig.~\ref{fig:platforms}(g), that is linked to the waveguide in the two  points, is proportional to $1+\cos 2\varphi$, where  $\varphi$ is a phase gained by photons travelling  between these points.  Thus, by tuning the phase $\varphi$, for example by changing the qubit resonance frequency, it is possible to control the decay rate. One can also realize configurations with braided coupling between giant atoms and the waveguide so that the atoms will be  coherently coupled to each other and at the same time protected from spontaneous decay into the waveguide~ \cite{Kannan2020}.}

A generalization of the giant atoms concept has been put forward in Ref.~\cite{Karg2019}, where it has been shown that coherent light-mediated coupling between two distant quantum systems can be realized when light interacts twice with each quantum system and the second interaction is the time reversal of the first. A proof-of-concept experiment has been reported in Ref.~\cite{Karg2020}, where a mechanical oscillator has been entangled with atomic spins located at the one meter distance due to the interaction mediated by a laser beam in a loop geometry.

More details on the specific experiments for different platforms can be found in Sec.~\ref{sec:applications}.
\subsection{Emerging waveguide QED platforms}\label{sec:outlook}
\begin{figure}[!t]
 \centering\includegraphics[width=0.48\textwidth]{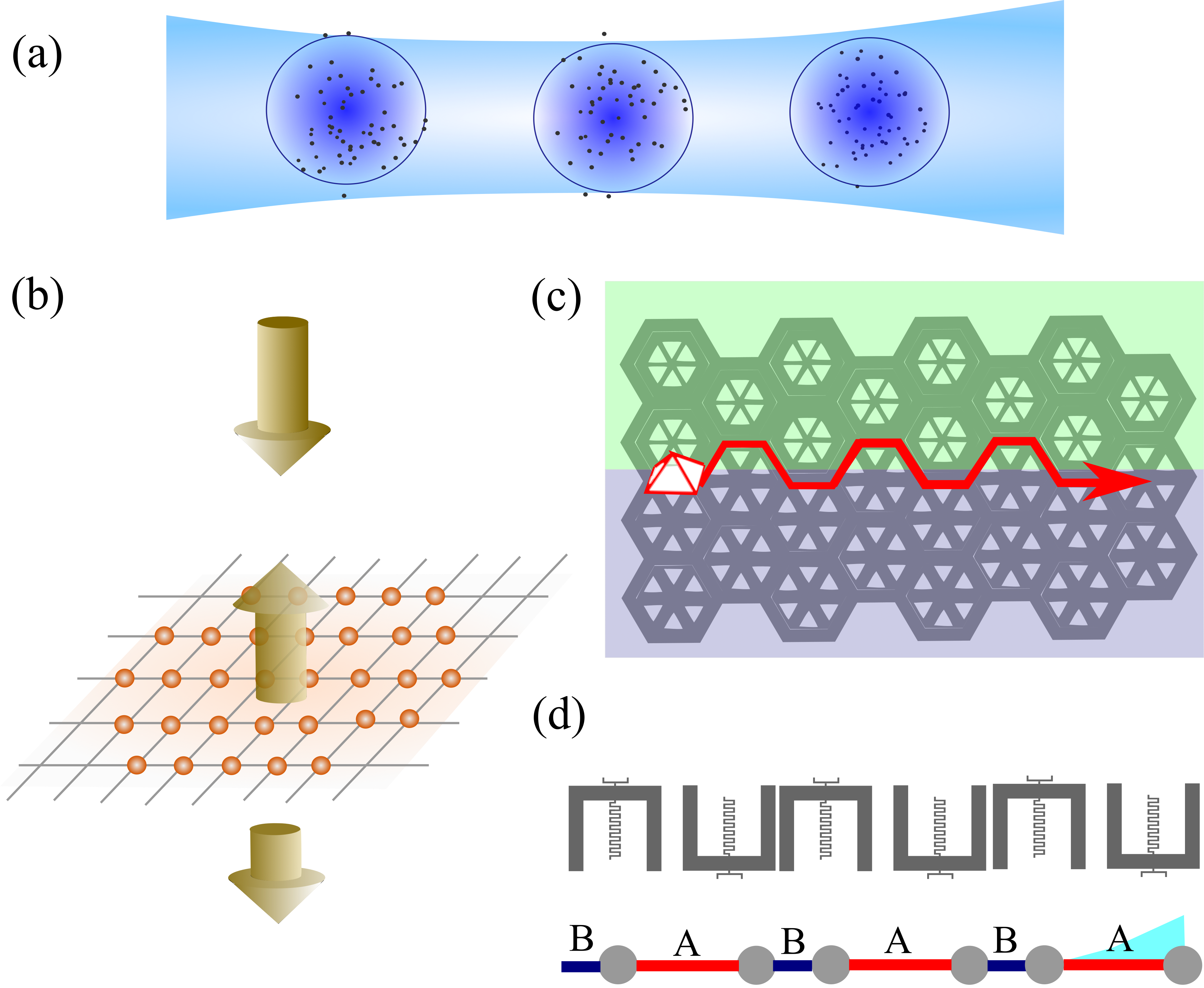}
 \caption{Schematic illustrations of potential future platforms for WQED. (a)  Three clouds of Rydberg atoms in the free space from Ref.~\cite{Stiesdal2021}. (b) Planar two-dimensional atomic array from Ref.~\cite{Rui2020} (c) 
Top: Array of superconducting LC resonators with alternating short and long spacings from Ref.~\cite{kim2020quantum}. Bottom schematic illustration of the topologically nontrivial Su-Schrieffer-Heeger model and the edge state, realized in this array.  (d). Topologically nontrivial photonic state, excited by a quantum dot and  propagating  between two photonic crystals from Ref.~\cite{Barik2018}.
  }\label{fig:superatoms}
\end{figure}

Despite tremendous achievements in waveguide QED technology, there is still  a room for improvement. We illustrate some of the emerging structures in Fig.~\ref{fig:superatoms}.

As demonstrated by the diagram in  Fig.~\ref{fig:platforms}(a), there seems to be a trade-off between the individual emitter coupling efficiency and the number of emitters, so that their product for state-of-the-art structures is roughly the same (see the shaded region). There is still a lack of structures with large number of resonant emitters $N\gtrsim 20$ that have at the same time high coupling efficiency on the order of unity so a lot of progress can be expected. One of the avenues to go in the large-$N$ large-$\beta$ direction could be offered by so-called Rydberg superatoms~\cite{Hofferberth2017,Hofferberth2018,Stiesdal2021}, see Fig. \ref{fig:superatoms} (a). Each superatom is formed by a cloud of thousands of individual atoms, e.g. $^{87}$Rb. Due to the Rydberg blockade, every cloud can absorb only one photon in a collective superradiant Dicke mode. Thus, the whole cloud acts as an effective two-level system that demonstrates characteristic two-\cite{Hofferberth2017} and three-photons \cite{Hofferberth2018} quantum correlations. Cascaded coupling of light to three clouds of Rydberg atoms has been demonstrated in Ref.~\cite{Stiesdal2021}. 
At the first glance, these three coupled atomic clouds, illustrated in Fig.~\ref{fig:superatoms}(a), have nothing in common with WQED, since they are trapped in free space and there is no waveguide at all. However, every cloud scatters almost all light in the forward direction. Thus, it is preferentially coupled just to one photon mode, and the  scattering directionality, being larger than $0.85$, plays the role of an effective $\beta$-factor. The structure from Ref.~\cite{Stiesdal2021} has been proposed for controllable substraction of up to $n=3$ photons from the input light pulse. The drawback of the setup is  unwanted scattering from the bright collective eigenstates of the cloud to its dark eigenstates. This process is more efficient than the forward emission by about an order of magnitude.

\begin{figure}[!t]
 \centering\includegraphics[width=0.48\textwidth]{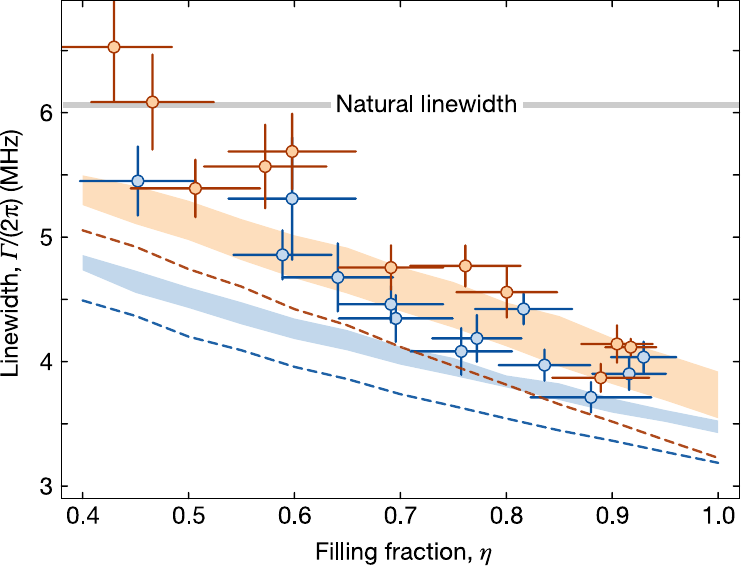}
 \caption{{Reproduced from Ref.~\cite{Rui2020}}. Linewidth of the optical resonance of the atomic array with the period $a/\lambda_0\approx 0.7$ depending on the array filling fraction. Decrease of the linewidth with the filling fraction indicates collective subradiant response.
   }\label{fig:Rui}
\end{figure}

Another idea of a ``waveguide-QED without a waveguide"  could be offered by ordered two-dimensional atomic arrays~\cite{Rui2020} in an optical lattice \cite{Bloch2012}, see  Fig.~\ref{fig:superatoms}(b).
If the spacing between atoms is smaller than the light wavelength, the array scatters light in far field only in the forward or the backward direction and light diffraction is not possible. In this case, the whole array could be viewed as an effective atom, coupled to photons propagating only in one dimension, perpendicular to the  array plane, a sort of an  atom-array ``antenna"~\cite{Yelin2017}. The parameters of such an effective atom can be controlled by changing the lattice  period. For example, subradiant behavior of the planar array of $^{87}$Rb atoms with the period $\approx 0.7\lambda_0$ has been recently demonstrated in Ref.~\cite{Rui2020}. The measured linewidth of an optical resonance depending on the filling factor of the lattice is shown in Fig.~\ref{fig:Rui}. As the filling factor increases, the measured linewidth becomes smaller and approaches the theoretical prediction $\Gamma_{\rm 2D}\approx 0.5\Gamma_{0} $, corresponding to this spacing, see also the discussion in Sec.~\ref{sec:2d}.

Constructive interference between atoms in an array enhances the coupling with light, that is beneficial for quantum memory applications. According to the study in Ref.~\cite{Manzoni2018}, the memory storage error is determined by the competition between the emission in the photonic mode propagating perpendicular to the array plane and the undesired diffraction, that is possible for the finite-size arrays. It has been predicted that the error is quickly suppressed with the number of atoms $N$ as $\log^2 N/N^2$ and an array of just $N=4\times 4$ atoms could have a storage error below  $1\%$, which is comparable to a disordered ensemble with an optical depth of around $600$. 

The studies of two-dimensional atomic arrays are now rapidly developing.
It is technologically possible  to assemble high-quality defect-free atomic lattices in arrays of tweezer microtraps~\cite{Barredo2016,Scholl2021,Ebadi2021} containing up to few hundreds of atoms. The limitation of this technique  that   the lattice spacing can not be subwavelength but is  on the order of several microns. Complicated atomic arrangements are also considered theoretically~\cite{Alaee2020,ballantine2020}.  
The unit cell with a quadrumers of atoms, each of those has the electric dipole optical transition, exhibits both electric and magnetic dipole response ~\cite{ballantine2020b}. Interference of electric and magnetic dipole emissions is constructive in the forward direction and destructive in the backward direction. As a result, the bilayer atomic array acts as a Huygens surface: it transmits light with a phase shift of $\pi$ while the light reflection is suppressed. The directional forward or backward scattering by atomic arrays, termed as the Kerker effect, has also been considered in Ref.~\cite{Alaee2020b}. Such a research direction is inspired by the recent dramatic progress in classical optics with conventional metamaterials~\cite{Kivshar2018}. The atomic arrays feature high quality resonances with vanishing inhomogeneous broadening and could be ideal for realization of complicated optical states. It seems to be only a matter of time before optical bound states of continuum \cite{Hsu2016} or high-quality subradiant states ~\cite{Koshelev2020} are realized on the atomic platforms. Collective subradiant states of the array have already been proposed to store and manipulate quantum correlations ~\cite{Jenkins2016,ballantine2020c} and engineer entanglement ~\cite{Zoller2019}.Quantum atom-made metasurfaces have  been also proposed for generation of highly entangled photon states~\cite{Bekenstein2020,Bettles2020} as will be discussed in a bit more detail in Sec.~\ref{sec:2d}.

So far, we have considered arrays of emitters coupled to conventional waveguides with linear light dispersion, or to free-space photons. However, emitters can be embedded in more complicated photonic structures. For example, the development of future WQED platforms can be inspired by topological photonics~\cite{Khanikaev2017,Ozawa2019} and we briefly review some of the considered systems below. 

First, it is possible to use the propagating topologically protected edge states of the two-dimensional photonic structure as the photonic modes, linking the quantum emitters~\cite{Barik2020,JalaliMehrabad2019,JalaliMehrabad2020}. Since the propagation of topological edge states is inherently unidirectional and robust against the backscattering on the imperfections, such structures could be beneficial for chiral quantum optics. One of the important recent milestones in this field is the demonstration of the on-chip coupling of a single semiconductor quantum dot to the topological states propagating along the boundary between two photonic crystals ~\cite{Barik2018}, see Fig.~\ref{fig:superatoms}(c). Generation of entangled photon pairs via spontaneous four-wave mixing in topological photonic crystals made of coupled ring cavities 
 and propagation of these pairs along the structure edge has been demonstrated in Ref.~\cite{Mittal2018}. This could be  potentially useful to protect quantum correlations.

 Second, it has been proposed to create topological edge states from atom-photon interactions. For example,  two-dimensional atomic arrays, subjected to perpendicular magnetic field, have been theoretically studied in Refs.~\cite{Bettles2017,Perczel2017}. Similarly to the conventional quantum Hall effect, the magnetic field leads to formation of single-photon topological edge states that propagate along the edges of the array and that are protected against the disorder. More recently, it has been proposed in Ref.~\cite{Perczel2020} to consider a lattice of nonlinear quantum emitters embedded in a photonic crystal slab. Again, this structure should feature band gaps induced by magnetic field, robust edge states, and also a nearly photonic flat band with a nonzero Chern number. Such flat band should be very sensitive to interactions and this proposal could be potentially used to probe the many-body fractional quantum Hall states in quantum optical setup.

In addition to atoms coupled by propagating edge states of photons and propagating atom-photon edge states, one can also study atoms embedded in the bulk of (topological) photonic crystals \cite{Song2018,Song2019,Mirhosseini2019,kim2020quantum}.
If the energy of quantum emitters is in the band of propagating states, subradiant and superradiant states can form, similarly to the conventional waveguide. On the other hand, if the emitter energy falls into the photonic band gap of the array it can act as a defect that leads to formation of localized photonic state bound to the emitter. This has been experimentally observed for a superconducting qubit coupled to a microwave metamaterial waveguide in Ref.~\cite{Mirhosseini2018}. 

\begin{figure}[t]
 \centering\includegraphics[width=0.48\textwidth]{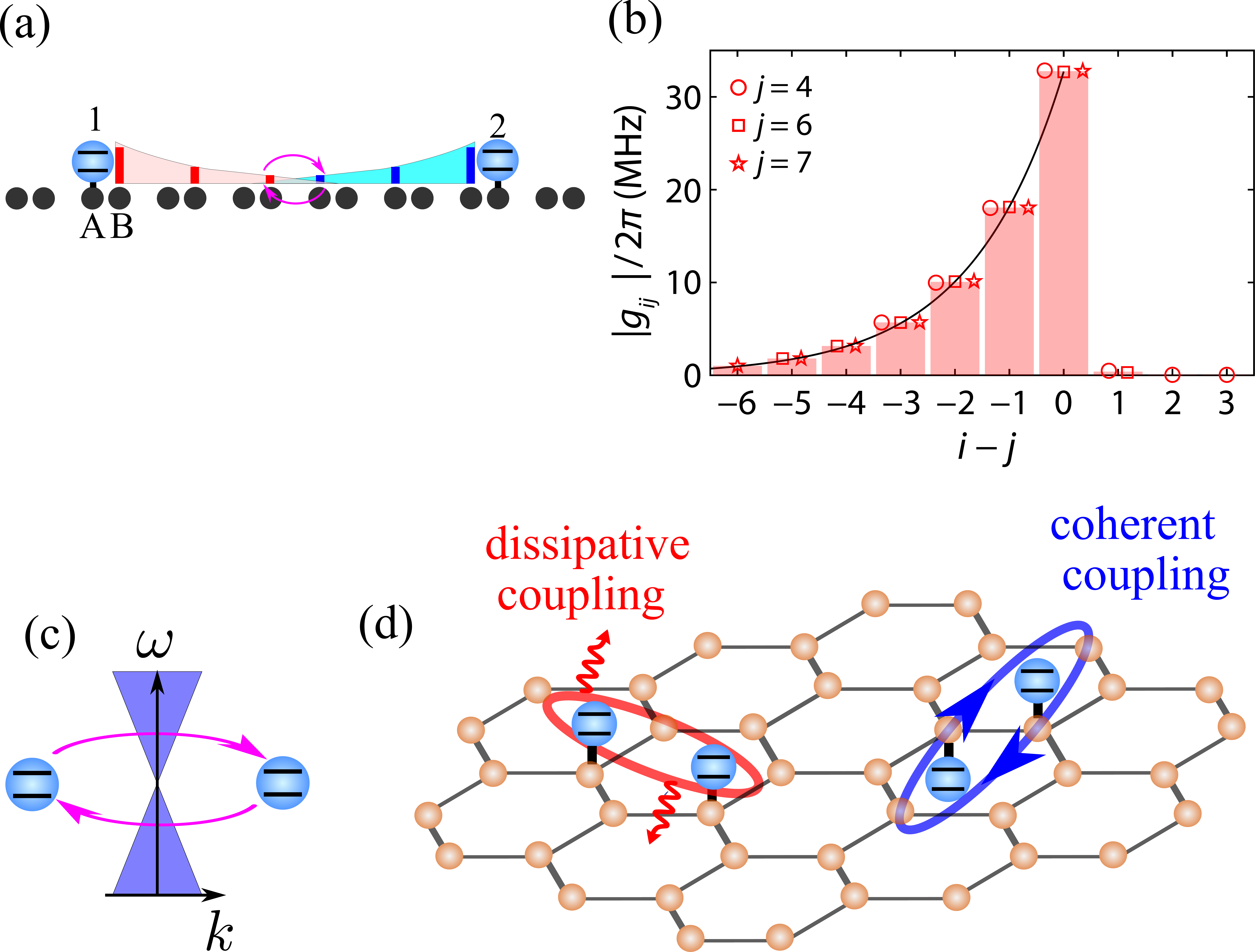}
\caption{(a) Schematics of two quantum emitters interacting via directional bound states in a Su-Schrieffer-Heeger lattice of coupled cavities. Vertical bars show the amplitudes of the photon wave function. (b) Reproduced from ~\cite{kim2020quantum}. Measured directional coupling between two emitters in the topological Su-Schrieffer-Heeger superconducting qubit array.
(c,d) Schematics of two emitters coupled via the photonic graphene bath and tuned to the Dirac point. Depending on whether emitters are in the same sublattice or different sublattices, the coupling is either dissipative or long-range coherent. Based on Ref.~\cite{Tudela2018}.} \label{fig:topo2}
\end{figure}

The situation becomes even more interesting when the emitter is embedded in a topologically nontrivial structure, such as the Su-Schrieffer-Heeger array of coupled cavities A-B=A-B=A-B\ldots ~\cite{Bello2019}, that is shown in Fig.~\ref{fig:superatoms}(d). The resulting bound states inherit the  topological features of the underlying array. First, they are directional, i.e. they decay to either to the left or to the right from the emitter depending on the sublattice the emitter is coupled to, see Fig.~\ref{fig:topo2}(a). Second,
the localized states have nonzero photon amplitudes only at one of the sublattices. The formation of photonic states bound to the emitters has been recently understood as a general feature of many topological photonic structures~\cite{Leonforte2021}. They have been termed vacancy-like states due to the following unifying feature: the photon amplitude is zero at the site, directly coupled to the emitter.

If the two emitters are coupled to different sublattices, the corresponding topological localized states can overlap and mediate their interaction,  
and the coupling will be directional as well. The proposal of Ref.~\cite{Bello2019} has been realized very recently in the array of coupled superconducting qubits~\cite{kim2020quantum}. Figure~\ref{fig:topo2}(b) shows the experimentally  measured  directional coupling depending on the relative position of the two qubits in the array $i-j$. The coupling is zero when $i>j$ and decays exponentially for $i<j$. The range of the interaction depends on the width of the band gap of the lattice, that is controlled by the differences of strong and weak couplings between the two qubits in the array. The smaller is the band gap the longer is the interaction range. Thus, one can expect very interesting physics also for the 2D setup, when the emitter is resonant with the photonic Dirac point.

Indeed, very unusual quantum optical features have been predicted for the emitter embedded in a honeycomb lattice of coupled cavities in the tight-binding model~\cite{Tudela2018}. Such structure features Dirac dispersion, similarly to the graphene, see Fig.~\ref{fig:topo2}(c). Novel effects appear even for such simple problems as spontaneous emission when the emitter is resonant with the Dirac point. Due to the vanishing density of photonic states at the Dirac point, within the Fermi Golden rule approximation, the emitter does not spontaneously decay at all. As a result, the non-Markovian effects start playing a decisive role in the spontaneous decay, and the decay kinetics remains very slow and in the infinite system the emitter population decays with time $t$ as $\propto 1/\ln^2 t$~\cite{Tudela2018}. When two emitters are resonant with the Dirac point, their interaction becomes strongly sensitive to the sublattices they are coupled to, see Fig.~\ref{fig:topo2}(d). If emitters are coupled to the same sublattice, the interaction is dissipative and collective subradiant states can form. If they are coupled to different sublattices, similarly to the case of Fig.~\ref{fig:topo2}(a), the coupling becomes coherent and can be long-ranged. The results for the discrete tight-binding model of Ref.~\cite{Tudela2018} have recently been also confirmed for a photonic crystal structures in  Ref.~\cite{Perczel2020b}. It has been predicted that such a system should feature long-ranged coherent light-mediated interactions between the emitters that are essential for exploring exotic many-body phases~\cite{Richerme2014}.

So far, most of the studies of the topological photonic structures focus on a single-photon regime. However,  Ref.~\cite{Bello2019} already considered many-body spin quantum phases emerging for array of emitters coupled to the Su-Schrieffer-Heeger array. Very recently, it has been predicted that hybridization of chiral photons in the topological two-dimensional cavity array with quantum emitters results in a whole zoo of interacting magnetic lattice models~\cite{debernardis2020lightmatter}.



\section{Light-matter interactions in  a waveguide}\label{sec:effects}
In this section, we present a general overview of light-matter interactions in a waveguide setup.

We start our consideration in Sec.~\ref{sec:symmetric} with the regular periodic arrays of atoms, symmetrically coupled to a waveguide. In Sec.~\ref{sec:input_output}, we present the effective Hamiltonian of the problem. Next, in Sec.~\ref{sec:single_photon}, we discuss a single-photon scattering and collective single-excited states of the atomic array, arising from waveguide-mediated interactions. Special attention is paid to long-living collective subradiant states in closely spaced atomic arrays and to the Bragg-spaced atomic arrays, where the period is an integer multiple of half light wavelength at the atomic resonance.
When measuring the reflected and transmitted intensities for low input power, the results are similar to those in the regime of linear classical optics. The reason is that the quantum correlations are not directly probed by intensity measurements. More precisely, the quantum scattering theory in a single photon regime yields results identical to those for a classical light scattered from e.g. a semiconductor quantum well~\cite{Ivchenko1994c} or a resonator coupled to a waveguide~\cite{Fan2002}. One more relevant but more exotic experimental realization is offered by arrays of M\"ossbauer nuclei, such as $^{57}$Fe with sharp scattering resonances in the $10$ keV spectral range~\cite{Rhlsberger2010}, see Ref.~\cite{Evers2020} for theoretical details on quantum optics with nuclei and a recent dedicated review in  \cite{Rhlsberger2021}. Thus, the  results in Sec.~\ref{sec:single_photon} for single-photon scattering spectra and eigenstates of waveguide-coupled atomic arrays could be applied to different setups with resonant scatterers coupled via a propagating photonic mode.

Light-matter interactions become especially interesting when more than one photon is present in the waveguide. Since a single atom can not resonantly absorb two identical photons at the same time due to the strong anharmonicity of atomic potential, there appears an effective photon-photon repulsion. This leads to nontrivial quantum correlations in the scattered light, photon bunching and antibunching. This is discussed in Sec.~\ref{sec:two_photons_rt}. We present a comprehensive overview of different theoretical techniques to consider the two-photon scattering. There has also been significant progress in theoretical studies of two-photon states in atomic arrays in the last couple of years, including the structure of two-photon subradiant states~\cite{Molmer2019,Albrecht2019} and an an existence of bound two-photon states~\cite{Zhang2020PRR},  we try to put these results in an universal perspective. We also briefly examine the ultrastrong coupling regime in Sec.~\ref{sec:ultrastrong}. In addition to the simplified model of idealized two-level atom, in Sec.~\ref{sec:real} we also consider a more realistic situation when an atom has a multilevel structure.
 
In the most part of this section, we discuss reciprocal symmetric  waveguides, when the atom is equally coupled to forward and backward propagating photons. However, it is  possible to make coupling  to forward and backward propagating photons asymmetric by breaking the electromagnetic reciprocity (e.g. applying external magnetic field, spin-polarizing the atoms, or using the nonlinearity). This very special chiral regime is reviewed in Sec.~\ref{sec:chiral}, where we proceed from the basics of chiral coupling (Sec.~\ref{sec:chiral_linear},Sec.~\ref{sec:rad:chiral}) to the collective polariton eigenstates (Sec.~\ref{sec:array:chiral}) to the  advanced experiments on tunable photon bunching and antibunching in this setup (Sec.~\ref{sec:Prasad}).
   
Section~\ref{sec:2d} considers a new promising platform of regular two-dimensional atomic arrays in the free space~\cite{Rui2020} that, as has been discussed in Sec.~\ref{sec:platforms}, shares a lot of similar concepts with the waveguide quantum electrodynamics. 

\subsection{Arrays with symmetric coupling}\label{sec:symmetric}
\subsubsection{General  formalism}\label{sec:general_formalism}
We start this section from a general model for light interaction with an array of atoms, that is embedded in an arbitrary structure with the dielectric permittivity $\eps(\bm r,\omega)$, where $\omega$ is the light frequency and $\bm r$ is the radius-vector. The linear electromagnetic properties of the dielectric environment can be characterized by the Green's function, satisfying the equation 
\begin{equation}\label{eq:Green}
\bm \nabla_{\bm r}\times \bm \nabla_{\bm r}\times \mathbf{G}(\bm{r},\bm{r}',\omega)=\left(\frac{\omega}{c}\right)^2 \eps(\bm r,\omega)
\mathbf{G}(\bm{r},\bm{r}',\omega)+\delta(\bm{r}-\bm{r}')\:.
\end{equation}
In vacuum, where $\eps=1$, the Green's function is given by
\begin{equation}\label{eq:G0}
G^{(0)}_{\mu\nu}(\bm r,\omega)=\left(\delta_{\mu\nu}+\left(\frac{c}{\omega}\right)^2\frac{\partial^2}{\partial x_\mu\partial x_\nu}\right)\frac{\e^{\rmi \omega r/c}}{4\pi r}\:,
\end{equation}
where $\mu,\nu=x,y,z$\:. For a waveguide, the Green's function can be separated into two parts, 
\begin{equation}\label{eq:G2parts}
\mathbf{G} = \mathbf{G}^{({\rm guid})} + \mathbf{G}^{({\rm leaky})}\:,
\end{equation}
corresponding to the interaction with guided and leaky modes. In particular, because of the translational symmetry, the wave vector along the waveguide axis $k_z$ is a good quantum number. Depending on whether $|k_z|$ is smaller or larger than $\omega/c$, the mode can either leak into free-space, or it is evanescent in the direction transverse to the waveguide,  corresponding to the two terms in Eq.~\eqref{eq:G2parts}. 
We now consider  interaction of light with an array of two-level atoms located  at the points $\bm r_m$ and having the same resonant frequency $\omega_0$. The light-atom coupling is treated in the dipole approximation, it is described by a Hamiltonian  $-\hat{\bm d} \cdot \bm E(\bm r_m)$, where $\bm E(\bm r_m)$ is the electric field at the atom and $\hat{\bm d}$ is the dipole momentum operator, 
$
\hat{\bm d}=\bm d \sigma+\bm d^*\sigma^\dag\:,
$
with
 $\bm d$ being the matrix element of electric dipole momentum between ground and excited states of the atom. Here and in the rest of the review, we consider a point-like atom, where this dipole approximation is reasonable. There also exist ``giant atoms'', based on superconducting qubits, that are connected to the waveguide in multiple distant points, see the review \cite{Kockum2021} and Ref.~\cite{Karg2019}. We also assume the Markovian approximation, which means that the photon degrees of freedom are fast and can be traced out. The effective Hamiltonian of the atomic array
 assumes the form 
\begin{equation}\label{eq:Heff-io0}
H_{\rm eff}=\sum\limits_{m=1}^N\omega_0\sigma_m^\dag \sigma_m^{\vphantom{\dag}}+
\sum\limits_{m,n=1}^N \sigma_m^\dag \sigma_{n}V_{mn}\:.
\end{equation}
We use the units with $\hbar=1$ and the Gaussian units system. The operator  $\sigma_m^\dag$ describes excitation of the atom $m$ and $\sigma_m\equiv |e_m\rangle\langle g_m|$ where  $|g_m\rangle$ and $|e_m\rangle$ are the ground and excited  states of atom $m$ and
\begin{equation}
V_{mn} =-4\pi\frac{\omega_0^2}{c^2}\bm{d}_m^*\cdot\mathbf{G}(\bm{r}_m,\bm{r}_n,\omega_0)\bm{d}_n.
\label{Sigma-renormalized}
\end{equation}
Derivation of Eq.~\eqref{Sigma-renormalized} and details of Green's tensor calculations for atoms near a realistic nanofiber waveguide can be found in Refs.~\cite{Kornovan2016,Pivovarov2018}.    The Hamiltonian Eq.~\eqref{eq:Heff-io0} describes the interactions between the atoms mediated by photons. It also assumes the  rotating wave approximation, that  holds provided that the array is excited  resonantly and the atom-photon coupling is reasonably weak.
An ultrastrong coupling regime, when the effective light-atom coupling constant is on the order of $\omega_0$ and the counter-rotating terms can not be ignored,   is considered in Sec.~\ref{sec:ultrastrong}.

The Hamiltonian Eq.~\eqref{eq:Heff-io0} assumes an especially simple form in  the fully one-dimensional case when 
the leaky part of the Green function is neglected and only one guided mode with the wave vector $k_z$ along the waveguide is taken into account in the guided part. The guided  term can be then presented as \cite{Saravi2017}
\begin{equation}\label{eq:Gguided}
G^{({\rm guid})}_{\alpha\beta}(\ve r, \ve r')=\rmi g_0\begin{cases} E_\alpha(\bm\rho)E_\beta(\bm\rho') \e^{\rmi k_z(z-z')}\:, &z>z'\:,\\E_\alpha (\ve \rho) E_\beta(\ve \rho') \e^{-\rmi k_z(z-z')}\:, &z<z'\:.\end{cases}
\end{equation}
where $\bm E(\bm \rho)$ is the electric field of the guided mode depending on  the transverse coordinates $\bm \rho=(x,y)$ and $g_0$ is a constant factor. As a result, Eq.~\eqref{eq:Heff-io0} reduces to
\begin{equation}\label{eq:Heff-io}
H_{\rm eff}=\sum\limits_{m=1}^N\omega_0\sigma_m^\dag \sigma_m^{\vphantom{\dag}}-\rmi \gamma_{\rm 1D}\
\sum\limits_{m,n=1}^N \sigma_m^\dag \sigma_{n}\e^{\rmi  k_z|z_m-z_n|}\:,
\end{equation}
where
\begin{equation} \label{eq:gamma1dGen}
\gamma_{\rm 1D}=4\pi \left(\frac{\omega_0}{c}\right)^2 \Im [{\ve d^*\cdot{ \mathbf G^{(\rm guid)} }(\ve r,\ve r, \omega_0)\ve d}]\equiv 
g_0 |\bm d \cdot \bm E(\bm\rho)|^2\:,
\end{equation}
see \cite{Gruner1996,AsenjoGarcia2017atom} for more details.
The non-Hermitian part of this Hamiltonian describes spontaneous decay due to the emission into the waveguide and $\gamma_{\rm 1D}$ is just  the spontaneous emission rate of an atom into the guided mode. 
Here and in the rest of the review, we use the small $\gamma$ letters for contributions to the imaginary part of complex eigenfrequencies, and capital  letters for the corresponding decay rates $\Gamma=2\gamma$, i.e. $\gamma_{\rm 1D}\equiv \Gamma_{\rm 1D}/2$ etc.

In this section, we consider the case of non-chiral light-atom  interaction, where emission to the left and to the right has the same probability. The chiral scenario   is analyzed in  Sec.~\ref{sec:chiral}.
The Hamiltonian in Eq.~\eqref{eq:Heff-io} explicitly demonstrates distant  long-ranged couplings between the atoms mediated by the waveguide mode.  
The non-Hermitian Hamiltonians Eq.~\eqref{eq:Heff-io0},\eqref{eq:Heff-io} are quite useful to understand  collective quasistationary eigenstates of the atomic array with one- or two-excitations, that will be considered in the following Sec.~\ref{sec:single_photon} and
Sec.~\ref{sec:two_photons_rt}.  Such eigenstates can be probed as resonances for the incident photons. The problem of photon scattering on the atomic array can be considered using the general input-output formalism, see Refs.~\cite{Blais2013,Caneva2015,Sorensen2018}. In this case, instead of using  the non-Hermitian Hamiltonian, one can also directly solve the Heisenberg equations for the atomic operators $\sigma_n$ or the master equation for the density matrix of the atomic array. In particular, Ref.~\cite{Sorensen2018} has addressed linear optical response in a general situation of multilevel atoms in an arbitrary dielectric environment. We also note that in this review we focus on the case of continuous input. We refer the reader to Refs.~\cite{Kiilerich2019,Kiilerich2020} and references therein for an application  of cascaded quantum theory
 \cite{Gardiner1993} to study interaction of quantum systems with the pulses of radiation.

In case with many excitations, both the eigenproblem and scattering  problem are quite difficult due to  the large size of the Hilbert space. In few particular cases, for example, when all the atoms are located in one point, the scattering problem can be solved analytically, as will be discussed in Sec.~\ref{sec:two_photons_rt}. A more general case can be attacked by the matrix product state (MPS) approach, that is an established and very powerful method to consider many-body effects in one-dimensional condensed matter systems \cite{Schollwock2011,Orus2014}. The MPS technique is based on the representation of the wavefunction of many-body quantum state as a product of auxiliary matrices $A$:
\begin{equation}\label{eq:MPS}
\psi(s_1,s_2,s_3\ldots)=A_{\alpha_0\alpha_1}(s_1)A_{\alpha_1\alpha_2}(s_2)A_{\alpha_2\alpha_3}(s_3)\ldots,
\end{equation}
where the indices $\alpha$ run in a finite range, $\alpha=1\ldots M$ and the indices $s_1,s_2\ldots$ describe quantum states of different particles 1,2\ldots In the case when $M = 1$, the wavefunction factorizes, which means that the particles are independent. In the case when $M > 1$, the quantum states of different particles become entangled with each other. It has been proven that the ansatz Eq.~\eqref{eq:MPS} convergences quite fast for the nondegenerate ground state of quantum systems with nearest-neighbor interactions, such as spin chains~\cite{Orus2014}. The Markovian MPS technique has been successfully applied to model atomic interactions in the waveguides formed by coupled cavities~\cite{Manzoni2017}. Photon scattering on an atomic array with the Langevin-MPS formalism has been considered in \cite{Roy2018}.
In order to go beyond the Markovian regime and take into account retarded long-ranged interactions, it has been proposed to discretize the problem in space and time~\cite{Grimsmo2015,Pichler2016}. The MPS technique can also be applied in the ultrastrong coupling regime~\cite{peropadre2013nonequilibrium}, see Sec.~\ref{sec:ultrastrong} for more details.
A detailed recent comparison of the MPS approach with another powerful technique, quantum trajectories method, can be found in \cite{regidor2020modelling}. This area is now rapidly developing and we expect that further powerful and practical calculation tools will soon become available.

\subsubsection{Single-excited states}\label{sec:single_photon}
\paragraph{Polariton eigenstates.}
\begin{figure}[t!]
\centering\includegraphics[width=0.45\textwidth]{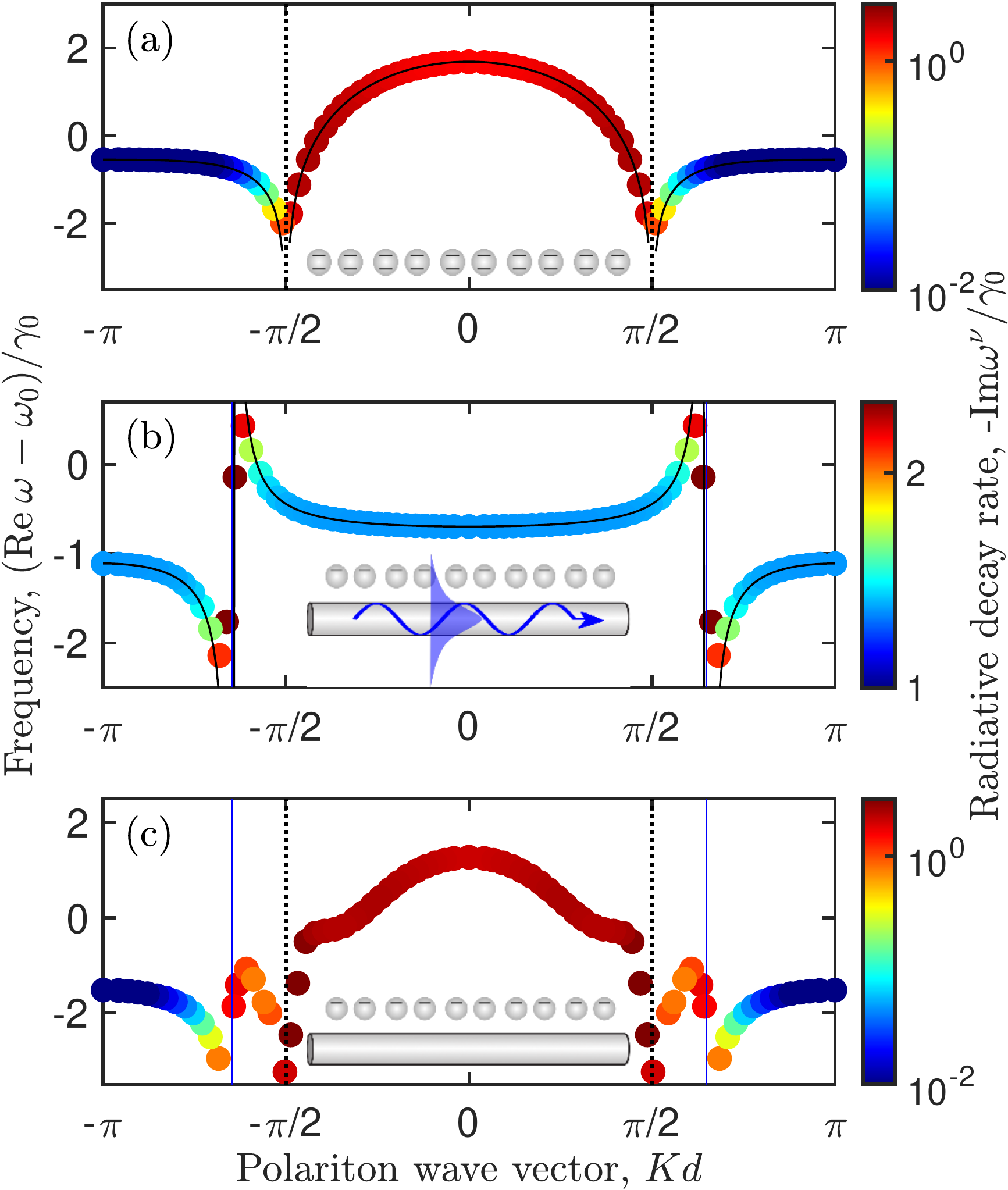}
\caption{Real and imaginary parts of the eigenmodes  calculated for (a) 
atomics array in free space (b) atomic array near the fiber waveguide neglecting  collective emission into the free space 
(c) full model. Vertical dotted lines in (a,c) show the light cone boundaries, $K=\pm \omega/c$. Vertical blue line in (b) shows the guided mode wave vector, $\pm \beta$. Black curves in (a,b) show the analytical results Eq.~\eqref{eq:polylog},  Eq.~\eqref{eq:Kd} for an infinite array. The eigenmodes have been calculated for $N=40$ atoms polarized perpendicularly to the fiber. The polariton wave vector has been extracted from the Fourier transform of the eigenmodes. Other parameters are given in text.
}
\label{fig:Kd}
\end{figure}

The effective  Hamiltonian Eq.~\eqref{eq:Heff-io0} commutes with the total number of excitations operator $\sum_{m=1}^N\sigma_m^\dag \sigma_m$. Thus, subspaces with different excitations numbers  can be analyzed separately. In this section, we consider  single-excited eigenstates. They can be found by projecting the full Hamiltonian Eq.~\eqref{eq:Heff-io0} onto the states $\sigma_n|0\rangle $, $H_{mn}=\langle m|H|n\rangle$. In Fig.~\ref{fig:Kd}, we present the real and imaginary parts of the eigenfrequencies $\omega$, found by diagonalizing the matrix $H_{mn}$, see \cite{Kornovan2016}. Three panels (a,b,c) correspond to (a) an array with $40$ atoms in free space (b) same  array interacting only via the waveguide mode, and neglecting collective coupling with the free space and (c) full calculation for an array  located close to the dielectric fiber waveguide. The parameters of calculation follow Ref.~\cite{AsenjoGarcia2017exp}: fiber permittivity is $4$, $\omega_0d/c=\pi/2$,  $\omega_0R/c=1.2$, where $d$ and $R$ are the array period and the waveguide radius, distance of the array from the fiber center is $\rho=1.5 R$, $k_z\approx 1.56 \omega_0/c$ is the guided mode wave vector. We now discuss these results in more detail.

The calculation in Fig.~\ref{fig:Kd}(a) has been performed including only the free-space Green function \eqref{eq:G0} into Eq.~\eqref{Sigma-renormalized} determining the effective Hamiltonian. The distribution of the real parts of the eigenfrequencies can be understood by comparing with the eigenfrequencies for an infinite array, where the eigenstates  are characterized with the wavevector  $K(\omega)$ and can be sought in the form $\sum_n\e^{\rmi Kn}\sigma_n^\dag |0\rangle$.  Their eigenfrequencies are well described by the following analytical expression~\cite{AsenjoGarcia2017exp} , that is shown by a black curve in  Fig.~\ref{fig:Kd}(a):
\begin{multline}\label{eq:polylog}
\Re \omega(K)=\omega_0+\frac{3\gamma_0}{4(\omega_0d/c)^3}\Re\sum_{\sigma=\pm}[\mathop{\mathrm{Li}_3}(\xi_{\sigma})\\-\rmi(\omega_0d/c)\mathop{\mathrm{Li}_2}(\xi_{\sigma})+(\omega_0d/c)^2 \ln(1-\xi_{\sigma})]\:,
\end{multline}
where $\xi_{\pm}=\exp[\rmi (\omega_0d/c)\pm\rmi Kd/c]$ and $\mathop{\mathrm{Li}}$ are the polylogarithm functions.

The radiative decay rate of the eigenmodes, denoted by color in Fig.~\ref{fig:Kd}(a),  strongly depends on the polariton wave vector along the array, $K$. Namely, the emission in the direction transverse  to the array  is suppressed when $|K|>\omega_0/c$, that is outside the light cone. In the infinite array, the decay rate for $|K|>\omega_0/c$ would have been exactly zero. In the finite array, emission is still possible at the array edge, in the longitudinal direction but the eigenstates are strongly subradiant. The most subradiant eigenstates are close to the Brillouin zone edge, $|K|\approx \pm\pi/d$. The spontaneous emission rate quickly decreases with the array size, as $-\Im \omega\propto \Gamma_0/N^3$~\cite{Zhang2020d}. Moreover, it  has been proven in Ref.~\cite{Zhang2020d}, that such decay law is universal. If the dispersion law close to the band edge behaves as $\omega(K)-\omega(\pi/d)\propto (K-\pi/d)^s$,  the emission rate decreases as $1/N^{s+1}$. This result has a transparent interpretation: the larger is the power $s$  the ``heavier``  are the polaritons and the harder it is for them to escape the array~\cite{Figotin2011,poddubny2020quasiflat}.

We now take into account interaction with the guided mode. In order to elucidate the role of the guided mode, we first use a simplified Hamiltonian \[H_{\rm eff}+(\delta\omega-\rmi\gamma) \sum_{n=1}^N\sigma^\dag_n\sigma_n\:,\] where $H_{\rm eff}$ is given by Eq.~\eqref{eq:Heff-io} and the $\delta\omega-\rmi\gamma$ term describes  the shift of the individual atom resonance  and the modification of its decay rate due to the interaction with non-guided mode. In another words, such model takes into account the Purcell factor for individual atoms, and their collective coupling through the waveguide mode, but ignores the collective emission into the free space.  The real part of the eigenfrequencies is well described by the dispersion law~\cite{Mahan1969}
\begin{equation}
\omega-\omega_0-\delta\omega+\rmi\gamma=\GO \frac{\sin \varphi}{\cos Kd-\cos \varphi}\:.
\label{eq:Kd}
\end{equation}
where $\varphi=k_zd$ is determined by the wave vector of the guided mode. Equation \eqref{eq:Kd} is relatively easy to obtain by looking for the eigenstates of Eq.~\eqref{eq:Heff-io} in the form $\sum_n\e^{\rmi Kn}\sigma_n^\dag |0\rangle$. The infinite sum over $n$ can be split into two parts, for $n<m$ and for $n\ge m$ and each part is just a geometric series. In the case when the period of the atomic array is much smaller than the wavelength, $\omega d/c\ll 1$, Eq.~\eqref{eq:Kd} can be approximately written as $K^2=(\omega/c)^2\eps(\omega)$
where \cite{Ivchenko1991}
\begin{equation}\label{eq:epsilon-eff}
\eps(\omega)=1+\frac{2\gamma_{\rm 1D}}{\varphi(\omega_0+\delta\omega-\omega-\rmi\gamma)}
\end{equation}
is the effective permittivity.  The more dense is the array the smaller is $\varphi$ and the stronger  is the resonance in the permittivity Eq.~\eqref{eq:epsilon-eff}.
We note, that while the permittivity Eq.~\eqref{eq:epsilon-eff} captures the enhancement of light-matter interaction due to the waveguide, it is a local characteristic. So, the long-ranged waveguide-mediated interactions, while being inherent to the Hamiltonian \eqref{eq:Heff-io}, are not directly manifested in the dense arrays (see also the discussion of Table~\ref{table:buzz} above).

The dispersion curve manifests a characteristic avoided crossing of the free light dispersion $K=\omega/c$ with the atomic resonance. The polaritonic band gap is located in the frequency range 
\begin{equation} \label{eq:gap}
\omega_0+\delta\omega-\gamma_{\rm 1D}\tan\frac{\varphi}{2}<\omega<
\omega_0+\delta\omega-\gamma_{\rm 1D}\cot\frac{\varphi}{2}\:.
\end{equation}
The radiative decay rate dependence on $K$ is very different from the free-space case. The largest decay rate corresponds to the states with $|K|\approx k_z$, that are in-phase with the guided mode. We note, that the eigenstates of the sole  guided Hamiltonian Eq.~\eqref{eq:Heff-io}  are also strongly subradiant for $K\approx \pm \pi/d$ and obey the save universal $-\Im \omega\propto 1/N^3$ scaling with the  array size. The only reason the decay rates are not small in Fig.~\ref{fig:Kd}(b) is the present of the constant free-space emission term $\gamma$.

In fact, as has been pointed out in Ref.~\cite{AsenjoGarcia2017exp}, using the constant term to describe  interaction with free-space  photons is only a  crude approximation. The results of full calculation, including exactly both free-space and guided modes and following Ref.~\cite{Kornovan2016, AsenjoGarcia2017exp}, are presented in Fig.~\ref{fig:Kd}(c). The result inherits the features of both Fig.~\ref{fig:Kd}(a) and Fig.~\ref{fig:Kd}(b). Inside the light cone the dispersion of the eigenmodes is mostly due to free-space interactions and the spontaneous decay rate is large.  There also is  a resonant feature  at the guided mode wave vector, shown by thin vertical blue line in  Fig.~\ref{fig:Kd}(c). It is absent
for an array in free-space and stems from the long-ranged waveguide-mediated couplings between the atoms. Close to the Brillouin zone edges, $K\approx \pm\pi/d$, the modes are strongly subradiant. As has been suggested in
Ref.~\cite{Kornovan2019, AsenjoGarcia2017exp}, these strongly  subradiant modes could be relevant for quantum memory applications. While for parameters of Ref.~\cite{AsenjoGarcia2017exp} the decay modes obey the usual $1/N^3$ scaling, the general results of Ref.~\cite{Zhang2020d}, mentioned above, suggest that by engineering the polariton  dispersion it should be possible to further suppress the decay rate. For example, it has been numerically found in Ref.~\cite{Kornovan2019} that the collective emission rate can be strongly suppressed for a certain lattice period $d/\lambda_0 \approx 0.24\times 2\pi c/\omega_0 $ and a specific fiber permittivity. 
The scaling $\sim 1/N^8$ has been extracted from the results of numerical calculation. By varying the lattice period $d$, one can achieve the degenerate band edge condition. Namely, the dispersion curve at the band edge appears to have quartic rather than quadratic dependence on the wave vector, \cite{Zhang2020d, Figotin2011}, resulting in a $1/N^5$ scaling. Moreover, by further tuning of the distance parameter $d(N)$ for each array size $N$, even stronger suppression of radiation can be observed due to destructive interference of two band edge modes \cite{Kornovan2021}. However, once the atoms are distributed in a non-periodic manner, even exponential decays of the emission rate become possible by forming the Bragg-mirror atomic cavities \cite{AsenjoGarcia2017exp}. 

We should also mention here that even a small disorder in the atomic array leads to significant modification of the collective decay rate. As it was shown in \cite{Kornovan2019}, the disorder $\sim 10^{-3}\lambda_0$ modifies the collective decay rate scaling to $1/N^{3.7}$.

\paragraph{Eigenmodes mediated by the interaction via the waveguide. }
We now consider in more detail the case, when the interaction with the free-space modes can be totally neglected and the coupling between the atoms in the array is fully determined by the interaction with the guided mode. We also assume for simplicity that the wavevector of the guided mode is equal to $k_z=\omega/c$. The effective Hamiltonian  Eq.~\eqref{eq:Heff-io} in the subspace of single-excited states $\sigma_n^\dag |0\rangle$, where $|0\rangle\equiv |gg\ldots g\rangle$  is the ground state of all the atoms, is then reduced to the following matrix ~\cite{Ivchenko2005,Caneva2015}
\begin{equation}\label{eq:Hmn}
H_{mn}=\omega_0\delta_{mn}-\rmi \gamma_{\rm 1D} \e^{\rmi\varphi |m-n|}\:,
\end{equation}
where $\varphi=\omega_0d/c$.
Such simplified  model   allows many instructive  analytical solutions. It is also directly applicable to arrays of superconducting qubits,  where the $\beta$-factor is close to unity and interaction with free-space modes can be ignored. 

\begin{figure}[t!]
\centering\includegraphics[width=0.47\textwidth]{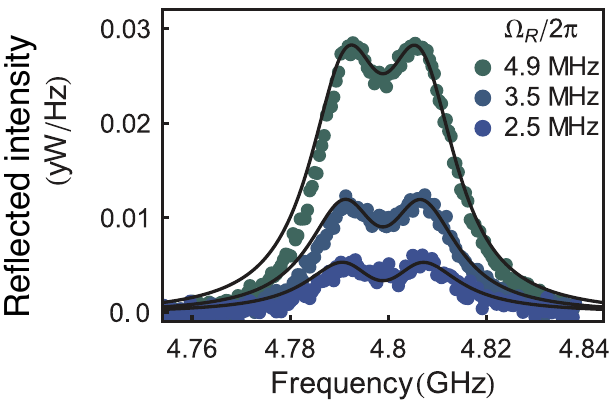}
\caption{Reproduced from Ref.~\cite{vanLoo2013}. Spectrum of light reflection from two superconducting qubits with the spacing $3\lambda(\omega_0)/4$ coupled to the waveguide. Different curves correspond to different pumping strengths characterized by the Rabi frequencies $\Omega_R$.}
\label{fig:vanloo}
\end{figure}

We start the analysis of the waveguide-mediated coupling,  from the illustrative case of $N=2$ qubits, where the matrix 
 Eq.~\eqref{eq:Hmn} assumes the form
 \begin{equation}
 H=\begin{pmatrix}
 \omega_0-\rmi( \gamma_{\rm 1D}+\gamma)
 & -\rmi \gamma_{\rm 1D} \e^{\rmi \phi}\\
 -\rmi \gamma_{\rm 1D} \e^{\rmi \phi} &
  \omega_0-\rmi( \gamma_{\rm 1D}+\gamma)
 \end{pmatrix}\:.
 \end{equation}
 The eigenfrequencies are given by
\begin{equation} \label{eq:wpm}
\omega_\pm=\omega_0 - \rmi\gamma -\rmi (\gamma_{\rm 1D}\pm  \gamma_{\rm 1D}\e^{\rmi \phi})\:.
 \end{equation}
and the eigenvectors correspond to symmetric and antisymmetric excitation, $[1,\pm 1]/\sqrt{2}$.  The splitting between the eigenfrequencies of the qubit array Eq.~\eqref{eq:wpm} can be observed experimentally by measuring the reflection spectra.
This has been first done for  two $3\lambda(\omega_0)/4$-spaced superconducting qubits, coupled to a waveguide, has been first measured in Ref.~\cite{vanLoo2013} and is illustrated in Fig.~\ref{fig:vanloo}.

\begin{figure}[tb]
\centering\includegraphics[width=0.47\textwidth]{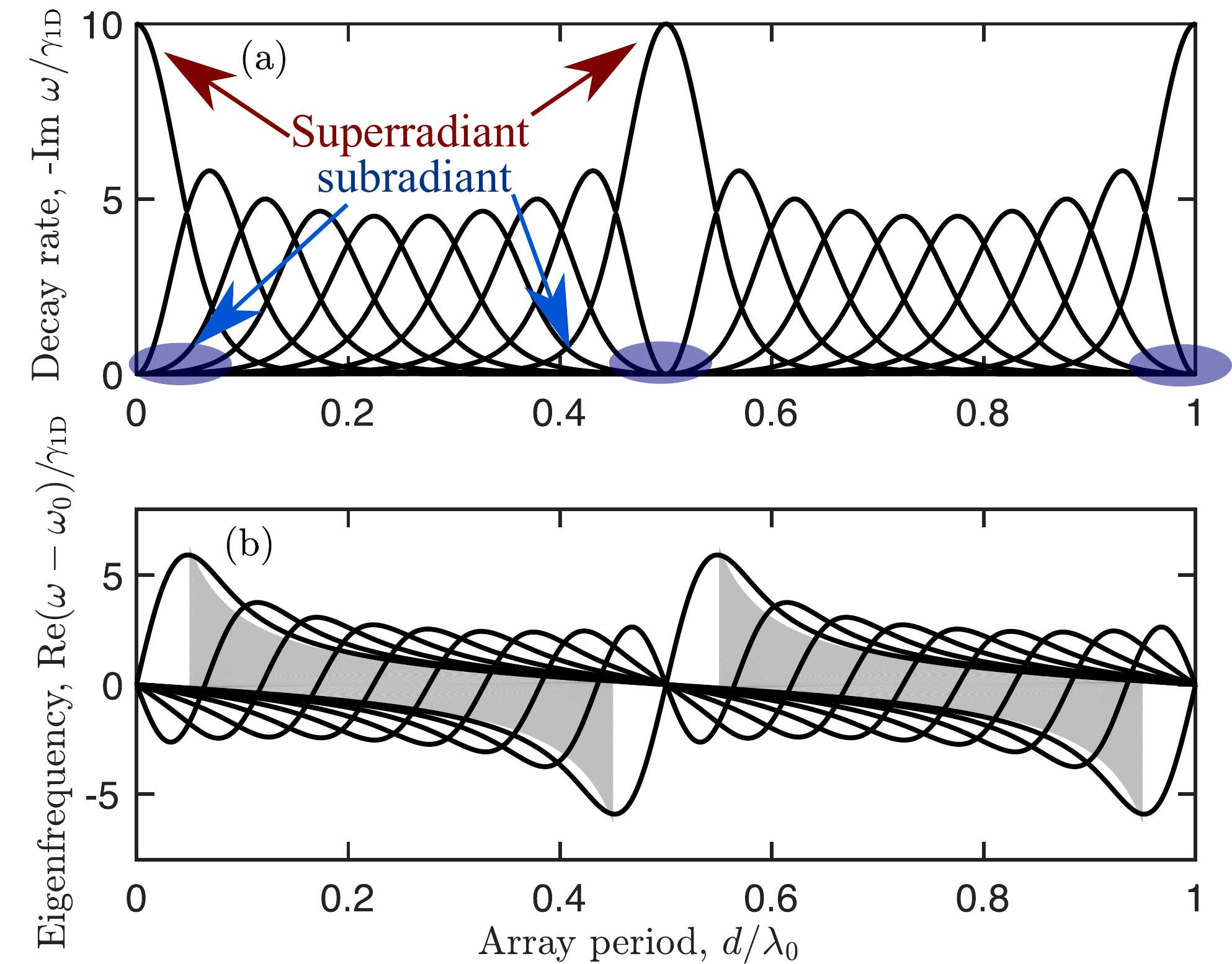}
\caption{Imaginary (a) and real (b) parts of the complex eigenfrequencies of the array of $N=10$ atoms coupled to a waveguide depending on the period of the array $d$. Shaded areas (b) show the edges of the polariton band gaps. Calculated by diagonalization of the Hamiltonian Eq.~\eqref{eq:Hmn} for vanishing nonradiative decay rate $\Gamma$. Each value of $d/\lambda$ corresponds to $N=10$ eigenvalues.}
\label{fig:subrad}
\end{figure}

The dependence of the real and imaginary parts of the eigenfrequencies 
$\omega^\nu$ for $N=10$ qubits on the array period $d$ is shown  in Fig.~\ref{fig:subrad}. The real parts of the $N=10$ eigenfrequencies in Fig.~\ref{fig:subrad}(b) periodically depend on $d$ and concentrate near the two band gap edges, given by Eq.~\eqref{eq:gap} for $\delta\omega=\gamma=0$ and $k_zd=\varphi$.
The matrix $H_{mn}$ is non-Hermitian but symmetric, which leads to the non-conjugated orthogonality condition $\sum_{n=1}^{N}P^\nu_nP^\mu_n=\delta_{\nu,\mu}$ for the eigenvectors $P^\nu_n$.
The exists an exact analytical representation of the eigenvectors  as a superposition of two polariton Bloch waves \cite{Voronov2007JLu}:
\begin{equation}
\label{eq:eigpol}
P_n\propto \rho \e^{\rmi Kn}+\e^{-\rmi Kn}\propto
\e^{\rmi K(n-N-1)}+\rho \e^{-\rmi K(n-N-1)},
\end{equation}
where $\rho=-(1-\e^{\rmi (\varphi-K)})/(1-\e^{\rmi (\varphi+K)})$
is the reflection coefficient of polaritons from the internal boundary of the array and $K$ is the polariton eigenvector found from Eq.~\eqref{eq:Kd} with $\delta\omega=\gamma=0$ and $k_zd=\varphi$. Two representations in Eq.~\eqref{eq:eigpol} are equivalent because an analog of the Fabry-Perot condition  $\rho^2\e^{2\rmi K(N+1)}=1$ holds at the eigenfrequencies $\omega^\nu$. 

 The eigenfrequencies $\omega_\nu$ are complex, and the decay rates of the corresponding modes is equal to $-2\Im \omega^\nu$. The center of mass of the eigenmodes does not depend on the spacing and is equal to $\omega_{0}-\rmi \gamma_{\rm 1D}
$~\cite{vladimirova1998ru}. In the limiting  case when $d\to 0$, we obtain a single superradiant mode with
$P_{{\rm SR},n}=1/\sqrt{N}$ and the eigenfrequency 
$
\omega_{\rm SR}=\omega_0-\rmi N\gamma_{\rm 1D}\:.
$. All the other $N-1$ modes are degenerate for $d=0$ with the eigenfrequency $\omega_0$. Their eigenvectors are found from the condition $\sum_{n=1}^NP_{n}=0$ and all these modes are dark, i.e. can not be excited by the waveguide mode. When the spacing between atoms increases, $d\ne 0$, the dark modes stop being degenerate and acquire finite radiative lifetime, as can be seen from  Fig.~\ref{fig:subrad}(a).  
When $d\ll \lambda$ ($\varphi\ll 1$), the radiative decay rate of the darkest subradiant modes is approximately given by 
\begin{equation}
\label{eq:sub1}
-\Im \omega^
\nu=\gamma_{\rm 1D}\frac{\pi^2\varphi^2 \nu^2}{8N^3},\quad \nu=1,2\ldots\ll N\:,
\end{equation} 
see \cite{vladimirova1998ru,Molmer2019}. Hence, the radiative decay rate is suppressed by the factor on the order of $\varphi^2\sim [d/\lambda(\omega_0)]^2$. Moreover, when the array size increases, the radiative decay rate further decreases as $1/N^3$~\cite{Albrecht2019,Zhang2020d} as has been discussed earlier in this section.
For any given value of 
$d/\lambda$, most of the 10 points in Fig.~\ref{fig:subrad}(a) are condensed near the abscissa axis and merge with each other, corresponding to strongly subradiant modes. The eigenvectors of subradiant modes can be found from Eq.~\eqref{eq:eigpol} as
\begin{equation}
\label{eq:Psingle}
P_n\approx (-1)^N\sqrt{\tfrac{2}{N}} \sin [(\pi-k)(n-\tfrac{1}{2})],\quad n=1,2,\ldots N,
\end{equation}
where $k=\pi-\pi/N,\pi-2\pi/N\ldots$. They are  standing waves with the wave vectors close to the edge of the Brillouin zone. The lifetimes of the eigenmodes depend periodically on the  array period, as shown in Fig.~\ref{fig:subrad}.
The situation with one superradiant mode and $N-1$ dark modes is also realized for the Bragg-spaced arrays with $2d/\lambda_0=1,2,\ldots $. In this case, the polariton band gap is enhanced, but should be calculated beyond the Markovian approximation. Physics of Bragg-spaced arrays is discussed in more detail later in this section.

It is instructive to note that the inverse to the shifted matrix Eq.~\eqref{eq:Hmn}
is a tri-diagonal one~\cite{poddubny2020quasiflat}, 
\begin{align}\label{eq:iH}
&\widetilde H\equiv (H-\omega_0)^{-1}\\=&\frac{1}{\gamma_{\rm 1D}}\begin{pmatrix}
 -\frac{1}{2}\cot \phi+\frac{\rmi}{2}&\frac1{2\sin\phi}&0&\ldots\\
 \frac1{2\sin\phi}&-\cot \phi&\frac1{2\sin\phi}&\ldots\\
0&  \frac1{2\sin\phi}&-\cot \phi&\frac1{2\sin\phi}\ldots\\
&&\ddots&\\
  \ldots& \frac1{2\sin\phi}&-\cot \phi& \frac1{2\sin\phi}\\
 \ldots&0& \frac1{2\sin\phi}& -\frac{1}{2}\cot \phi+\frac{\rmi}{2}
 \end{pmatrix}\nonumber\:.
  \end{align}
  Identity  \eqref{eq:iH} allows one to reduce  the infinite-range Hamiltonian Eq.~\eqref{eq:Hmn} to a usual tight-binding ``Hamiltonian'' $\widetilde H$  with the nearest-neighbor couplings with and the eigenvalues $1/(\omega-\omega_0)$\:.
While the Eq.~\eqref{eq:iH} is not  obvious,
it can be easily checked that it is compatible with the  polariton dispersion law $\omega(K)$ Eq.~\eqref{eq:Kd} in the infinite structure (for $\delta \omega=\gamma=0$). Namely, according to Eq.~\eqref{eq:Kd}   $1/[\omega(K)-\omega_0]$ is proportional to $\cos K-\cos \varphi$. The $\cos K$ dispersion law corresponds to a tight-binding model with nearest neighbor couplings~\cite{bernevig2013}, which is exactly Eq.~\eqref{eq:iH}. So the infinite-range Hamiltonian Eq.~\eqref{eq:Hmn} could be viewed as a tight-binding Hamiltonian \eqref{eq:iH} in disguise. The distinction between long-ranged and tight-binding situations becomes more clear  in the Bragg structure, with $\sin\varphi=0$, where Eq.~\eqref{eq:iH} is not applicable. Thus, the Bragg-spaced array seems to present the most clear manifestation of the long-ranged interaction.

The only nonzero imaginary elements of the matrix Eq.~\eqref{eq:iH} are at the corners,  for $n=m=1$ and $n=m=N$. This is why the radiative decay rate vanishes for an infinite array: radiative losses are only through the array edges.
It can be seen from Eq.~\eqref{eq:Psingle} that the edge values of the wave function are on the order of $P_1=P_N\propto N^{-3/2}$. The radiative decay rate is obtained in the first order of perturbation theory in the imaginary matrix elements $\Im \widetilde H_{n,m}$ of Eq.~\eqref{eq:iH}. Namely, it is proportional to $\Im \widetilde H_{1,1} P_1^2\propto N^{-3}$. This is an alternative way to obtained   the $\sim N^{-3}$ scaling  of the radiative decay rate.

The approach based on the non-Hermitian Hamiltonian Eq.~\eqref{eq:Hmn} 
can be also generalized to the non-Markovian case.  To this end the  phases $\varphi|m-n|\equiv \omega_0 |z_m-z_n|/c$ should be replaced by $\omega |z_m-z_n|/c$.  Physically, the Markovian approximation assumes that the light flight time is much smaller than all the other relevant time scales in the system. So  it is typically valid for  photon scattering on a single atom or on an array of closely-spaced atoms. However, non-Markovian effects are still possible in this case. For example, they arise due to the finite bandwidth of the infinite photon wavepacket, as has been studied in Ref.~\cite{Ciccarello2018}.

\paragraph{Bragg-spaced arrays. }
{
The  situation when the atomic resonance frequency and array period $d$  satisfy the resonant Bragg condition,
\begin{equation}\label{eq:Bragg_cond}
d=\frac{m}{2}\lambda(\omega_0),\quad m=1,2\ldots\:,
\end{equation}
deserves a special attention.
In this case, the incident wave exhibits not only resonant reflection from each individual atom, but also the Bragg diffraction: waves reflected from different atoms   interfere constructively. Calculation in Fig.~\ref{fig:subrad}(b) indicates that the width of the polariton band gap increases when the array period approaches the resonant Bragg condition Eq.~\eqref{eq:Bragg_cond}. 

Bragg diffraction in arrays of resonant scatterers has been studied in very different setups.} Probably, historically the first platform is presented by natural crystals, such as iron, where sharp resonances with the widths on the order of neV exist for $\gamma$-rays ($\hbar\omega\approx 14$~keV) exhibiting M\"ossbauer scattering on the nuclei. Such crystals are experimentally studied since the 1960s, see the reviews  Refs.~\cite{kagan1999,hannon1999}.  One has also considered artificial Bragg lattices for $\gamma$-rays, made from alternating layers of different isotopes~\cite{chumakov1993}. This field has recently experienced a lot of progress \cite{Rhlsberger2010,Rhlsberger2012,Haber2017,Haber2019} with the advent of high-brilliance synchrotron radiation sources , see also a  review\cite{Rhlsberger2021}. While initially the researchers have mainly studied the angular dependence of the reflectivity instead of its spectral properties~\cite{chumakov1993}, modern technologies have enabled high-resolution spectroscopic demonstration of Bragg reflection form nuclear multilayers \cite{Haber2016}. 
In the 1990s, it has been independently proposed to use Bragg-spaced lattices of semiconductor quantum wells ~\cite{Ivchenko1994,Ivchenko1994c} and optical lattices of cold atoms ~\cite{Deutsch1995} for light.   Some other examples of Bragg-spaced lattices with resonant scatterers include ring resonators ~\cite{yanik2004}, metallic gratings with plasmonic resonances~\cite{taubert2012} and dielectric cylinders with Mie resonances~\cite{Rybin2015}. A detailed comparison between cold atom systems, semiconductor lattices and M\"ossbauer  isotopes can be found in the review \cite{Ivchenko2013}. { It has also been theoretically suggested to consider Bragg lattices of atoms \cite{Vahid2016} and superconducting qubits coupled to the waveguide ~\cite{greenberg2020waveguide}. The 
modification of Bragg conditions for scattering of light from an array of atoms into the guided modes of a waveguide has been analyzed in Ref.~\cite{olmos2021bragg}.
 Large Bragg reflection from atomic arrays trapped near a one-dimensional waveguide have already been demonstrated experimentally in the groups of J.~Laurat \cite{Corzo2016} and E.~Polzik  \cite{Polzik2016}. These experiments are reviewed in more detail in Sec.~\ref{sec:Bragg}.

\begin{figure}[t!]
\centering\includegraphics[width=0.45\textwidth]{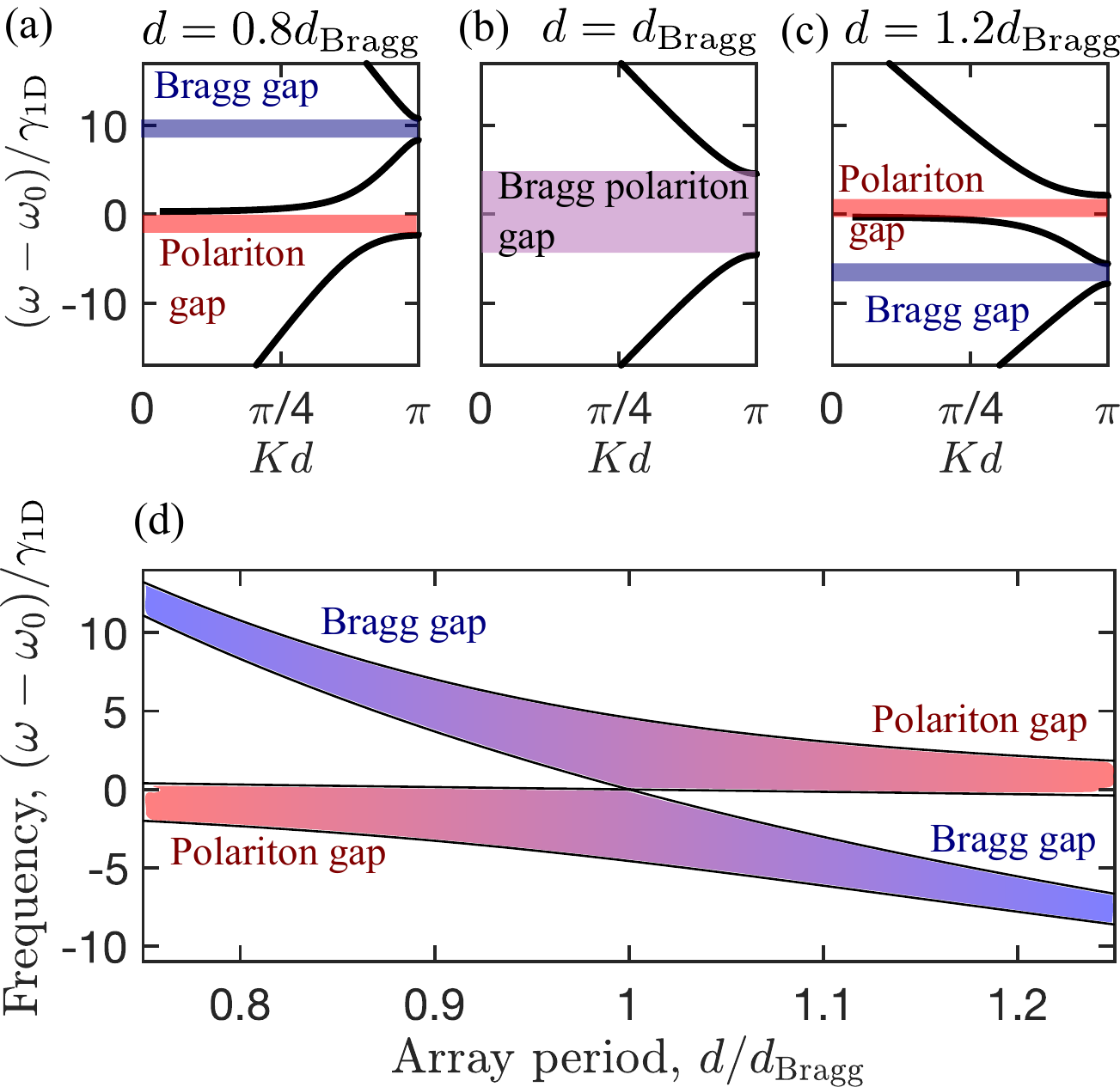}
\caption{Polariton dispersion in the atomic arrays with different periods. Panels (a,b,c) correspond to three different values of $d/d_{\rm Bragg}$, indicated in the graph. Bragg and polariton band gaps are indicated.
Panel (d) shows the dependence of the band gap positions on the period of the array.
Calculation has been performed for $\gamma_{\rm 1D}/\omega_0=0.03$ and $\gamma=0$. }\label{fig:Bragg}
\end{figure}

We examine the polariton dispersion law in the close-to-Bragg regime in Fig.~\ref{fig:Bragg}.  Figures~\ref{fig:Bragg}(a),(b),(c) present the polariton dispersion law $\omega(K)$ calculated for the periods close to the Bragg value for $m=1$. The first panel corresponds to the situation when the period is smaller than the Bragg value. The dispersion features two band gaps: the polariton band gap Eq.~\eqref{eq:gap}  below the atomic resonance, and the usual photonic band gap at the frequency satisfying the Bragg condition $\omega d/c=\pi$. Figure~\ref{fig:Bragg}(c) presents an opposite scenario where the Bragg band gap is located below the polariton one. In the Bragg case, illustrated in Fig.~\ref{fig:Bragg}(b), the two band gaps fuse with each other and form a wide Bragg polariton band gap around the atomic resonance. The polariton dispersion law Eq.~\eqref{eq:Kd} in the vicinity of the resonance can be approximately described by the following equation 
\begin{equation}
\frac{Kd}{\pi}-1=\pm \sqrt{\left(\frac{\omega-\omega_0}{\omega_0}\right)^2-\left(\frac{\Delta_{\rm Bragg}}{\omega_0}\right)^2},
\end{equation}
where  is the half-width of the polariton band gap
\begin{equation}
\Delta_{\rm Bragg}=\sqrt{\frac{2\gamma_{\rm 1D}\omega_0}{\pi}}\:.
\end{equation} 
The gap half-width $\Delta_{\rm Bragg}$ exceeds the radiative linewidth of the atomic resonance of $\gamma_{\rm 1D}$ by the large factor $\sim\sqrt{\omega_0/\gamma_{\rm 1D}}$. 
Hence, the light incident upon the Bragg-spaced array will exhibit a strong reflection in the wide spectral range $\omega_0-\Delta_{\rm Bragg}<\omega<\omega_0+\Delta_{\rm Bragg}$. However, this Bragg band gap will be manifested in reflection only if the number of atoms of the array is large enough, exceeding 
\begin{equation}\label{eq:trans}
N^*\sim \frac{1}{m} \sqrt{\omega_0/\gamma_{\rm 1D}}\:.
\end{equation}
Indeed, the phase gained by light between two atoms in the Bragg-spaced array is an integer multiple of $\pi$. So, at the first glance, the distance does not matter and the Bragg-spaced array is equivalent to the array with $d=0$. However, such analysis assumes the validity of the Markovian approximation when the time of flight of photons through the array $Nd/c$ is smaller than the inverse lifetime of the superradiant mode $1/(N\gamma_{\rm 1D})$. When the total length of the structure exceeds the wavelength in $\sim \sqrt{\omega_0/\gamma_{\rm 1D}}/m$ times, as specified by Eq.~\eqref{eq:trans}, the time of flight of photons can no longer be ignored ~\cite{Poshakinskiy2012} and the waveguide-mediated interaction between the atoms stops being instantaneous. Namely, for $N\gg\frac{1}{m} \sqrt{\omega_0/\gamma_{\rm 1D}}$ the reflection coefficient is close to unity inside the Bragg band gap and quickly decays outside the gap.

}

\begin{figure*}[t!]
\centering\includegraphics[width=0.8\textwidth]{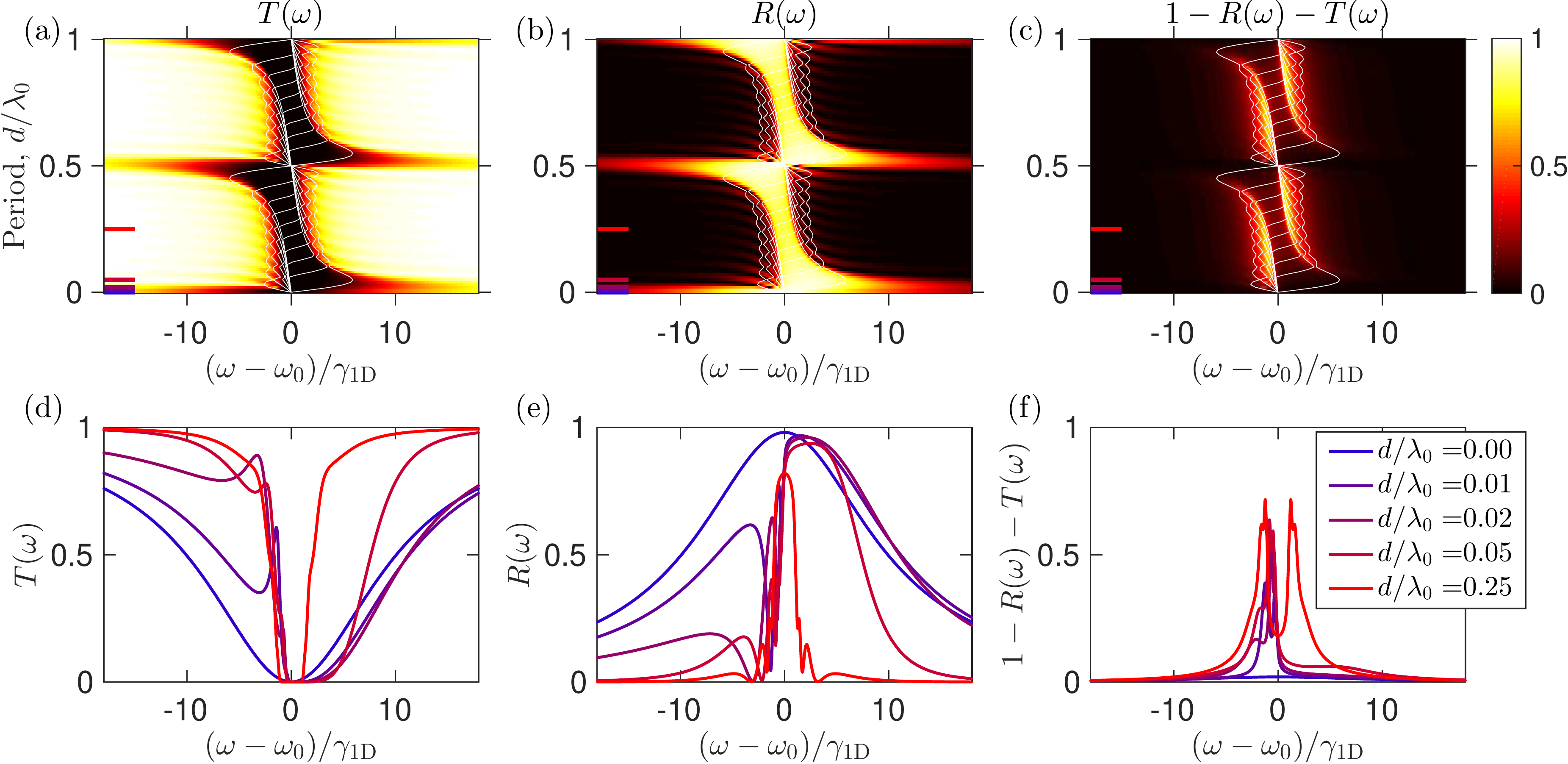}
\caption{Transmission (a,d), reflection (b,e) and absorption spectra (c,f) of the array of $N=10$ atoms coupled to a waveguide depending on the period of the array $d$. Bottom panels are calculated for four specific periods, indicated by corresponding dashes in the upper panels. Thin wide lines in (a--c) show the real parts of the polaritonic eigenfrequencies $\omega^
\nu$. Calculation has been performed for $\gamma=0.1\gamma_{\rm 1D}$ and $\gamma_{\rm 1D}/\omega_0=10^{-3}$.}
\label{fig:rta-period}
\end{figure*}

In general, the collective eigenmodes considered above can be observed as resonances in the reflection and transmission spectra.
The amplitude reflection coefficient can be found from the Hamiltonian \eqref{eq:Hmn}  as
\begin{equation}\label{eq:rNM}
r=-\rmi \GO\sum\limits_{m,n=1}^NG_{mn}(\omega)\e^{\rmi \omega (z_n+z_m)/c},
\end{equation}
and the transmission coefficient $t$ is obtained by replacing $z_m$ in Eq.~\eqref{eq:rNM} with $-z_m$ and adding unity.
Here, $G\equiv [\omega-H]^{-1}$ is the matrix Green's function of atomic excitations.
This follows from Eqs.~\eqref{eq:r-io},\eqref{eq:t-io} in Appendix~\ref{sec:input_output} in the low excitation regime, when nonlinear terms in Eq.~\eqref{eq:sigmam2} can be ignored.
An explicit answer for the Green's function of the finite periodic array can be found in Ref.~\cite{Voronov2007JLu}.
The Green's function  can be presented as an expansion over the eigenmodes:
\begin{equation}\label{eq:sumG}
G_{mn}=\sum\limits_{\nu=1}^N \frac{\psi^\nu_m \psi^\nu_n}{\omega-\omega^\nu}
\end{equation}
and has resonances at the eigenfrequencies $\omega_\nu$. Another equivalent way to calculate the reflection and transmission coefficients is presented by the transfer matrix method, see Appendix~\ref{app:transfer}.

 We present in Fig.~\ref{fig:rta-period} dependence of the reflection, transmission and absorption spectra on the period of the $10$-atoms array. The spectral features reflect the eigenfrequencies of the array, shown before in Fig.~\ref{fig:subrad}: positions of the features correspond to real parts of the eigenfrequencies, shown by thin wide lines in Fig.~\ref{fig:rta-period}(a--c).
Spectral widths of the features correspond to the imaginary part.

When the distance between atoms is much smaller than the light wavelength, or equal to a Bragg value
\eqref{eq:Bragg_cond} reflection and transmission coefficients assume an especially simple form, 
\begin{align}\label{eq:rtN}
r_N&=\frac{\rmi N \GO}{\omega_0-\omega-\rmi( \gamma+N\GO)},\\\nonumber t_N&=1+r_N=\frac{\omega_0-\omega - \rmi\gamma}{\omega_0-\omega-\rmi (\gamma+N\GO)}\:.
\end{align}
The  transmission and reflection spectra have a symmetric Lorentzian shape. The half-width at half-maximum of the transmission dip (reflection peak) is equal to $\gamma+N\gamma_{\rm 1D}$, and it scales linearly with the number of atoms $N$. This is the manifestation of the radiative decay rate of the collective Dicke superradiant state where the atoms are excited symmetrically. 
It is  instructive to stress here, that the resonant transmission coefficient through the Bragg-spaced array, $|t_N(\omega_0)|^2$ is given by  $\gamma^2/(\gamma+N\GO)^2$. Such non-exponential dependence is because of the formation of the collective superradiant state. It can obviously not be captured by a concept of OD which assumes independent light interaction with all the atoms. However, Eqs.~\eqref{eq:rtN} for the Bragg-spaced array are valid  in the Markovian approximation, for $N\ll N^*$, while  in long arrays and the Lorentzian spectrum is saturated. 

Detuning of the  spacing from zero or a Bragg value leads to suppression of  reflection and the optical spectra acquire narrow resonant features corresponding to the excitation of subradiant modes. The reflection is at minimum at the anti-Bragg condition $d=\lambda(\omega_0)/4$, when the interference between waves reflected from different atoms is destructive. Suppression of the reflection can be used for the selective radiance \cite{AsenjoGarcia2017exp}. 

{Another important effect which should be mentioned here is a disorder. Indeed, introducing small disorder in the Bragg array leads to a disturbance of a Bragg condition Eq.~\eqref{eq:Bragg_cond},  suppresses the reflection and modifies the transmission. For strong disorder  collective effects are quenched. The coupling efficiency $\beta$ and the optical depth (OD) become main constituents of the transmission. }

%

\subsubsection{Two-photon scattering}
\label{sec:two_photons_rt}

\paragraph{Model and historical overview.}
We start with a general discussion of  a quantum problem with two or more photons scattering  on an ensemble of  atoms coupled to a waveguide.  Since a single two-level atom can not be excited by two identical photons at the same time, the photon-photon interactions become crucial. This problem  has become a perfect testing ground for different theoretical techniques, and we try to review the development below. 

 The study of photon-photon interactions and nonlinearity of the Maxwell's equations in vacuum due to the excitation of virtual electron-positron pairs is a cornerstone problem of quantum electrodynamics, see for example the  reviews~\cite{Liang_2012,scharnhorst2019photonphoton}. However, in relativistic quantum electrodynamics the solution is obtained perturbatively since the electron-photon interaction constant $\alpha=e^2/\hbar c\approx 1/137$ is a small parameter. On the other hand, in  the nonrelativistic quantum optical problem there exists an exact analytical solution in all orders in the light-atom coupling parameter $g$, provided that the rotating wave approximation remains valid, the array has zero spacing, $d=0$ and the free-space dipole-dipole coupling is neglected.
One of the reasons why this is possible is  that when atoms are located in the same point light excites only symmetric Dicke states of the type
\begin{equation}\label{eq:Dicke_states}
\frac{1}{N}\sum\limits_m \sigma_j^\dag |0\rangle,\quad \frac{1}{N^2}\sum\limits_{m,n} \sigma_j^\dag \sigma_m^\dag |0\rangle\:,
\end{equation}
etc.

 Historically, the quantum Dicke problem in rotating wave approximation has been first diagonalized exactly by Rupasov and Yudson in 1984 using the Bethe ansatz technique, that was initially developed in the context of condensed matter physics ~\cite{Yudson1984,Yudson1985,Yudson2008}. Rupasov and Yudson have noticed, the similarity between the Dicke problem and  the Kondo problem of electrons with linear dispersion interacting with a single impurity, that has been independently solved in 1980 by Andrei \cite{Andrei1980} and Vigman \cite{Vigman1980} using the Bethe ansatz. The Kondo problem in turn has certain similarities to the problem of one-dimensional boson gas with contact interaction, that was solved by Lieb and Liniger in 1963 by an analogous Bethe ansatz approach \cite{Lieb1963}, see also recent reviews of the Bethe ansatz ~\cite{Batchelor2007,Faddeev_2013}. {The Kondo problem with linear dispersion is also related to the spin-boson model \cite{Leggett1987}, that is in turn equivalent to the problem of two-level system interacting with photons  without rotating wave approximation, see also Sec.~\ref{sec:ultrastrong} on the ultrastrong coupling regime.} The details of the Bethe ansatz used in \cite{Yudson1984} are given in Appendix ~\ref{app:2scat-electron}.  
Much later, the same answer as in Refs. ~\cite{Yudson1984,Yudson1985} has been obtained by Shen and Fan \cite{Fan2007,Shen2007} for the particular case of the two-photon scattering. Shen and Fan have solved the  Schr\"odinger equation  directly in the Hilbert subspace with only two excitations. We present an analogous derivation in Appendix
~\ref{app:2scat-kspace}. Next, \cite{Liao2010} have  considered a  related  system with a two-level atom replaced by a nonlinear cavity.   Later on, this approach has been extended to several photons using the path integral formalism \cite{Shi2009,Shi2011,Shi2015} that is discussed in Appendix \ref{app:2scat-func}. The scattering has been also analyzed using the conventional input-output theory of quantum optics~\cite{Blais2013,Caneva2015}. In both approaches 
 \cite{Shi2009,Shi2011,Caneva2015}, the photon degrees of freedom are effectively traced out and the problem is solved in the atomic subspace of the full Hilbert space. The disadvantage of a such technique is that it is valid only in the Markovian approximation. Such approximation seems reasonable for closely spaced atoms, but can fail in a large array. Namely, the Markovian approximation sets all the phases gained by light when traveling the distance $d$ between any two atoms $\omega_kd/c$ to $\omega_0d/c$, and the introduced phase error $|\omega_k-\omega_0|d/c$ can become important for a large spacing \cite{Baranger2013,Fang2014}. However, the path integral formalism also allows to take into account the non-Markovian effects~\cite{Shi2015}.

The photon scattering problem on an atomic array has also been attacked by the diagrammatic Green function techniques ~
\cite{Pletyukhov2012,Baranger2013,Laakso2014,Fang2014,Kocabas2016,Schneider2016}. These can also be separated into two types. The first type is based on the electron representation, being inspired by the original Feynman approach from the quantum electrodynamics. In such a technique, a photon absorption is viewed as a transition of an electron from a lower atomic state $|1\rangle$ to the upper one $|2\rangle$, as described by the Hamiltonian of the following type:
$
 a_k c_2^{\phantom{\dag}}c_1+H.c.,
 $
 where $c_1, c_2$ are corresponding electron creation operators. The perturbation series for $N$ closely spaced atoms can be summed exactly and the original answer by Rupasov and Yudson can be recovered, see Appendix ~\ref{app:2scat-electron}. However, this approach fails for spatially separated atoms due to an appearance of extra diagrams ~\cite{Kocabas2016} and a closed-form solution can not be obtained. The most practical technique to date for the photon scattering, in our opinion, is the Green's function approach in the exciton representation, developed by Zheng and Baranger in ~\cite{Baranger2013,Fang2014}. In this representation, the absorption of a photon by an atom leads to a creation of an exciton. It can work both for closely spaced and spatially separated atoms and is not restricted by the Markovian approximation. Another advantage is that it naturally handles multilevel atoms, and the two-level atom case is recovered as a particular limiting case.
  Very recently, the Green's function technique has also been generalized to the multi-photon scattering~ \cite{piasotski2021diagrammatic}.
The model of multilevel atom strongly coupled  to the waveguide  photons is very similar  to the so-called irreversible quantum graph model, where a propagating wave is coupled to an oscillator~\cite{Smilansky2004,Smilansky2006}.

\paragraph{Photon-photon correlations for $d=0$.}
\label{sec:yudson}
In case when the atomic array has zero spacing $d=0$, the two-photon scattering problem can be solved exactly [see Appendices~\ref{app:2scat-kspace}--\ref{app:2scat-exciton} for different equivalent derivations]. We consider coherent state of light  $\exp(-\alpha^2/2+\alpha a_{\eps/c}^{\dag})|0\rangle$ incident from the left upon the atomic ensemble.
Here, $\varepsilon$ is the frequency and we assume that the  excitation amplitude is weak,  $\alpha \ll 1$. 
The scattered two-photon state can be then presented as \cite{Poshakinskiy2016}
\begin{multline}\label{eq:psi2}
\psi_{\rm scat}=\e^{-\alpha^{2}/2} \Bigl\{|0\rangle+\alpha t(\eps)a_{\eps/c}^{\dag}|0\rangle+\alpha r(\eps)a_{-\eps/c}^{\dag}|0\rangle
\\+\tfrac{\alpha^{2 }}{2}\bigl[t^{2}(\eps)a_{\eps/c}^{\dag,2}
+r^2(\eps)a_{-\eps/c}^{\dag,2}+2r(\eps)t(\eps)a_{\eps/c}^{\dag}a_{-\eps/c}^{\dag} \bigr]|0\rangle
\\+{\frac{\rmi \alpha^{2}}{4}}\int\limits_{-\infty}^\infty \frac{\rmd\omega}{2\pi} M(\eps-\omega,\eps+\omega \leftarrow\eps,\eps)\\\times
(a_{(\eps+\omega)/c}^{\dag}+a_{-(\eps+\omega)/c}^{\dag})(a_{(\eps-\omega)/c}^{\dag}+a_{-(\eps-\omega)/c}^{\dag})|0\rangle
\Bigr\}\:.
\end{multline}
Here, we use the notation $M(\omega_{1}',\omega_{2}'\leftarrow\omega_{1},\omega_{2})$ for the  matrix element describing
the incoherent  scattering process where the two incident photons have  the frequencies $\omega_1$,$\omega_2$ 
and the two scattered photons have the frequencies  $\omega_1'$,$\omega_2'$.
The first line in Eq.~\eqref{eq:psi2} describes the superposition of the vacuum state and the states where a single photon is either reflected or transmitted. Second line describes independent coherent scattering of the two photons. The last two lines in Eq.~\eqref{eq:psi2} describe correlated  incoherent scattering of two photons. One of the two scattered photons has the frequency $\eps-\omega$ and the other one has $\eps+\omega$. The total energy $2\eps$ is conserved and equal to that of the incident photon pair. Such process is characterized by the following scattering matrix, derived in Appendix~\ref{app:2scat-Bethe}:
\begin{multline}\label{eq:M-2phot-1}
M(\omega_{1}',\omega_{2}'\leftarrow\omega_{1},\omega_{2})=4N\GO^2s(\omega_1)s(\omega_2)s(\omega_1')s(\omega_2')\\\times
\frac{(\eps-\omega_0+\rmi \gamma)[\eps-\omega_0+\rmi(N \GO+\gamma)]}{\eps-\omega_0+\rmi (N-1)\GO+\rmi\gamma}\:,
\end{multline} 
where $s(\omega)=1/[\omega_0-\omega-\rmi(\gamma + N\GO)]$.
The four factors $s$ in the first line of 
Eq.~\eqref{eq:M-2phot-1} describe the resonances of incident and scattered photons with the single-excited superradiant Dicke state at $\omega=\omega_0-\rmi(\gamma + N\GO)$. The factor in the second line describes the two-photon resonance, when the average energy of two incident (or two scattered) photons is equal to   the double-excited Dicke state in Eq.~\eqref{eq:Dicke_states}, $\eps=\omega_0 -\rmi\gamma -\rmi (N-1)\GO$. Since double-excited  states are only present for $N>1$ atoms, this two-photon resonance in Eq.~\eqref{eq:M-2phot-1} cancels out with the corresponding term in the numerator for $N=1$.
\begin{figure}[t!]
\centering\includegraphics[width=0.35\textwidth]{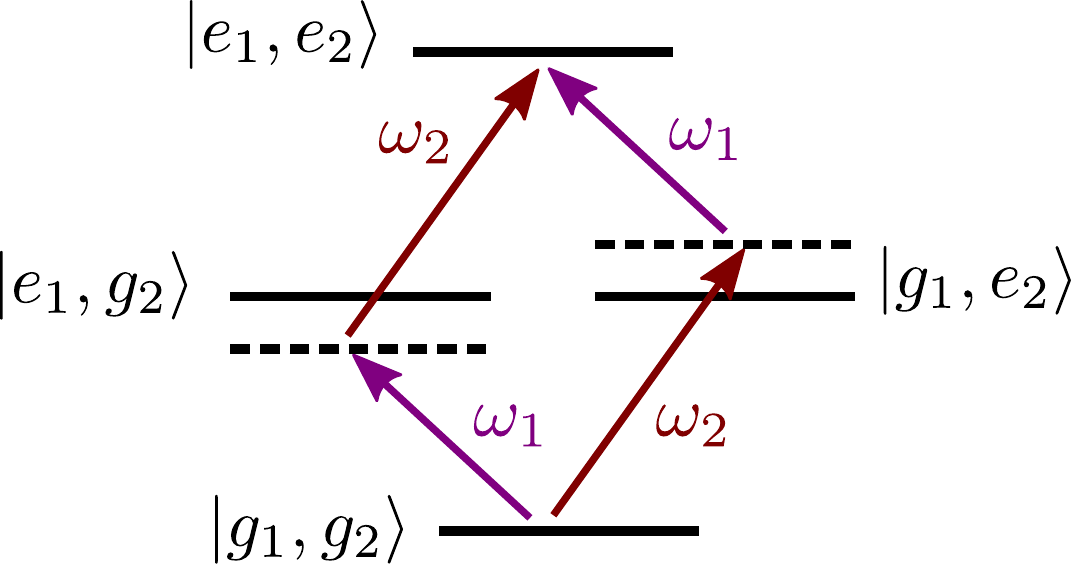}
\caption{Two destructively interfering pathways for absorption of two photons by two different atoms.
}
\label{fig:2-interference}
\end{figure}
Equation \eqref{eq:M-2phot-1}  also shows that  the incoherent scattering is partially suppressed when the array is excited exactly at the atomic resonance, so that the absolute value of the factor $\eps-\omega_0+\rmi(\gamma + N \GO)$ in the numerator  is at minimum.
Such suppression can be qualitatively understood by analyzing the destructive interference of two different quantum pathways for a related process, absorption of two photons by two different atoms ~\cite{Scully2004}, schematically illustrated in Fig.~\ref{fig:2-interference}.
The process goes through an intermediate virtual state, where only one of the two atoms is excited, either the first one or the second one. The sum of matrix elements of these two processes is proportional to 
\[
\frac1{\omega_1-\omega_0}+\frac1{\omega_2-\omega_0}
\]
and vanishes for $(\omega_1+\omega_2)/2=\eps=\omega_0$.

The wavefunction in Eq.~\eqref{eq:psi2} also allows one to calculate coherent photon reflection and transmission coefficients up to the linear order in the incident light power $\sim c\alpha^2/L$. The transmission coefficient is given by
\begin{align}\label{eq:Tcoh}
T_{\rm coh}&=|t|^2-\frac{c\alpha^2}{L}\Im \left[M(\eps,\eps\leftarrow \eps,\eps)
t^*(t^*+r^*)\right]\:,
\end{align}
where $t$ and $r$ are coherent single-photon amplitudes of transmission and reflection coefficients, respectively. Coherent reflection coefficient $R_{\rm coh}$ is obtained by replacing $t$ by $r$ in Eq.~\eqref{eq:Tcoh}.
It can be checked that in a case of vanishing non-radiative decay, the energy flux conservation holds
$
R_{\rm coh}+T_{\rm coh}+I_{\rm incoh}=1\:,
$
where  \begin{multline}\label{eq:I_incoh}
I_{\rm incoh}=\frac{c\alpha^2}{L}\int\limits_{-\infty}^\infty\frac{\rmd \omega}{2\pi}
|M(\eps-\omega,\eps+\omega \leftarrow\eps,\eps)|^2\:.
\end{multline} is the total incoherent scattering rate.

An insight in the two-photon scattering process can be obtained from the time-dependent second-order photon-photon correlation functions 
\begin{equation}
g_\nu^{(2)}(\tau)=\frac{\langle a_\nu^\dag (0)a_\nu^\dag (\tau)a_\nu (\tau)a_\nu (0)\rangle}{|\langle a_\nu^\dag (0) a_\nu(0) \rangle|^2}
\end{equation}
where $\nu=\rightarrow,\leftarrow$.  Zero-time correlation functions 
$g_\nu^{(2)}(0)$  are equal to the ratio of the probability of two photons being emitted together to the probability of their independent emission and thus determine the emission statistics.
Calculating the expectation values  with the help of Eq.~\eqref{eq:psi2} we find 
\begin{multline}\label{eq:g2gen}
g_\rightarrow^{(2)}(\tau)\\=\left|1+\frac{\rmi}{2t^2(\eps)}\int\limits_{-\infty}^\infty \frac{\rmd\omega}{2\pi} \e^{-\rmi \omega \tau}M(\eps-\omega,\eps+\omega \leftarrow\eps,\eps)\right|^{2}\:,
\end{multline}
in transmission geometry. The correlation function $g_\leftarrow^{(2)}(\tau)$ for reflected photons is obtained by replacing $t(\eps)$ with $r(\eps)$ in Eq.~\eqref{eq:g2gen}. 
The integrals can be straightforwardly evaluated analytically using the Cauchy theorem. 

\begin{figure}[t!]
\centering\includegraphics[width=0.45\textwidth]{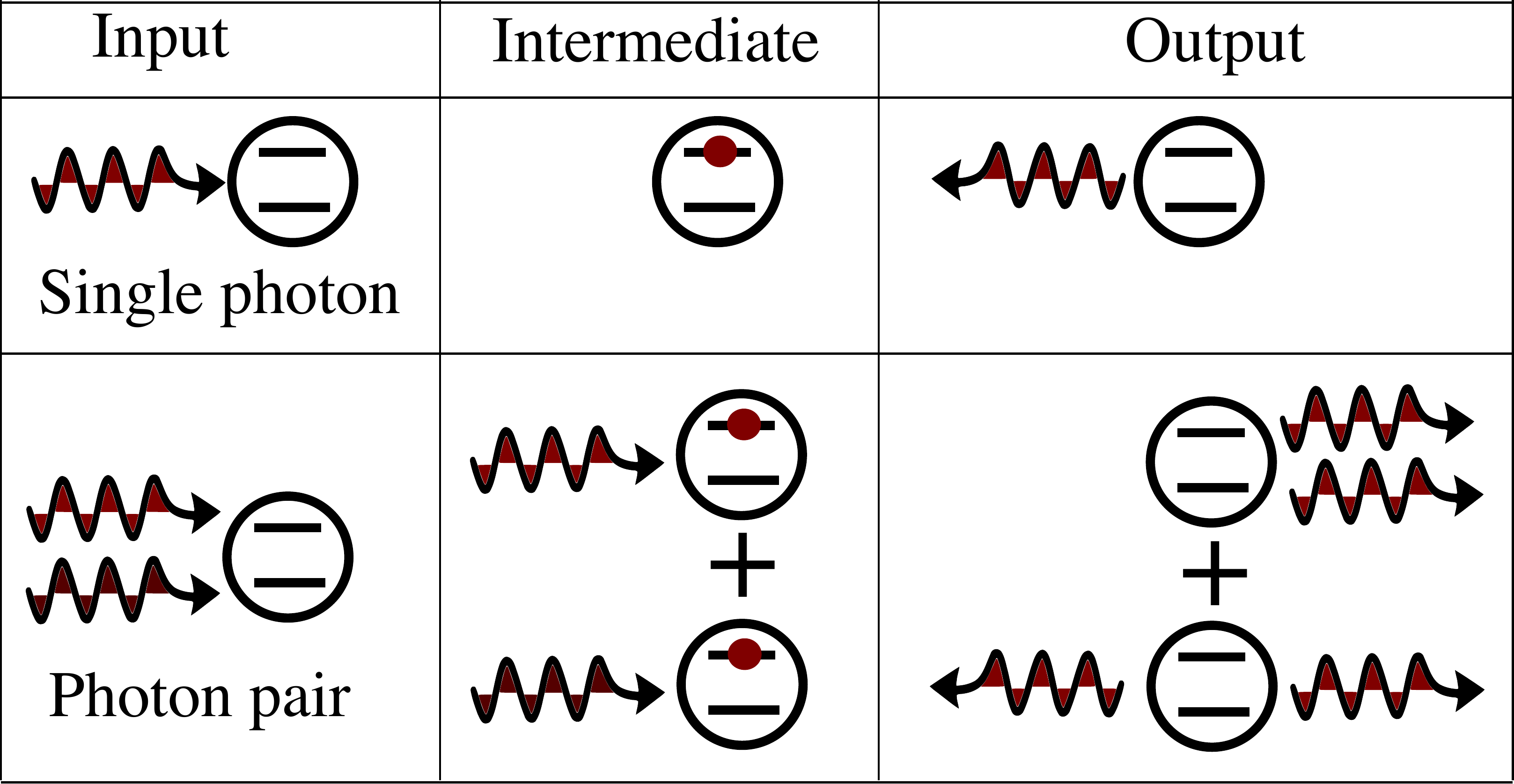}
\caption{Schematic illustration of single- and two- photon scattering on a two-level atom under resonant excitation. }
\label{fig:g2table}
\end{figure}

{
The calculation results are most easily  interpreted in case of just $N=1$ atom excited at the resonance, $\eps=\omega_0$. It is then straightforward to show from Eq.~\eqref{eq:g2gen} that 
$g_\rightarrow^{(2)}(0)$ is diverging (photon bunching) and $g_\leftarrow^{(2)}(0)=0$ (antibunching).
The bunching  occurs because the atom becomes transparent after having absorbed  a photon. Thus, 
a single photon can not pass through the atom,  while a pair of photons can pass.
The antibunching in reflection geometry stems from the fact that a single  two-level atom can not accommodate two photons at the same time and hence can not  emit two photons simultaneously.  Scattering of single- and two- photons on an atom are schematically illustrated in Fig.~\ref{fig:g2table}, where we show an incident state,an intermediate virtual state, after one of the photons has been absorbed, and output states. 
The two-photon output state, given by Eq.~\eqref{eq:psi2}, is an entangled state of two photons propagating to the right and two photons propagating in opposite directions.

It is also instructive to analyze the equal-time photon-photon correlation function $g^{(2)}(0)$ depending on the number of atoms in the array. For resonant excitation ($\eps=\omega_0$)  it is given by
\begin{align}\label{eq:g2N}
g_{\leftarrow}^{(2)}(0)&=\left(\frac{1-1/N}{1-\Gamma_{\rm 1D}/(\Gamma+N\Gamma_{\rm 1D})}\right)^2\:,\\
g_{\rightarrow}^{(2)}(0)&=\left(\frac{1-\Gamma_{\rm 1D}/\Gamma}{1-\Gamma_{\rm 1D}/(\Gamma+N\Gamma_{\rm 1D})}\right)^2\:.
\nonumber
\end{align}
The  dependence of photon-photon correlation functions Eq.~\eqref{eq:g2N} on the atom number $N$
and the ratio of the decay rates is plotted in  Fig.~\ref{fig:g2N}.
One atom in reflection geometry demonstrates full antibunching $g_{\leftarrow}^{(2)}(0)=0$ for any value of $\Gamma>0$, see Fig.~\ref{fig:g2N}(a), since it can not host two photons. However, this antibunching is fully suppressed already for $N=2$ atoms, $g_{\leftarrow}^{(2)}(0)\approx 1$. The naive physical explanation is very simple: the array of $N>1$ atoms can host two photons so the photon blockade is not manifested. In transmission geometry the dependence on $N$ is weak. The transmitted photons are bunched (anti-bunched) for small (large) values of $\Gamma\ll \Gamma_{\rm 1D}$, see Fig.~\ref{fig:g2N}(b).

\begin{figure}[t!]
\centering\includegraphics[width=0.48\textwidth]{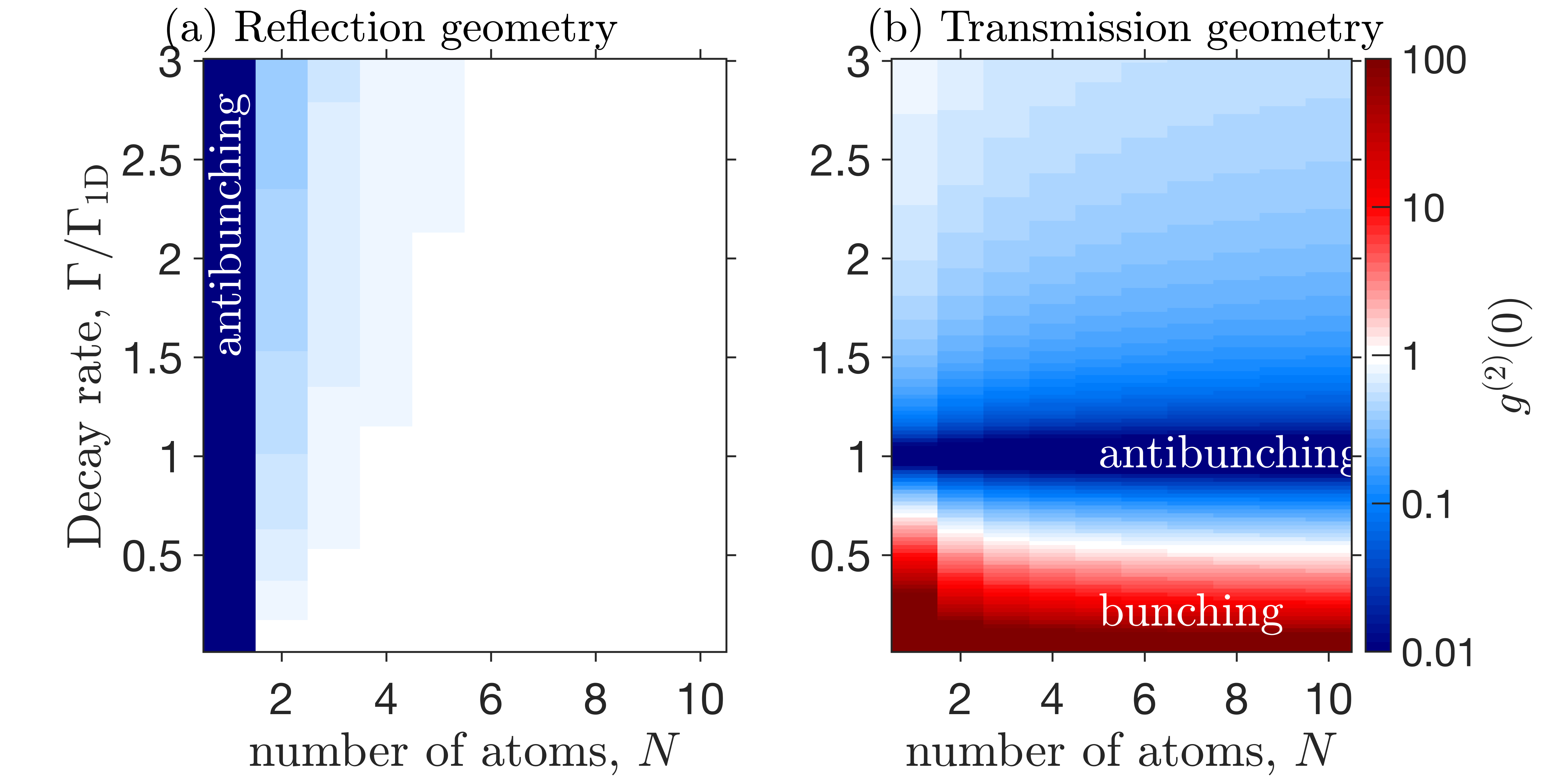}
\caption{Dependence of the photon-photon correlation function $g^{(2)}(0)$ on the number of atoms $N$ and on the ratio of nonradiative and radiative damping rates $\Gamma/\Gamma_{\rm 1D}$ calculated with Eq.~\eqref{eq:g2N}. Panels (a) and (b) correspond to the reflection and transmission geometry. Calculated for the light incident at the resonant frequency $\eps = \omega_0$ of the two-level atoms ($U\to \infty$).}
\label{fig:g2N}
\end{figure}

\paragraph{Array with non-zero spacing, $d>0$.}
In the case of non-zero spacing,  it is not possible to write an explicit analytical solution   of type \eqref{eq:M-2phot-1}, but the scattering can be considered using the generalization of the Green's function method \cite{Baranger2013} described in detail in Appendix~\ref{app:2scat-exciton}. The two-photon wavefunction is obtained by replacing the last term in Eq.~\eqref{eq:psi2} by~\cite{Poshakinskiy2016}
\begin{equation}
\sum\limits_{\mu,\nu=\pm}\int\limits_{-\infty}^\infty \frac{\rmd\omega}{2\pi} M_{\mu\nu}(\omega,2\eps-\omega \leftarrow\eps,\eps)
a_{\mu\omega /c}^{\dag}a_{\mu (2\eps-\omega)/c}^{\dag}|0\rangle\:,
\end{equation}
where the scattering kernel is given by
\begin{equation}\label{eq:Mgen}
M_{\mu\nu}(\omega_1',\omega_2')= -2\rmi \GO^2\sum_{m,n=1}^{N} s_n^{\mu}(\omega_1')s_n^{\nu}(\omega_2')Q_{nm}s_m^+(\eps)s_m^+(\eps) \end{equation}
with $s_m^\pm(\omega) = \sum_m G_{mn}(\omega) \e^{\pm\rmi (\omega/c) z_n}$ and 
 \begin{equation}\label{eq:QSigma}
 Q =\Sigma^{-1}, \quad  \Sigma_{nm}=\int\limits_{-\infty}^\infty\frac{\rmd\omega}{2\pi} G_{nm}(\omega)G_{nm}(2\eps-\omega)\:.
 \end{equation}
 Eq.~\eqref{eq:Mgen} is valid beyond the Markovian approximation. However, in the Markovian approximation it can be simplified further. Specifically, the poles $\eps$ of the matrix $Q$ in Eq.~\eqref{eq:QSigma} correspond to the complex energies of the double-excited states 
 \begin{equation}\label{eq:2photon-ansatz}
|\psi\rangle=\sum\limits_{m,n=1}^N \psi_{mn} \sigma_m^\dag \sigma_n^\dag |0\rangle
\end{equation}
($\psi_{mn}=\psi_{nm}$) of the effective non-Hermitian atomic Hamiltonian
Eq.~\eqref{eq:Heff-io}.
The corresponding two-photon Schr\"odinger equation for the amplitude $\psi_{nm}$
can be obtained by substituting the ansatz Eq.~\eqref{eq:2photon-ansatz} into the general
Schr\"odinger equation $H_{\rm eff}|\psi\rangle=2\eps |\psi\rangle$. More details are presented in Appendix~\ref{app:2scat-exciton}.

\begin{figure}[t!!]
 \centering\includegraphics[width=0.4\textwidth]{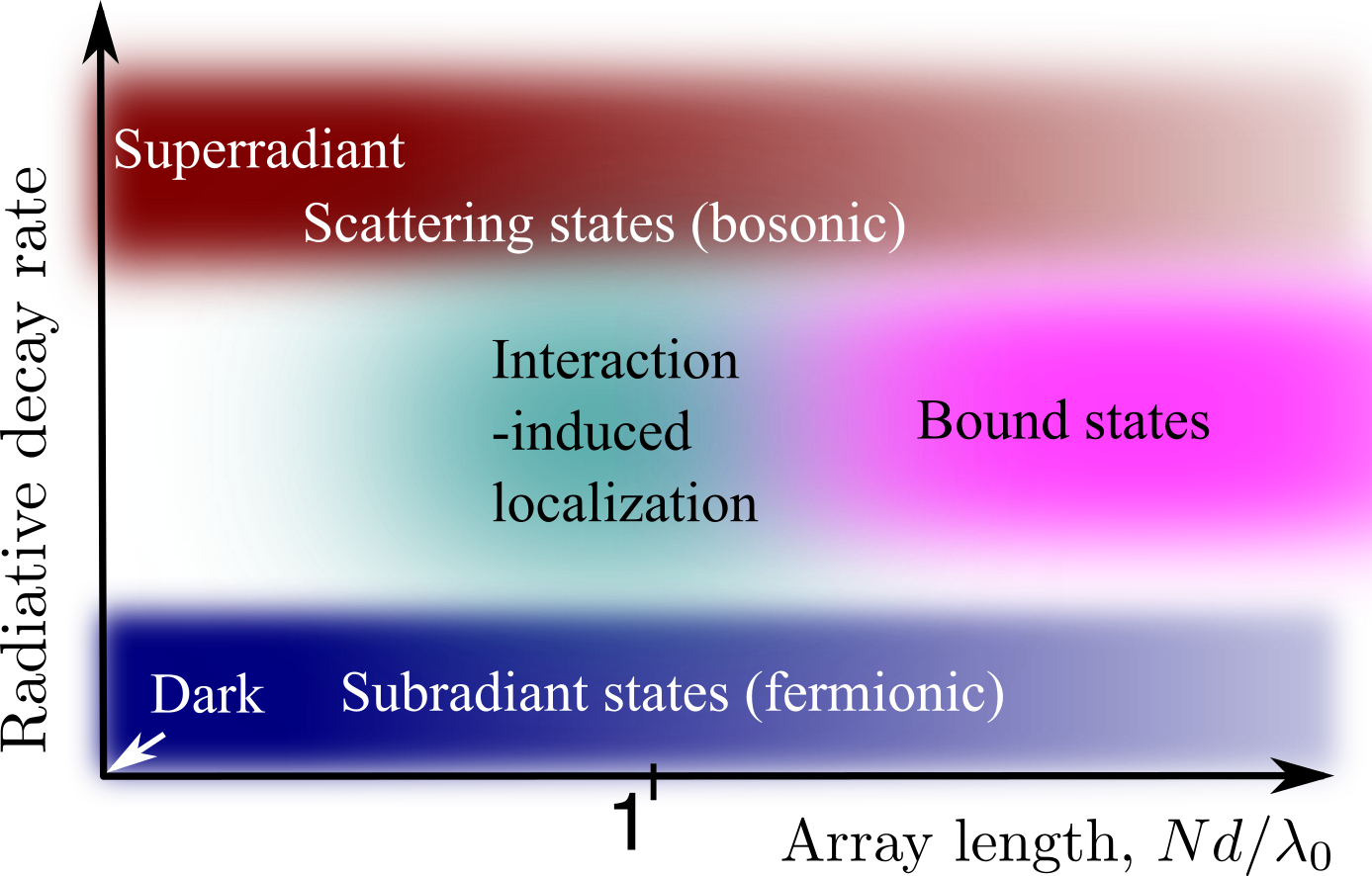}
 \caption{Schematic phase diagram showing different possible types of two-polariton states in an array of atoms coupled to a waveguide depending on their radiative decay rate and on the ratio of the array length $Nd$ to the wavelength at the atom frequency. Wavefunctions for these states  are shown in Fig.~\ref{fig:all2phot}.
 }\label{fig:diagram-2phot}
\end{figure}

\begin{figure*}[t]
 \centering\includegraphics[width=\textwidth]{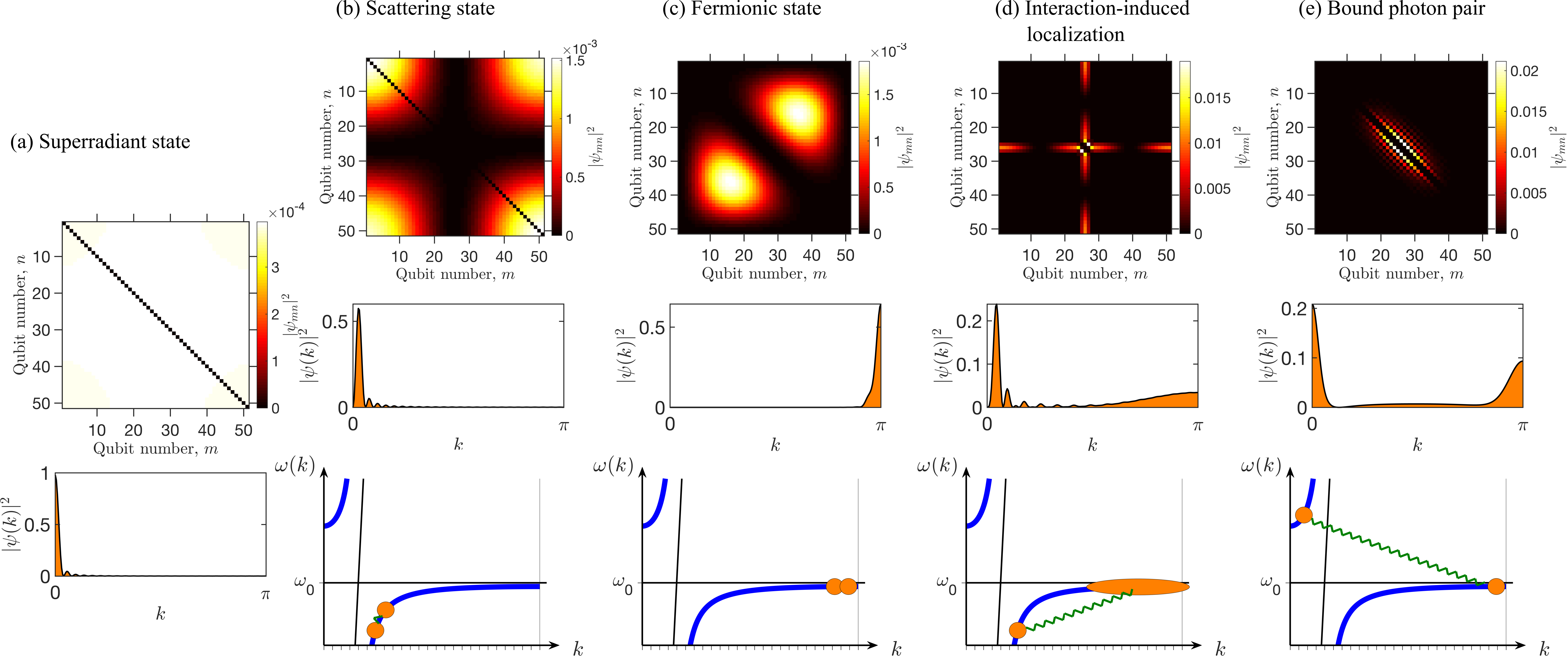}
 \caption{Examples of different double-excited states in the finite array of $N=51$ two-level atoms coupled to a non-chiral waveguide. Top row: real space two-polariton wavefunctions 
 $|\psi_{nm}|^2$. Second row: Fourier transforms  $|\psi(k)|^2$. 
Panels in  bottom row in (b--e) illustrate  the wave vectors of two polaritons corresponding to the two-polariton wave functions in the upper panel (not in scale). 
 Calculation has been performed for the array period $\omega_0d/c=0.01$ (a,b,d) and $\omega_0d/c=0.4$ (c,e).
 The complex energies of the shown states are $(\eps-\omega_0)/\gamma_{\rm 1D}$=$8.4-48.9\rmi$, $-5.0-1.0\rmi $,$-0.2-4\times 10^{-6}\rmi$,$1.9-2\times 10^{-6}\rmi$ for the panels (a--e), respectively.
 }\label{fig:all2phot}
\end{figure*}

We will now discuss  the spatial structure of the double-excited states that is surprisingly diverse. On the applied side, these are the states  responsible for spatial and time dependence of the two-photon correlations and spatial entanglement. For example,  recently observed tunable bunching and antibunching \cite{Prasad2020} are enabled by the correlated double-excited states, see also following Sec.~\ref{sec:Prasad}. On the more fundamental side, we will now show that  the double-excited states uncover
drastically different regimes of particle-particle interaction.  Namely, eigenstates of the same array of atoms in a waveguide manifest at the same time fermionization, interaction-induced localization,  quantum Hall phases with topological edge states, Hofstadter butterfly and quantum chaos. The main reason why this seemingly simple two-body problem is so rich is  the strongly nonlinear and nonparabolic polariton dispersion, shown in Fig.~\ref{fig:Kd}, with the slow polariton group velocity decreasing near the resonance frequency. The interactions between the particles with the polaritonic dispersion are very different from the more commonly studied  parabolic dispersion case~\cite{Girardeau,Lieb1963}. 

We try to present a general simplified phase diagram distinguishing domains of different double-excited states depending on their radiative lifetime and on the array length in Fig.~\ref{fig:diagram-2phot}.  Figure~\ref{fig:all2phot} also shows the characteristic wavefunctions of these states. The real space  joint  two-polariton probabilities $|\psi_{nm}|^2$ are shown in the top row. Second row presents
the Fourier transforms $|\psi(k)^2|\equiv \sum_{m=1}^N|\psi_m(k)|^2/N$, where $\psi_m(k)=\sum_{n=1}^N\psi_{mn}\e^{\rmi kn}$ and the bottom row illustrates these Fourier transforms schematically.  We will now discuss these states in detail.

 Similarly to the single-photon states, if the distance between atoms is vanishing, the only double-excited state probed by light is the symmetric superradiant state (top left corner of Fig.~\ref{fig:diagram-2phot}). The wavefunction $\psi_{nm}$ for this state is constant if $n\ne m$ and $\psi_{nn}=0$ due to the photon blockade, as can be seen in Fig.~\ref{fig:all2phot}(a). The double-excited superradiant state has the complex  eigenfrequency $\eps=\omega_0-\rmi (N-1)\gamma_{\rm 1D}-\rmi \gamma$ that can be also found from the resonance of Eq.~\eqref{eq:M-2phot-1}.  
 
 When the distance between atoms becomes nonzero, other states with more interesting spatial profile become accessible by light. Probably, the simplest one is the scattering state of two polaritons, shown in Fig.~\ref{fig:all2phot}(b). Typically, scattering  states are realized when the wave vectors of both polaritons $k$ are much smaller than the edge of the Brillouin zone $\pi$, see  schematics in the bottom panel of Fig.~\ref{fig:all2phot}(b). In this case, both polaritons have relatively high group velocity, so the role of their interaction is weak. Such two polaritons can be thought of as quasi-independent. The wave function of a single-polariton state in a finite array is a standing wave $P_n^{(\xi)}$ with the wave vector $k_\xi=\pi \xi/N$, given by Eq.~\eqref{eq:Psingle}. Hence, the wavefunction of the two-polariton scattering state is approximately described by a symmetrized product of two standing waves, slightly modified by the interaction that sets $\psi_{nn}=0$, namely 
 \begin{equation}\label{eq:bosons}
 \psi_{mn}\approx \frac1{\sqrt{2}}(  P_{m}^{(\xi)}P^{(\xi')}_{n}+P_{m}^{(\xi')}P^{(\xi)}_{n})(1-\delta_{mn})\:.
 \end{equation}
For example, the state in Fig.~\ref{fig:all2phot}(b) corresponds to $k_\xi=k_{\xi'}=\pi/N$.

In the case of zero spacing between atoms $d=0$,  all the states excepting the superradiant one are fully dark and degenerate, $\eps=\omega_0-\rmi \Gamma$ (bottom left corner of Fig.~\ref{fig:diagram-2phot}) and they become subradiant for $d>0$.
Contrary to well-known single-excited subradiant states, the spatial structure of double-excited subradiant states has been revealed only very recently, in Refs.~\cite{Albrecht2019,Molmer2019}. We remind that single-excited subradiant states  are just standing waves with the wave vectors $k$ close to the edge of the Brillouin zone, see Eq.~\eqref{eq:Psingle}. The polaritonic dispersion law given by Eq.~\eqref{eq:Kd} in the vicinity of the Brillouin zone edge is parabolic, $\omega(k)-\omega(\pi)\approx-\phi \gamma_{\rm 1D}(k-\pi)^2/8$ with $\varphi=\omega_0d/c$, see also Fig.~\ref{fig:Kd}. The two polaritonic excitations described by the Schr\"odinger equation ~\eqref{eq:Sh2} exhibit contact repulsion due to the photon blockade. The problem of interacting bosons with parabolic dispersion and contact repulsion is well known in the condensed matter physics and has been solved by Lieb and Liniger \cite{Lieb1963} by means of the Bethe ansatz. It has been shown that strong repulsion between bosons  emulates Pauli exclusion principle. Specifically, the two-particle wavefunction is proportional to the antisymmetric combination of two single-particle wave functions, i.e.
\begin{eqnarray}\label{eq:fermions}
  \psi_{mn}\propto\frac1{\sqrt{2}}\begin{cases}
  P_{m}^{(\xi)}P^{(\xi')}_{n}-P_{m}^{(\xi')}P^{(\xi)}_{n},&(m \ge n)\\
  -(P_{m}^{(\xi)}P^{(\xi')}_{n}-P_{m}^{(\xi)}P^{(\xi')}_{n}),&(n\le m)\:.
 \label{ansatz}
  \end{cases}
\end{eqnarray}
(compare with the scattering state Eq.~\eqref{eq:bosons}).
While the wavefunction Eq.~\eqref{eq:fermions} still has bosonic symmetry, $\psi_{nm}=\psi_{mn}$,  the corresponding probability distribution  $|\psi_{nm}|^2$ in the real space is the same as for non-interacting spinless fermions. This is termed as fermionization.
Top panel of Fig.~\ref{fig:all2phot}(c) shows the probability distribution for the most-subradiant state calculated numerically in the array of $N=51$ qubits. Its spatial structure is well captured by the ansatz Eq.~\eqref{eq:fermions} with {$k_\xi=\pi-\pi/N$, $k_{\xi'}=\pi-2\pi/N$}. Bottom panel of  Fig.~\ref{fig:all2phot} (c) illustrates schematically the origin of this fermionic state as a result of interaction of two polaritons with the wave vectors close to the edge of the Brillouin zone.
 
 There also exists an interesting mesoscopic regime when the length of the array is on the order of its wavelength, $N\omega_0d/c\sim 1$. It has been predicted that the interaction between the two polaritons can make one of them localized, even though all single-polariton states are delocalized and the structure has no disorder~\cite{Zhong2020}. Specifically, the first polariton forms a standing wave that drives localization of the second polariton in the node (or in the antinode) of this wave. Since the two polaritons are indistinguishable, at the same time the second polariton drives localization of the first one and the two-polariton wave function can be approximately described by the ansatz 
 \begin{equation}\label{eq:Janet}
 \psi_{mn}\propto \psi_{\rm loc}(n)\psi_{\rm free}(m)+\psi_{\rm loc}(m)\psi_{\rm free}(n)\:.
 \end{equation}  
Here, the state $\psi_{\rm loc}(n)$ is localized just at several atoms and the state $\psi_{\rm free}(n)$ is a standing wave. The localization is so strong because of the low group velocity at large wave vectors. In another words, polaritons with $k\gg \omega_0/c$ have large effective mass and are easy to be localized by interaction, see  Ref.~\cite{Zhong2020} for more details.  
The situation becomes even more interesting when the standing wave has multiple nodes. In this case the
the first polariton experiences both the lattice potential and the periodic standing wave potential, induced by the interaction with the second polariton and determined by $|\psi_{\rm free}(n)^2|$. The one-dimensional problem of the particle in  such  potential with two periods is similar to the Aubry-Andr\'e-Harper model that can be in turn mapped to the two-dimensional quantum Hall problem on a lattice~\cite{Kraus2012,Poshakinskiy2014}. Thus, it turns out that the WQED setup hosts an analogue of the  topological quantum Hall phase, that arises solely due to the  interactions, without any applied magnetic field. Such phase manifests an analogue to the Hofstadter butterfly and also topological two-polariton edge states, when one of the polaritons is localized at the edge of the structure and another one forms a standing wave~\cite{Poshakinskiy2020}. 

Yet, another type of two-polariton states realized in relatively long structures with the thickness of many wavelengths, is a state where two polaritons form a bound pair that can propagate as a whole and is characterized by a certain center-of-mass momentum $K$. In the finite array, such a pair forms a standing wave
 \begin{equation}\label{eq:bound2}
 \psi_{mn}\approx \cos \left(K\frac{m+n}{2}\right)\psi_{\rm bound}(|m-n|)\:,
 \end{equation}  
 where the relative motion wavefunction $\psi_{\rm bound}(|m-n|)$ decays exponentially with distance. An example of such state is shown in Fig.~\ref{fig:all2phot}(e). There exist two types of bound two-photon states~\cite{Zhang2020PRR}. The first type corresponds to one of the two polaritons in the upper polaritonic branch and one in the lower polaritonic branch, as shown in the lower panel of Fig.~\ref{fig:all2phot}(e). Its center of mass dispersion is very sensitive to the ratio of the array period to the light wavelength at the atomic resonance $d/\lambda(\omega_0)$. Namely, there exists a ``magic value" $d=\lambda(\omega_0)/12$ where the center-of-mass dispersion depends on the wave vector near the edge of the Brillouin zone as $(K-\pi)^4$, and the quadratic term vanishes~ \cite{poddubny2020quasiflat}. This means that the bound pair acquires infinite mass and it is hard for photons to escape the array, so their radiative lifetime increases dramatically~\cite{Zhang2020PRR}. Another type of bound pair states is formed by both polaritons in the upper branch. Generalization of two-photon states from Fig.~\ref{fig:all2phot} for the three-photon case has recently been done in Ref.~\cite{Zhong2021three}.

We also note, that the considered two-body problem is in general not integrable, which means, in particular, that neither of the simple ansatzes  Eq.~\eqref{eq:bosons}-- Eq.~\eqref{eq:bound2} is exact. The intermediate regime, when neither of these ansatzes holds, corresponds to an interaction-induced quantum chaos~\cite{Poshakinskiy2021Chaos}. By this we mean that the two-polariton wave function becomes highly irregular in the real space and also occupies a large region of the reciprocal space. 
\subsubsection{Ultrastrong coupling regime}\label{sec:ultrastrong}
While in the majority of cases the rotating wave approximation is justified, i.e. the characteristic energy of the qubit-photon interaction $g$ is much less than the qubit transition frequency $\omega_0$, the break down of this approximation has recently been demonstrated in the circuit QED systems based on superconducting qubits~\cite{niemczyk2010circuit,forn2017ultrastrong}, where $g/\omega_0>0.1$ has been demonstrated.  From the theory side, departure from the rotating wave approximation Eq.~\eqref{eq:RWA} leads to the Hamiltonian
\begin{align}
    H=\sum_m\omega_0 \sigma_z^{(m)}+\frac{1}{\sqrt{L}}\sum_{k} \omega_{k}a_{k}^{\dagger}a_{k}^{\vphantom{\dag}}+\sum_{m,k} g_k \sigma_x^{(m)} (a_{k}+a^{\dagger}_{k}), \label{eq:Ham_USC1}
\end{align}
where $\sigma_x^{(m)}=\sigma_m^{\vphantom{\dag}}+\sigma^{\dagger}_m$. The account for the antiresonant terms, $\sigma^{(m)}a_k,\sigma^{\dagger (m)}a_k^{\dagger}$ lifts the conservation of the total number of excitations. Thus, the Hilbert space of the solutions can no longer be factorized to the blocks with the fixed number of excitations. The analytical solution as of today has been obtained only for the simplest case of single qubit and single photonic mode, the celebrated Rabi model~\cite{braak2011integrability}. The ground state of the Rabi model is a squeezed vacuum state comprising multiple photonic Fock states. The modification of the ground state is the distinct feature of the ultrastrong coupling regime that persists in the multi-mode and multi-spin case and may lead to the cavity mediated phase transitions~\cite{Ashida2020}. It should be noted that the Hamiltonian~\eqref{eq:Ham_USC1} is not gauge invariant. To restore the gauge invariance, an additional term corresponding to the photon occupation number and coupling strength should be added~\cite{FriskKockum2019}. While at moderate coupling strengths, this term can be neglected, its omission in the ultrastrong coupling regime can lead to the unphysical phase transitions.

For the multi-mode case, the system resembles the spin-boson model~\cite{Leggett1987} and its multiple spin counterparts. The central quantity in the spin boson model is the spectral distribution of the coupling strength $J$:
\begin{align}
    J(\omega)=\frac{2\pi}{\sqrt{L}}\sum_{k}|g_k|^2\delta(\omega-\omega_k)
\end{align}
The specific shape of the spectral distribution function $J(\omega)$ depends on a specific geometry of the waveguide. One particularly explored case corresponds to the so-called Ohmic bath, $J(\omega)=\alpha\omega f(\omega/\omega_c)$, where $f(x)$ is the cut-off function which decays quickly as $x>1$, $\omega_c$ is the cut-off frequency usually defined by the waveguide band width, and $\alpha$ is the dimensionless coupling constant. It has been appreciated that the spin-boson model supports various quantum phase transitions as a function of the coupling strength~\cite{le2010quantum}. Moreover, for certain waveguide dispersions and coupling coefficients $g_k$, the model can be directly mapped to the Kondo problem~\cite{blume1970spin} having exact analytical solution. We note, that the applicability of the two-level model for  superconducting qubits remains a subject of  discussions. For example, it has been predicted in Ref.~\cite{Kaur2021} that the intrinsic multilevel structure of the qubits drastically restricts the validity of the spin-boson paradigm. 

The properties of the spin-boson model become especially interesting in the presence of disorder. In the absence of disorder at the threshold, when atom-photon coupling strength exceeds a certain threshold, all spins become aligned forming a ferromagnetic phase, while being in analogue of paramagnetic phase below threshold. The  disordered multimode Dicke model features also a quantum spin-glass phase, where a random linear combination of the cavity modes becomes superradiant~\cite{Goldbart2011,Sachdev2011,Rotondo2015b}.

There are two general approaches to the theoretical treatment of the WQED regime. Within the first approach, the waveguide Hamiltonian is written in the real space representation, with the subsequent application of the matrix product states to find the eigenspectrum of the system~\cite{peropadre2013nonequilibrium,sanchez2014scattering, wall2016simulating,Mahmoodian2019rwa}. Alternatively, one may introduce the unitary transformation, which asymptotically transforms the Hamiltonian to the one with conserving number of excitations~\cite{shi2018ultrastrong,sanchez2019single,ashida2021nonperturbative}. This resembles the polaron transformation widely used in the theoretical treatment of electron-phonon interaction in condensed matter~\cite{silbey1984variational}. For a single-qubit case of Hamiltonian~\eqref{eq:Ham_USC1}, the transformation operation $U_p$ reads
\begin{align}
    U_p=\exp\left[-\sigma_x\sum_{k}(f_{k}a_{k}^{\dagger}-f^*_{k}a_{k})\right].
\end{align}
Under the transformation, the ground state transforms to $|GS\rangle=U_p|0\rangle$, and $|0\rangle$ is the vacuum state. The parameters $f_k$ are obtained by the minimization of the ground state energy yielding the equations:
\begin{align}
    f_k=\frac{1}{\sqrt{L}}\frac{g_k}{\Delta_r+\omega_k},\quad \Delta_r=\omega_0 \e^{-2\sum |f_k|^2},
\end{align}
The transformed Hamiltonian reads
\begin{align}
    &H'=\Delta_r\sigma_z+\sum_k a_k^{\dagger}a_k^{\vphantom{\dag}}-2\Delta_r(\sigma\hat{A}^{\dagger}+\sigma^{\dagger}\hat{A})-\nonumber\\&2\Delta_r\sigma_z \hat{A}^{\dagger}\hat{A}+E_0+\mathcal{O}(f^3), \label{H_polaron}
\end{align}
where $\hat{A}=\sum f_k a_k$.
It can be seen that up to the terms quadratic in coupling constant, the transformed Hamiltonian conserves the number of excitations. Therefore, this Hamiltonian can be treated by projection to the subspace with fixed number of excitation similar to the case of WQED in the rotating wave approximation. 
The polaron picture, described by the Hamiltonian~\eqref{H_polaron} is particularly useful to gain the physical insight on the origin of the peculiar effects occurring in the ultrastrong regime. Since the ground state $|GS\rangle$ comprises the Fock states with non-zero photon occupation, the system hosts virtual photonic excitations even in the ground state. These can be realized by the non-adiabatic change of the coupling constant resulting in the photon emission from the vacuum state in the ultrastrong coupling regime~\cite{sanchez2019single}. Moreover, the ultrastrong coupling leads to the inelastic (Raman) scattering of the single photons from the WQED system~\cite{sanchez2014scattering}, and even conversion of the single incoming photon to the multiple photons of lower energy~\cite{belyansky2021frustration}. Elastic scattering also gets substantially modified in the USC regime. In~\cite{shi2018ultrastrong} the expressions were obtained  for the coherent elastic reflection and transmission coefficients $r_k,t_k$:
\begin{align}
    t_k=1+r_k,\quad r_k=\frac{i(\omega_k+\Delta_r)\mathrm{Im}\Sigma(\omega_k)}{(\omega_k-\Delta_r)\Delta_r-(\omega_k+\Delta_r)\Sigma(\omega_k)}. \label{eq:RT_USC}
\end{align}
The self energy $\Sigma(\omega)=\delta_L(\omega)-i\gamma_{\rm 1D}(\omega)$ includes the Lamb shift $\delta_L(\omega)$ and the renormalized decay rate $\gamma_{\rm 1D}(\omega)$ given by
\begin{align}
    \delta_L(\omega)=2\Delta_r^2\mathcal{P}\int \frac{\rmd k}{2\pi} \frac{f_k^2}{\omega-\omega_k},\\
    \gamma_{\rm 1D}(\omega)=\Delta_r^2 f_{k'}^2 |\partial \omega _k /\partial k|_{k=k'}^{-1}\:.
\end{align}
In the weak coupling limit, $\delta_L\approx 0$ and $ {\gamma_{\rm 1D}(\omega)\approx \gamma_{\rm 1D}(\omega_0)\equiv \gamma_{\rm 1D}}$. Thus,  the expressions~\eqref{eq:RT_USC} reduce to the conventional expressions for the qubit reflection and transmission [Eqs.~\eqref{eq:rtN} for $N=1$].

\subsubsection{Multilevel atoms}\label{sec:real}

So far, we have considered an idealized situation of a two-level atom coupled to a single propagating waveguide mode. The two-level approximation is reasonable for superconducting qubits but many modern WQED experiments are realized with $^{133}$Cs or $^{87}$Rb atoms, which have a complex degenerate multilevel structure. 

In this section, we consider a cesium atom initially prepared at its hyperfine structure level $F_g = 4$ of the ground state, see Fig. \ref{fig:decay_multilevel} (a). We assume a single photon propagating through the nanofiber with a frequency $\omega$ close to the atomic resonance frequency $\omega_0$ of the transition $|F_g = 4 \rangle \rightarrow |F_e = 5\rangle$ in the $D_2$-line. Here, $F_g$ and $F_e$ are the total angular moments of $6S_{1/2}$ ground state and $6P_{3/2}$ excited state, respectively. Consideration of the chosen transition allows one to avoid additional influence of the hyperfine structure of the excited state due to the selection rules. Therefore, the magnetic sublevels $e_{M'}$ and $g_M$ of the hyperfine excited $F_e = 5$ and the ground $F_g = 4$ states form a closed set.  
In order to simplify the consideration, in atomic physics experiments, a multilevel system  shown in Fig.~\ref{fig:decay_multilevel}(a) can be reduced to a two-level system. Fistful, this can be done by applying magnetic field that leads to splitting of Zeeman sublevels along the magnetic field direction, see Ref. \cite{Sayrin2015} . In this case, all transition have different frequencies, see Fig.~\ref{fig:decay_multilevel}(b). As a result, a propagating photon is resonant to only one transition. Another way to reduce the number of considered levels in the atomic system is based on the optical trapping technique, see Ref. \cite{Scheucher2016}. In this case, one can transfer atoms from all Zeeman sublevels to one (edge one $g_M = -4$, $g_M = 4$ or a middle one $g_M = 0$), see Fig. \ref{fig:decay_multilevel}(c). Then, for each possible photon polarization, the atom can be considered as two-level one.

\begin{figure}[t]
\centering\includegraphics[width=0.5\textwidth]{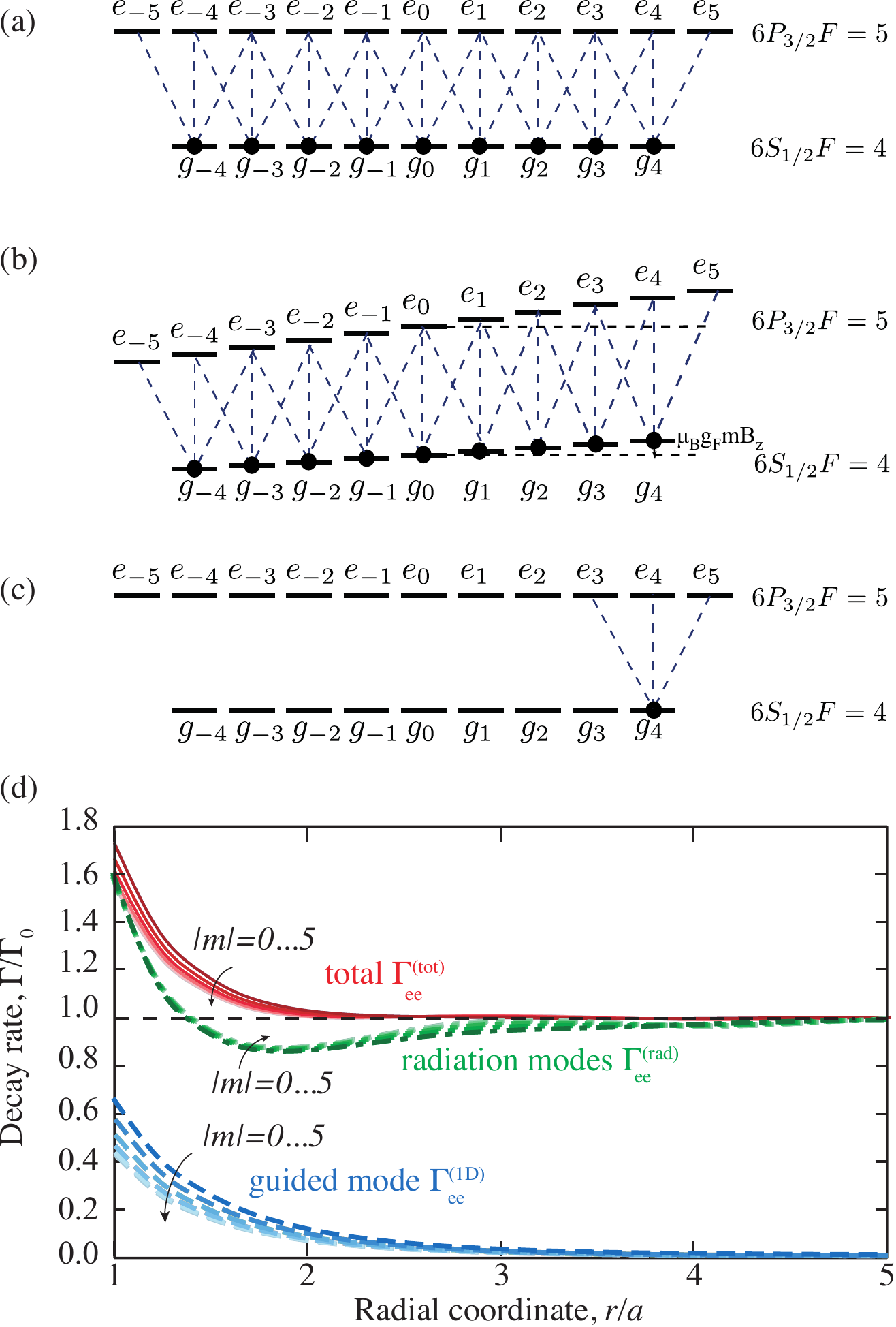}
\caption{(a-c): Schematics of $|6S_{1/2} F = 4\rangle \rightarrow |6P_{3/2}F = 5\rangle$ transition in $D_2$ line of $^{133}$Cs atom. (b) Level structure of the $|F_g = 4\rangle \rightarrow |F_e = 5\rangle$ of $^{133}$Cs in magnetic field. (c) Atoms trapped in the edge state $|F_g = 4, g_M = 4\rangle$. (d) Spontaneous emission rates for various magnetic sublevels of the excited state $|6P_{3/2} F' = 5\rangle$ of $^{133}$Cs atom. Blue dashed, green dash-dotted and red solid lines show the rates of emission into guided modes, radiation modes, and total decay rates as functions of the atom distance from the fiber center. Different lines of each plot correspond to different Zeeman sublevels of the excited state. The fiber radius was chosen as $a = 200$~nm. The wavelength of $D_2$ line of $^{133}$Cs is $\lambda_0 = 852$ nm. The refractive indexes of the fiber and the vacuum clad are $n_1 = 1.45$ and $n_2 = 1$, respectively. The decay rates are normalized to the free-space decay rate $\Gamma_0$. }
\label{fig:decay_multilevel}
\end{figure}

The case when no magnetic field is applied and the optical trapping technique is not used is more involved. For the first time, spontaneous emission of a multilevel atom in the vicinity of an optical waveguide was calculated by Fam Le Kien \textit{et al.} in Ref. \cite{LeKien2005}. It was shown that the multilevel structure of a real atom modifies its decay rate. Here, we briefly summarize results of this paper. 

The total decay rate of one Zeeman sublevel of the excited state is given by the sum $\Gamma^{\rm tot}_{ee'} = \Gamma_{ee'}^{\text{1D}} + \Gamma_{ee'}^{\text{rad}}$, where $\Gamma_{ee'}^{\text{1D}} $ and $\Gamma_{ee'}^{\text{rad}}$ describe spontaneous emission into the guided mode and into the radiation modes, respectively.
In Fig. \ref{fig:decay_multilevel} (d), one can see the spatial dependence of the spontaneous emission rates for various magnetic sublevels $|e\rangle = |6P_{3/2}F' = 5\rangle$ into the guided modes, radiation modes and both types of modes. 
The calculation demonstrates that the efficiency of the emission into the waveguide mode quickly decays with the distance from the atoms to the waveguide surface. For the atoms being located exactly at the surface the total spontaneous decay rate increases by the Purcell factor  $\sim 1.5$ with respect to the free space value $\Gamma_0$, and the fraction of emission into the waveguide mode is about $\beta\sim 0.3$. When the distance from the atoms to the surface becomes  larger than the fiber radius, the total decay rate is not much different from that in the free space,  and the $\beta$ factor drops below 10\%.

The presence of the off-diagonal elements such as $\Gamma^{\rm tot}_{ee'}$ with $e \ne e'$ is a characteristic difference from  the case of two-level atoms. They describe the decay rate of the cross-level coherence and arise only in the framework of a multilevel atom model. The knowledge of both diagonal and off-diagonal types of decay characteristics is important for the studies of absorption and emission properties of the multilevel atom.

Despite the simplicity and versatility of a two-level approximation, some quantum information applications such quantum memory, slow light, quantum computing etc. can be realized only in multi-level atomic schemes. To describe these processes, one needs to go beyond two-level approximation. 
We consider as an example  an array of $\Lambda$-type three-level atoms trapped along an optical nanofiber, see Fig.~\ref{fig:EIT_3-level}. We assume that only one ground state $|g\rangle$ is populated. Thus, adding only one additional level to the ground state $|s\rangle$ changes the collective decay rate into the waveguide. Indeed, for two-level atoms trapped near a waveguide with a spacing between atoms $d = \lambda_0/2$, the collective decay rate can be found as $ \Gamma + N\Gamma_\text{1D}$, see Eq. (\ref{eq:rtN}). However, for $\Lambda$-type three-level atoms, the collective decay rate reads  $\Gamma + N\Gamma_\text{1D}/2$. Here, the factor $1/2$ comes from presence of two channels of decay in the fiber mode for $\Lambda$-configurated atoms, see \cite{Pivovarov2021}. 

In the rest of this section, we discuss the propagation of guided light under the condition of electromagnetically induced transparency (EIT), described in \cite{AsenjoGarcia2017atom}. 
We assume that the transition $|e\rangle \rightarrow |g\rangle$ is coupled to the guided mode and the orthogonal transition $|e\rangle \rightarrow |s\rangle$ is excited by an external to the nanofiber, classical and uniform control field with the Rabi frequency $\Omega_c = 2d_{es}E_c/\hbar$. Here, $d_{es}$ is the dipole moment of the transition $|e\rangle \rightarrow |s\rangle$ and $E_c$ is the control field amplitude. Applied control field transfers the guided photon from the ground state $|g\rangle$ to a superposition of states $|g\rangle$ and $|s\rangle$, and forms a so-called dark state. Guided photon transfer to the dark state leads to two main consequences, important for quantum communications. First, it prevents the photon losses due to the long lifetime of the state $|s\rangle$. Second, it results in reduction of the group velocity and the possibility to slow down the light.

\begin{figure}[t]
\centering\includegraphics[width=0.43\textwidth]{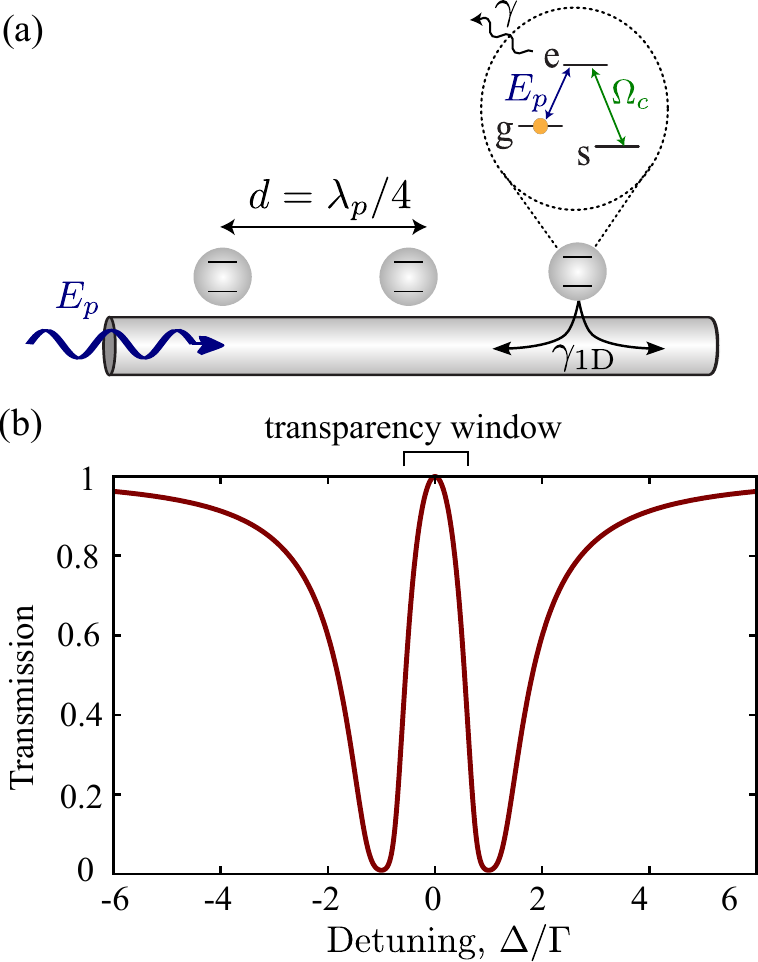}
\caption{Reproduced from \cite{AsenjoGarcia2017atom}. (a) Schematics of the electromagnetically induced transparency effect. The transition $|g\rangle \rightarrow |e\rangle$ is coupled to the guided photon mode, and the transition $|s\rangle \rightarrow |e\rangle$ is driven by classical strong control field with Rabi frequency $\Omega_c$, that is external to the nanofiber. (b) Calculated  transmission spectrum $|t_N|^2$, with the characteristic transparency window. The number of atoms is $N = 5$, the lattice period is $d = \lambda_p/4$, where $\lambda_p$ is the wavelength of the probe photon, and the control field $\Omega_c = 2\Gamma$, the guided decay rate $\gamma_{\text{1D}} = 0.5\gamma$.  }
\label{fig:EIT_3-level}
\end{figure}
When the control field is present, the interaction part of the atom-photon coupling Hamiltonian Eq.~(\ref{eq:RWA})  should be modified as follows:
\begin{align}\label{eq:RWA_lambda}
H_{\rm atom-phot}=\sum\limits_{i=1}^N&\Bigl(\frac{1}{\sqrt{L}}\sum\limits_{k} g_{k}\e^{-\rmi k z_i} a_k^\dag \sigma_{ge}^{(i)} \\&+ \frac{\Omega_c}{2}\sigma_{se}^{(i)} + {\rm H.c.} \Bigr)\:.\nonumber
\end{align}
The energy shift and the dissipation stemming from a coherent interaction between the atoms $i$ and $j$ of the array can be found from the Hamiltonian (\ref{eq:Hmn}) and correspond to $-\gamma_{\text{1D}}\sin(k_{\text{1D}}|z_i - z_j|)$ and $\gamma_{\text{1D}}\cos(k_{\text{1D}}|z_i - z_j|)$, respectively. The transmission coefficient of the atomic array affected by the external control field can be found as
\begin{equation}
t_N = \prod_{\xi}^{N}\frac{\Delta(\Delta + i\gamma ) - \Omega_c^2/4}{\Delta(\Delta +  i\gamma + \lambda_{\xi}) - \Omega_c^2/4},
\label{eq:t_EIT}
\end{equation}
where $\Delta = \omega - \omega_0$ is detuning of the guided photon frequency $\omega$ from the atomic resonance $\omega_0$, and $\lambda_{\xi} = \Delta_{\xi} + i\gamma_{\text{1D}}^{(\xi)}$ with $\Delta_{\xi}$ and $\gamma_{\text{1D}}^{(\xi)}$ corresponding to the energy shift and decay into the waveguide of the eigenstate $\xi$. Figure \ref{fig:EIT_3-level} shows the transmittance spectrum of the array of $N = 5$ atoms separated by a distance $d = \lambda_p/4$, where $\lambda_p$ is the wavelength of the guided probe photon. One can see an appearance of the characteristic transparency window  around to the atomic transition frequency.
 
From Eq.~(\ref{eq:t_EIT}), one can find an effective wavevector of the polaritonic excitation. Indeed, after the propagation of the light through the array of $N$ atoms, the transmission coefficient acquires as phase factor $t_N\propto \e^{ik_{\text{eff}}Nd}$, where $k_{\text{eff}}$ is a complex number which characterises both transmission and absorption. Expansion of Eq.~(\ref{eq:t_EIT}) in series over $\Delta$ gives:
\begin{equation}
k_{\text{eff}} = -\frac{i}{Nd}\sum_{\xi}^{N}\frac{4\lambda_{\xi} }{\Omega_c^2}\left[\Delta + \frac{4\Delta^2}{\Omega_c^2}(\lambda_{\xi} + 2i\gamma ) + ... \right].
\end{equation}
This expression is valid for any linear and isotropic quasi-1D structure. Almost all configurations have $N$ eigenstates $\lambda_\xi$ that makes calculation of $k_{\text{eff}}$ nontrivial. 
Therefore, for a chain of atoms near a waveguide, the effective polaritonic wavevector scales differently with the number of atoms and depends on the interatomic distance. However, as has been shown in \cite{AsenjoGarcia2017atom}, the group velocity at the atomic resonance $v_g(\Delta=0) = (dk_{\text{eff}}/d\Delta)^{-1}|_{\Delta = 0} = \Omega_c^2d/(4\gamma_\text{1D})$ is not affected by a specific atomic configuration. More details on light propagation through an array of atoms with complex-multilevel structure under the EIT-condition and in the presence of undirectional coupling can be found in Ref.~\cite{LeKien2015}.}

Three-level atoms driven by two light beams also enable
 amplification~\cite{Astafiev2010a} and cross-Kerr nonlinearity~\cite{Hoi2013},  see also Ref.~\cite{Roy2020} for the theoretical details.


\subsection{Chiral atomic arrays}\label{sec:chiral}
So far, we have considered a situation where an atom is symmetrically coupled to forward- and backward-propagating photons in the waveguide. However, the  complex structure of light polarization in the vicinity of  a waveguide interface results, under applied transverse magnetic field, in the chiral (directional) coupling. Namely, the strength of atom interaction with forward and backward propagating photons becomes different which provides the grounds for novel non-reciprocal \cite{Scheucher2016} and cascaded quantum systems \cite{Carmichael1993, Stannigel2012}.
The systems with broken forward/backward propagation symmetry are  now actively studied in the domain of {\it chiral quantum optics} \cite{Lodahl2017}. We start by discussing the microscopic origins of the spin-momentum locking in nanophotonic waveguide in Sec.~\ref{sec:chiral_linear}. Next, we consider directional coupling of a single atom to the waveguide mode in Sec.~\ref{sec:rad:chiral}. Section~\ref{sec:array:chiral} is devoted to the polariton excitations in the array of chirally coupled emitters. To conclude this section, we discuss recent experiments on tunable photon bunching and antibunching in the chiral setup, Sec.~\ref{sec:Prasad}.
\subsubsection{Spin-momentum locking}\label{sec:chiral_linear}
\begin{figure}
\includegraphics[width=0.4\textwidth]{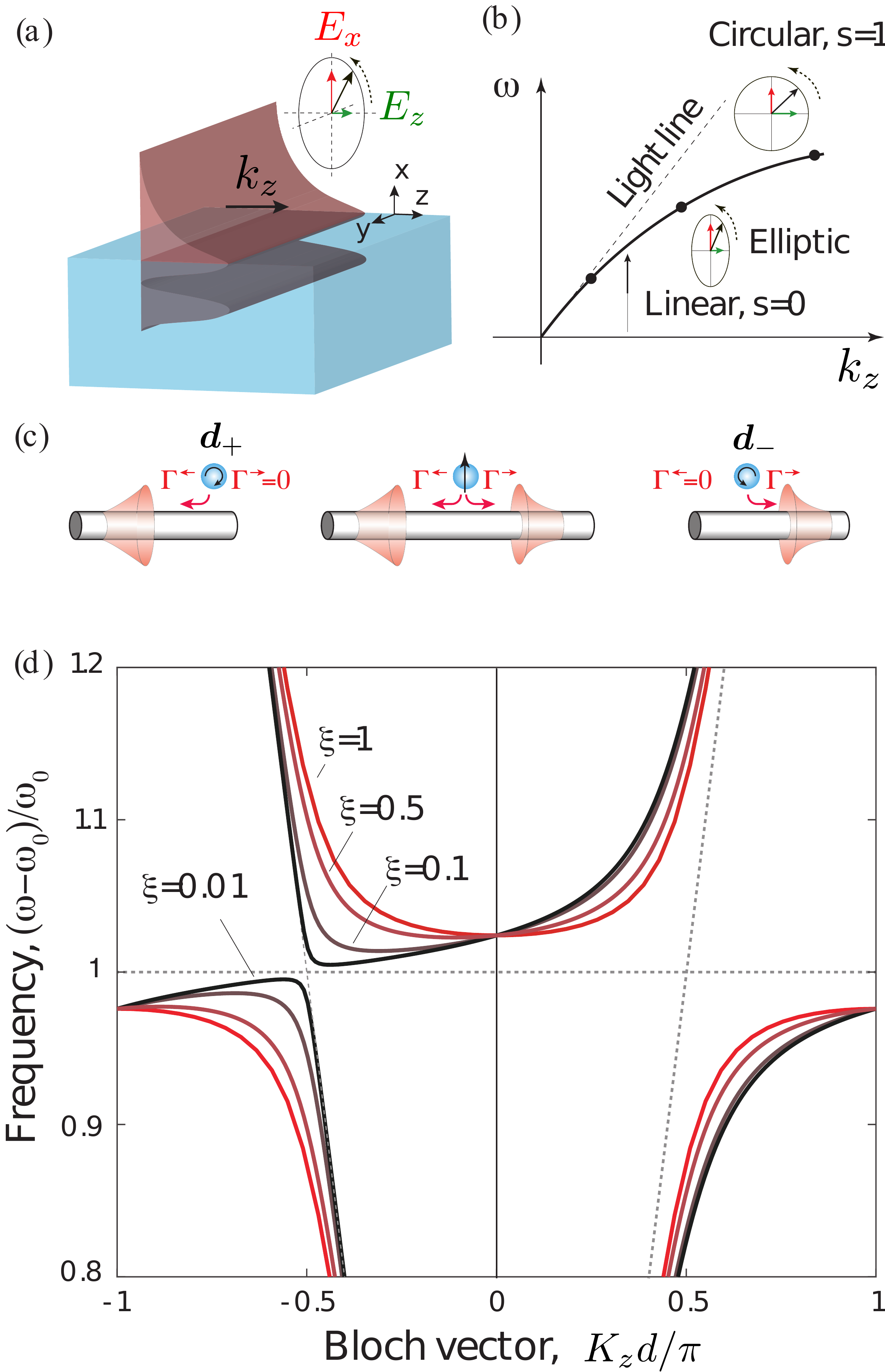}
\caption{ (a) The evanescent field at a waveguide interface. (b) Schematics of the  waveguide mode dispersion with different electric field polarization states. (c) Directional coupling of the emission for right and left circularly polarized atomic dipole transition. (d) Polariton dispersion in a chiral waveguide depending on the coupling asymmetry $\xi$.
Calculation has been performed  following Eq.~\eqref{eq:disp:chiral} for a regular array of circularly polarized $\ve d_-$ atoms  and separated with the phase $\varphi=\omega_0d/c=\pi/2$.  }\label{fig:chiral:1}
\end{figure}

Nanophotonic waveguides provide a unique platform for reaching the directional emission of photons due to spin-momentum locking effect, thus, realizing the one-way interactions between the quantum emitters. 
The spin-momentum locking can be understood by analyzing the coupling of circularly polarized optical transition  to the  guided mode of the planar waveguide, as shown in Fig.~\ref{fig:chiral:1}(a--c). The main observation is that the polarization of the guided mode is in general elliptical. Indeed, the electric field $\bm E$ outside the waveguide is a transverse plane wave, i.e. $\bm k \cdot \bm E=0$, where $\bm k=k_x\bm e_x+k_z\bm e_z$ is the wave vector. Since the guided wave is by definition evanescent outside the waveguide, the wave vector component $k_x=\sqrt{(\omega/c)^2-k_z^2}$, transverse to the waveguide surface, is pure imaginary, $k_x=\rmi \kappa $. Thus, the guided wave assumes the form $\ve E(x,z)\sim \exp{(-\kappa x)}\exp{(ik_z z)}$. The polarization state of the field is fully defined by the dispersion of the mode $k_z(\omega)$ and varies from linear polarization close to a light line $k_z=\omega/c$ to circular polarization for strongly evanescent waves when  $k_z\to \infty$, see Fig.~\ref{fig:chiral:1}(b).
We  can introduce the polarization parameter $s\in[0,1]$ such that $\ve E(x,z)=E_0/\sqrt{s^2+1}(\ve e_x-is\ve e_z)\e^{\rmi k_zz}$,  with $s=0$ and $s=1$ corresponding to linear and circularly polarized field, respectively.
Crucially, the sign of circular polarization is determined by the sign of the wave vector $k_z$, it is opposite for forward and backward going waves, $s\propto \sign k_z$. Such spin-momentum locking is a universal feature of guided and surface waves~\cite{Sinev2017}. It exists for planar waveguides, for nanofibers, for surface plasmon polaritons~\cite{ginzburg2013,Spitzer2018}, see also the  reviews on  chiral quantum optics \cite{Lodahl2017} 
and spin-momentum locking \cite{Aiello2015}. Intuitively, spin-momentum locking can be understood as a "photonic wheel" effect~\cite{Aiello2015}. The wheel rotation (light circular polarization) leads to to the translational motion along  the surface under the wheel (wave vector $k_z$) and oppositely rotating wheels travel in opposite directions. However, more general considerations show that the intrinsic origin of the directional excitation of the guided mode is related to particular angular momentum of the photon emitted by the atom \cite{Mitsch2014, Lamprianidis2021} rather than to its helicity.  

 \subsubsection{Directional atom-waveguide coupling}\label{sec:rad:chiral}
The spin-momentum locking drives  asymmetric coupling of transverse circularly polarized emitters with the dipole matrix elements 
$\ve d_\mp=d_0/\sqrt{2}(\ve e_x\mp i\ve e_z)$ to the waveguide mode. Microscopically, circularly polarized emitters can be realized by applying magnetic field, that leads to Zeeman splitting of optical transitions and changes selection rules. The emitters can be 
natural atoms (see also Sec.~\ref{sec:real})
as well as solid state emitters, quantum dots
\cite{Sllner2015} or quantum wells \cite{Spitzer2018}. 
An alternative approach is based on charged quantum dots. There, the spin of an extra electron can control the circular polarization of the charged exciton transition~\cite{Yilmaz2010,javadi2018spin}.

In the fully chiral setup ($|s|=1$), the circularly polarized emitter will be coupled either only to forward-  or only to backward- propagating waves. 
In the case of general elliptic polarization, the couplings can be characterized by emission rates of forward ($\Gamma^\rightarrow\equiv 2\GOR$) and backward ($\Gamma^\leftarrow\equiv 2\GOL$) propagating photons, that are proportional to $|\bm E^*(k_z)\cdot\bm d|^2$ and 
$|\bm E^*(-k_z)\cdot\bm d|^2$, respectively. Explicitly, the emission rates  are given by \cite{Gruner1996,AsenjoGarcia2017atom}
\begin{equation}\label{eq:GORL}
\GOR=\GO\dfrac{(1\pm |s|)^2}{(s^2+1)},\quad 
\GOL=\GO\dfrac{(1\mp |s|)^2}{(s^2+1)},\\
\end{equation}
where the total decay rate $\GO=\GOR+\GOL$ is  found from the Green function, see Eq.~\eqref{eq:gamma1dGen}.
If one takes into account only one circularly polarized mode with $s=1$, and uses Eq.~\eqref{eq:Gguided} for the guided part of the Green function, one can find that  the rate $\GOR$ ($\GOL$) is equal to zero for $\ve d_-$ ($\ve d_+$) which means fully chiral one-way coupling. We also introduce the  interaction asymmetry parameter $\xi=\GOL/\GOR=(1\mp|s|)^2/(1\pm|s|)^2$ that varies from $0$ to infinity for ideal right or left coupling, correspondingly. 
From now, we assume that the Zeeman splitting is large enough so that left- and right-circularly polarized transitions can be  spectrally separated. We  restrict the consideration to left-circularly polarized transitions and, thus, the asymmetry parameter will vary from $\xi=1$ for symmetric coupling to $\xi=0$ for fully asymmetric (chiral) coupling. 

The realistic experimental values of the asymmetry parameter vary depending on the particular quantum platform. In
\cite{Ramos2014}, the authors provide estimations of  the asymmetry parameter for Rb atoms near the fiber obtaining the limits of $10^{-3}<\xi<1$. In \cite{Corzo2016, Mitsch2014}, the estimated level of asymmetry extracted from the experimental spectra was found $\xi=0.083$ and $\xi=0.087$, correspondingly. Alternatively, the asymmetry of Rydberg atoms spin states coupling with a phonon mode was estimated as $\xi\sim 1/400$ in \cite{Vermersch2016}. The recently proposed  experimental concept of directional coupling in superconducting circuits \cite{Guimond2020} provided  the directional $\beta$-factor more than $99\%$, which  corresponds to $\xi$ parameter of at least  $\xi\sim 1/100$.

It is also instructive to consider scattering of guided photons on the asymmetrically coupled atom. The scattering is characterized by the  amplitude reflection coefficient $r$ and forward and backward amplitude transmission coefficients $t_{\rightarrow,\leftarrow}$, given by \cite{Lodahl2017}
\begin{align}\label{eq:rt1oneway}
r&=\frac{2\rmi \sqrt{\GOR\GOL}}{\omega_0-\omega-\rmi (\gamma+\GO)},
\nonumber\\
t_{\rightarrow/\leftarrow}&=1+\frac{2\rmi \GORL}{\omega_0-\omega-\rmi (\gamma+\GO)}\:,
\end{align}
where the constant $\gamma$ describes all other decay channels.
The absolute values of transmission coefficients $t_\rightarrow$ and $t_\leftarrow$ are the same, and for vanishing losses $\gamma=0$ we obtain the energy conservation law $|r|^2+|t_{\rightarrow/\leftarrow}|^2=1$\:.
In the symmetric case, when $\GOR=\GOL=\GO/2$ the reflection and transmission coefficients Eq.~\eqref{eq:rt1oneway} reduce to Eqs.~\eqref{eq:rtN} with $N=1$. In the  fully chiral setup, $\GOL=0$, and for vanishing losses, $\gamma=0$, the reflection coefficient vanishes and the transmission coefficient is equal to $1$ by the absolute value. At the resonance, we obtain $t_{\rightarrow}=-1$, so that light obtains $\pi$ phase shift when resonantly passing an atom.

There exist other possibilities to realize directional atom-waveguide interactions, not relying on spin-momentum locking. Instead, one could use nonlinearity that breaks time-reversal symmetry and leads to nonreciprocal photon transmission~\cite{Roy2010,Roy2013b,Shi2015b,Roy2017b}. To this end, the nonlinear structure also needs to lack the $z\to -z$ mirror symmetry. The quantum nonreciprocity for two superconducting qubits coupled  to a waveguide has been recently demonstrated in ~\cite{Fedorov2018}. While at lower power the structure behaved reciprocally, increase of power has led to nonreciprocal transmission driven by quantum nonlinearity. At even larger power, the reciprocity has been restored due to the saturation of the qubit transitions.
\begin{figure}[t]
    \centering
    \includegraphics[width=0.48\textwidth]{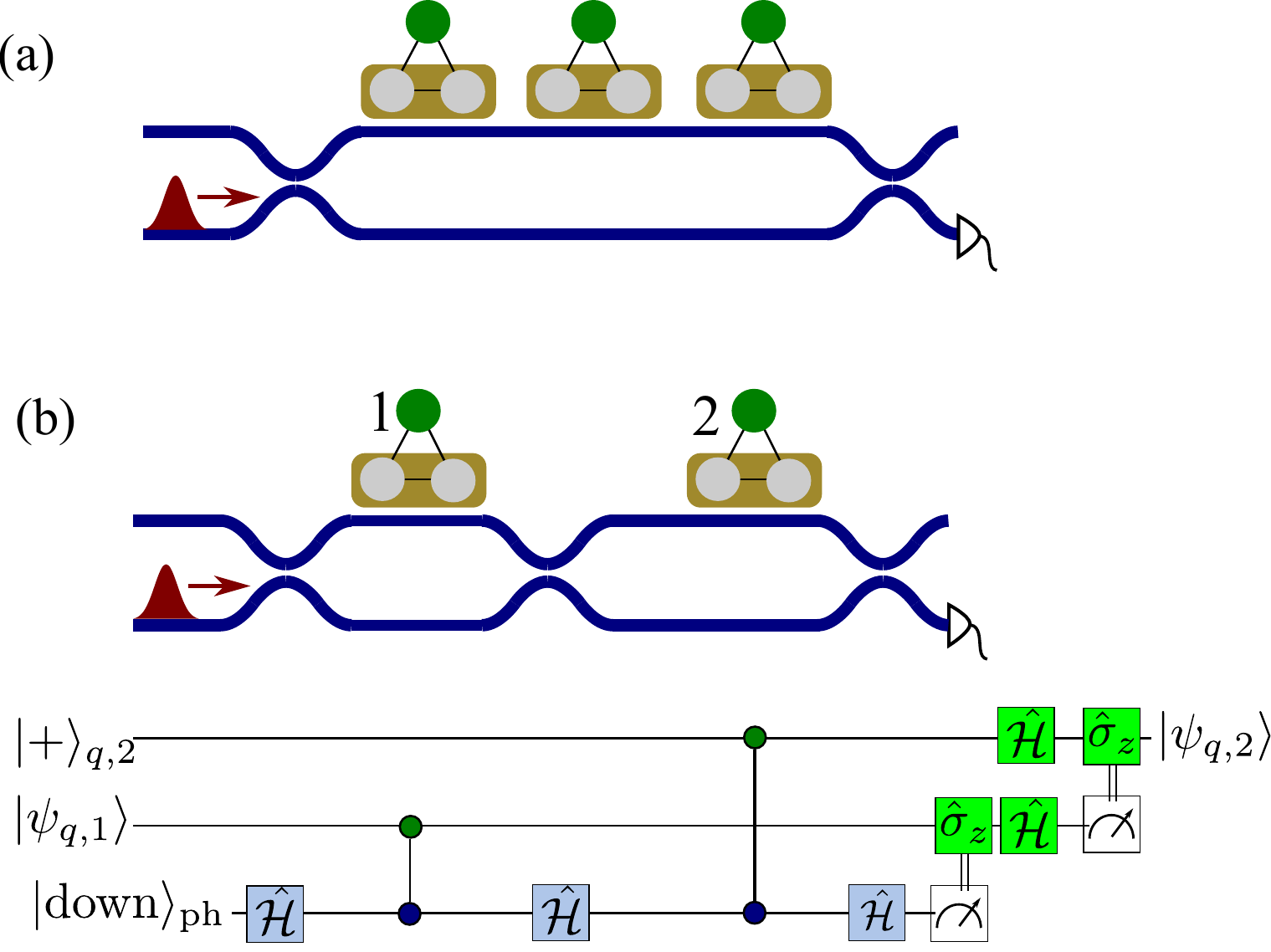}
    \caption{Illustration of the protocol proposed in Ref.~\cite{Guimond2020}. for generation of a GHZ state (a) and quantum state transfer (b) with an array of superconducting qubits coupled to two waveguides. Bottom panel in (b) shows the  quantum circuit, and is described also in Appendix~\ref{app:Guimond}. }
    \label{fig:Guimond}
\end{figure}
Moreover, there exist structures, that, while being reciprocal in the single-photon regime and not directly belonging to the traditional domain of chiral quantum optics, also rely on unidirectional atom-photon interactions. Specifically, one can have $|t_\leftarrow|^2=|t_\rightarrow|^2=1$ and $r=0$, so that the atom scatters only in the forward direction. Such system has been considered in Refs.~\cite{Guimond2020,Gheeraert2020}, where each effective atom has been formed by two identical waveguide-coupled qubits spaced by a quarter of the wavelength. Because of this $\lambda/4$ spacing, the photons reflected from first and second qubits interfere destructively, and the backscattering is suppressed. The array of such qubit dimers illustrated in Fig.~\ref{fig:Guimond}(a) 
has been proposed in Ref.~\cite{Guimond2020} for generation of complex quantum states such as the GHZ state and the 1D cluster state. The structures under consideration consist of the qubit dimers (dark yellow rectangles) coupled to two waveguides (shown in blue). Each dimer also interacts with an additional stationary qubit, shown by the green color. When the stationary qubit is in its ground state, the dimer transmits a photon with a $\pi$-phase shift, and when the stationary qubit is excited, the phase shift is equal to zero.  Judiciously linking the waveguides by beamsplitters and using photons in the waveguides as ``flying qubits'', one can then realize complex quantum states in the stationary qubit array.

For example, the proposed protocol to generate the GHZ state shown in Fig.~\ref{fig:Guimond}(a) starts by initialization of the stationary qubits in the product state 
$|+\rangle_1\otimes |+\rangle_2\ldots$, where $|+\rangle_n$ is the Hadamard state,
$|\pm\rangle\equiv (|1\rangle\pm|0\rangle)/\sqrt{2}$,  and sending one photon in the lower waveguide. As is described in more detail in Appendix~\ref{app:Guimond}, upon the  conditional detection of the transmitted photon in one of the waveguides, the stationary qubit array ends in one of the GHZ states
$|+\rangle_1\otimes |+\rangle_2\ldots\pm |-\rangle_1\otimes |-\rangle_2\ldots$.
Slightly  more complicated protocol with three beam splitters,
shown in Fig.~\ref{fig:Guimond}(b), enables the transfer of an arbitrary quantum state $|\psi_q\rangle$ of the qubit $1$ to the qubit $2$. In this case, photon scattering realizes an effective controlled-Z gate between the distant qubits,
thereby enabling universal quantum computation. A larger array, with  more dimers, separated by the beam splitters, allows one to generate a one-dimensional photon cluster state.

A concept based on unidirectional scattering can be implemented even in free space without any waveguide. Namely, it has been proposed in Ref.~\cite{Grankin2018} to couple an atom to an auxiliary two-dimensional bilayer atomic array that acts as a ``quantum antenna" providing  unidirectional photon emission.  A related experimental demonstration has already been made in  Ref.~\cite{Stiesdal2021} for  three clouds of Rydberg ${}^{87}$Rb atoms. Each cloud preferentially scattered photons in the forward direction, realizing cascaded coupling.

 \subsubsection{Arrays of chirally coupled emitters}\label{sec:array:chiral}

Waveguide-mediated chiral coupling between emitters can be considered using an effective non-Hermitian Hamiltonian Eq.~\eqref{Sigma-renormalized} with traced out electromagnetic field.
Combining Eqs.~\eqref{eq:Gguided} and \eqref{eq:GORL} we obtain 
\begin{align}
\label{eq:chiral_coupling_const}
V_{mn} = -\rmi \gamma_{\rm 1D}\delta_{mn}+
\begin{cases} 
	-\rmi\GOR e^{\rmi \omega (z_m-z_n)/c}\ \text{for}\ z_m>z_n,\\[2ex]
	-\rmi\GOL e^{\rmi\omega (z_n-z_m)/c}\ \text{for}\ z_m<z_n\:.
\end{cases} 
\end{align}
where  $z_m$ and $z_n$ are emitter coordinates along the waveguide. 
Using this Hamiltonian, one can obtain the dispersion of polaritonic Bloch waves, $\psi_m\propto \e^{\rmi K_zm}$ in a periodic array atoms with the spacing $d$~\cite{fedorovich2020disorder,Calajo2022}: 
\begin{equation}\label{eq:disp:chiral}
\omega-\omega_0=\dfrac{\GO}{1+\xi} \left[ \text{cot}\left(\dfrac{K_zd-\varphi}{2}\right)+\xi \text{cot}\left(\dfrac{K_zd+\varphi}{2}\right)\right]\:,
\end{equation}
where $\varphi=\omega/c$. The two terms in the square brackets describe avoided crossing of forward and backward propagating photons dispersion with the atomic resonance. For purely symmetric coupling $\xi=1$, the dispersion relation transforms into Eq.~\eqref{eq:Kd}. The dependence of dispersion on the asymmetry parameter  $\xi$ is plotted in Fig.~\ref{fig:chiral:1} and it demonstrates strongly unidirectional character of polariton propagation for $\xi\to0$, that is $\omega(K_z)\ne \omega(-K_z)$. 
\begin{figure}
\includegraphics[width=0.4\textwidth]{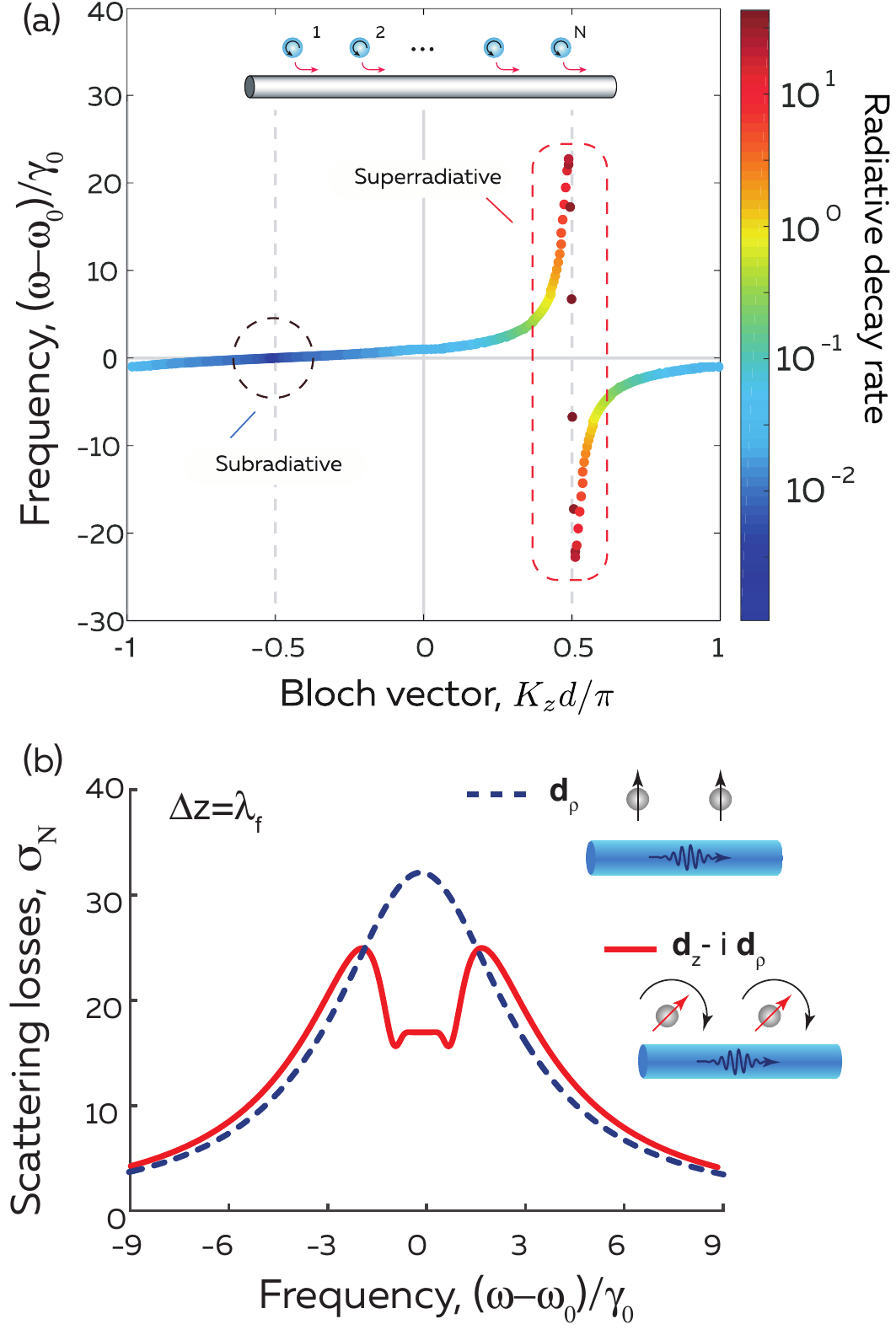}
\caption{(a) {Reproduced from Ref.~\cite{fedorovich2020disorder}}. The frequencies and radiative rates of polariton eigenstates of a finite chain regularly spaced atoms chirally coupled   through a waveguide mode with $\xi=2\cdot 10^{-5}$ depending on the corresponding Bloch wave vector. Radiative decay rate is normalized to the decay rate of individual atom $\GO$. Calculation has been performed for $N=400$ atoms and the anti-Bragg spacing  $\varphi=\omega_0d/c=\pi/2$. (b) {Reproduced from Ref.~\cite{Kornovan2016}}. The scattering losses spectrum for an array of $N=200$ atoms with  linear (black dashed line) and purely circular (red solid line) dipole transitions at the Bragg spacing condition $\varphi=\omega_0d/c=\pi$.
The losses are normalized to a resonant value of losses for photon scattering on a single atom.
 }\label{fig:chiral:2}
\end{figure}

The effects of chiral coupling can be also observed in the spectrum of eigenmodes of the finite atomic array.
Figure~\ref{fig:chiral:2} (a) compares polariton energy spectra calculated by numerically diagonalizing the Hamiltonian Eq.~\eqref{eq:chiral_coupling_const} for a fully chirally coupled arrays of $N=400$ atoms in Ref.~ \cite{fedorovich2020disorder}. As  discussed in Sec.~\ref{sec:single_photon}, the eigenstates are just standing waves with the frequencies satisfying the dispersion law for the infinite structure Eq.~\eqref{eq:disp:chiral}. Hence, the calculated eigenfrequencies for the finite structure lie on the dispersion curve for the infinite structure that is strongly asymmetric in the chiral case.  Similarly to the symmetric case of Fig.~\ref{fig:Kd},  the radiative decay rate is at maximum for the eigenstates with the wave vector $K_z$  closest to the wave vector of light $\pm \omega_0/c$
and when the polariton wave vector is strongly detuned from the light wave vector, the polariton states become subradiant. We also refer the reader to Ref.~\cite{Olmos2020} for a detailed theoretical analysis of superradiant chiral emission from atomic arrays into the nanofiber. It is predicted there that 
near-perfect chirality can be achieved already for arrays containing 10 to 15 atoms, by phase matching a superradiant collective guided emission mode via an external laser field. 

The effect of the chirality in atom-photon coupling  near a nanofiber can be clearly seen in the spectrum of the photons passing through the waveguide. Figure~\ref{fig:chiral:2}(b) shows the loss spectrum of the photons scattered in free space while passing the regular array of $N=200$ atoms with linear or purely circular dipole transitions~ \cite{Kornovan2016, AsenjoGarcia2017exp}. One can see that at the Bragg condition $\varphi=\omega_0d/c=\pi$ the spectrum strongly changes its profile and a pronounced dip appears due to the suppressed Bragg interference. The chiral origin of atom-photon interaction in such a setup is crucial for quantitative explanation of the photon reflection  spectra observed in experiments~\cite{Corzo2016} (see Sec.~\ref{sec:Bragg} for more details).

\subsubsection{Photon bunching and antibunching in a chiral waveguide}\label{sec:Prasad}
Here, we examine the correlations between photons in a waveguide chirally coupled to an array of closely spaced atoms. We consider a fully chiral setup, $\xi=\GOL=0$. In this case, the photons are transmitted by atoms one by one. The reflection is absent and the transmission coefficient through $N$ atoms is just a product of transmission coefficients of individual atoms\:
\begin{equation}
\label{eq:Ntchiral}
t_N=t_1^N,\quad t_1=1+\frac{2\rmi \gamma^\rightarrow}{\omega_0-\eps-\rmi (\gamma+\gamma^\rightarrow)}\:.
\end{equation}
Here, the radiative decay rate $\gamma^\rightarrow$ is linked to the matrix element of the atom coupling to the right-going photon mode $g$ as $\gamma^\rightarrow=g^2/(2c)$. 

The photon-photon correlation function for a single resonantly excited two-level atom chirally coupled to the waveguide is given by
\begin{equation}\label{eq:g21chiral}
g_1^{(2)}(0)=\frac{(\gamma+\gamma^\rightarrow)^2(\gamma-3\gamma^\rightarrow)^2}{(\gamma-\gamma^\rightarrow)^4}\:.
\end{equation}
It can be obtained by standard input-output techniques of quantum optics~\cite{Kojima2003,Koshino2004}, the problem is also  quite  similar to photon reflection from a one-sided cavity ~\cite{Rice1988}.

Equation~\eqref{eq:g21chiral} demonstrates that the photon-photon correlations are very sensitive to the ratio of the decay rates $\gamma$ and $\gamma^\rightarrow$ and it is possible to realize both bunching and antibunching. Specifically, for $\gamma=0$ one has $g_1^{(2)}(0)=9$. Increasing the value of $\gamma$ leads to an even stronger bunching. The value of  $\gamma=\gamma^\rightarrow$ when the single-photon transmission is suppressed, $t_1=0$, corresponds to a perfect bunching, $g_1^{(2)}(0)=\infty$. Further increase of $\gamma$ leads to the perfect antibunching at $\gamma=3\gamma^\rightarrow$.
However, in practice the coupling of a single natural atom to a waveguide is very weak,  $\gamma^\rightarrow/\gamma\sim 1\%$. This means that photons transmitted through one atom are almost uncorrelated. The correlations can be enhanced by either putting an atom in cavity \cite{Dayan2008,Kimble2009,Scheucher2016} or by increasing the number of atoms $N$\cite{Prasad2020}.   For the artificial atoms, such as superconducting  qubits or quantum dots, the coupling can be much stronger, see also Fig.~\ref{fig:platforms}.

\begin{figure}[t!]
\centering\includegraphics[width=0.43\textwidth]{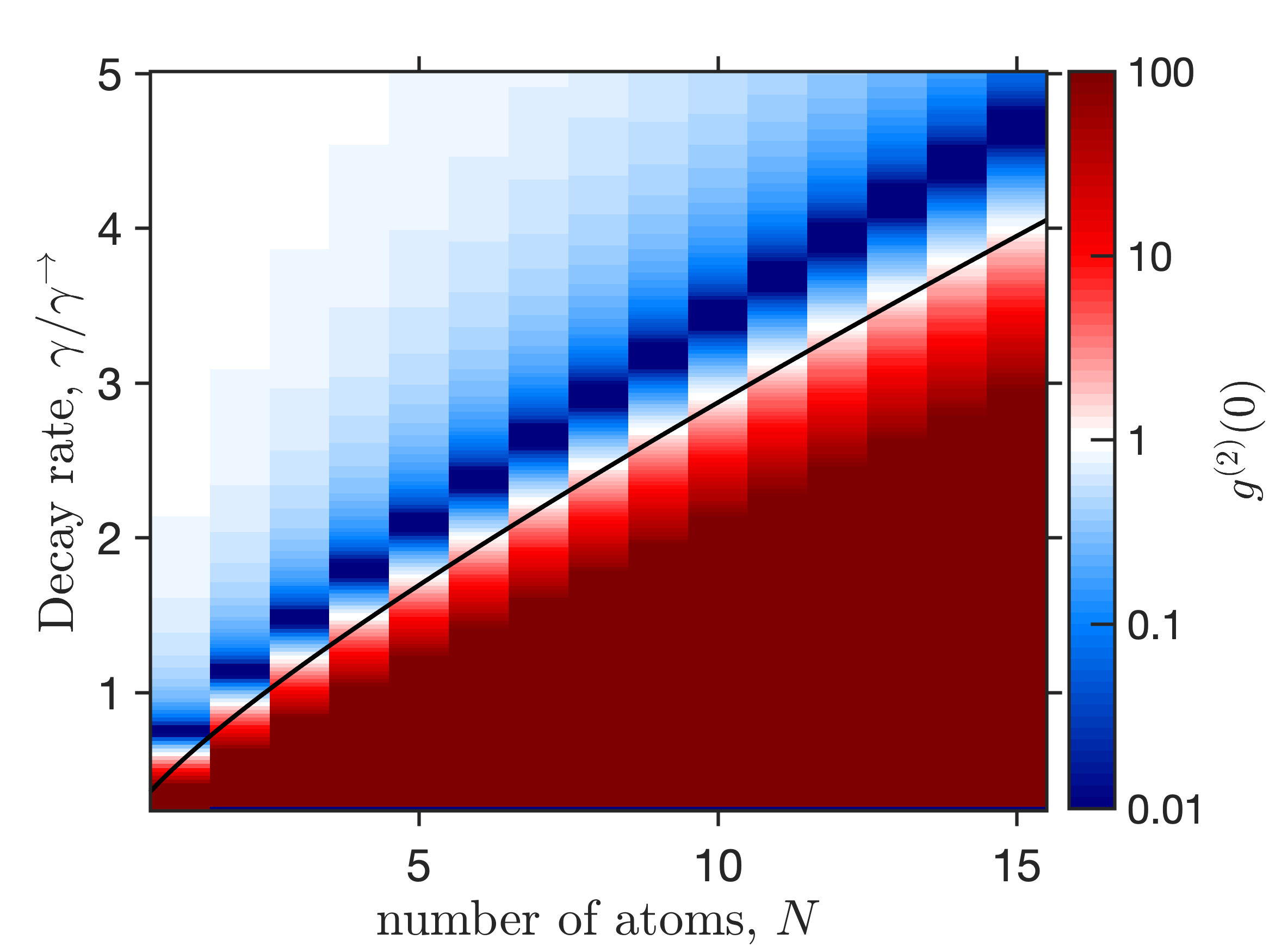}
\caption{Dependence of the photon-photon correlation function $g^{(2)}(0)$ in a chiral waveguide on the number of atoms $N$ and on the ratio of nonradiative and radiative damping rates $\gamma/\gamma^\rightarrow$.  Black curve shows the threshold value $N^*(\gamma)$  given by Eq.~\eqref{eq:Nstar-chiral} when $g_{N*}^{(2)}(0)=1$. Calculation has been performed  using Eq.~\eqref{eq:g2Nchiral} for light incident at the resonant frequency $\eps=\omega_0$ at the two-level atoms ($U\to \infty$). }
\label{fig:g2-chiral-N}
\end{figure}

\begin{figure}[t!]
\centering\includegraphics[width=0.4\textwidth]{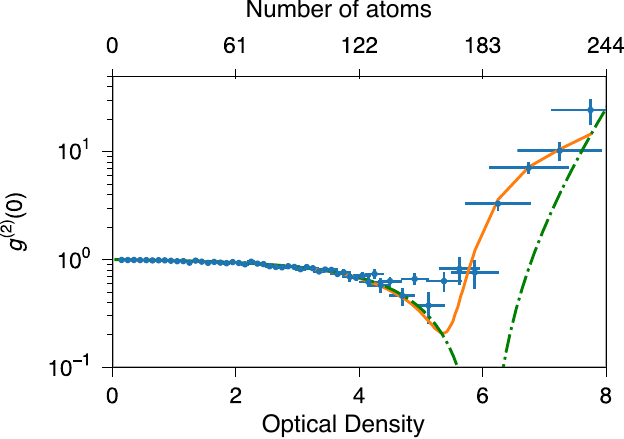}
\caption{{Reproduced from Ref.~\cite{Prasad2020}}.
Experimentally measured (crosses) and calculated photon-photon correlation functions depending on the number of atoms in the array and optical density. Solid orange and dash-dotted green curves have been calculated without and with taking into account the experimental uncertainty in the optical density. The parameters of experiment in Ref.~\cite{Prasad2020} correspond to $\gamma/\gamma^\rightarrow\approx 130$.}
\label{fig:prasad}
\end{figure}

Before proceeding to the results of the experiment \cite{Prasad2020}, we will first briefly discuss the theoretical problem of  photon pair scattering in the chiral setup.   Most interesting effects occur  when the number of atoms reaches $N\sim \gamma/\gamma^\rightarrow\sim 100$, but calculations are  significantly more involved than those for  non-chiral Dicke problem~\cite{Yudson1984}, discussed in Sec.~\ref{sec:two_photons_rt}, even when the inter-atomic spacing is zero. The reason is that all the atoms in the non-chiral problem are equivalent to each other, while in the chiral case they are ordered from left to right and hence are not equivalent. However, the problem can be still solved exactly by means of the Bethe ansatz \cite{Ringel2014,Lodahl2018}. 
An important milestone in theoretical research has been achieved  in Ref.~\cite{Mahmoodian2020} where the Dicke problem has been 
considered  for a chiral waveguide with multiple photons and multiple atoms. It has been demonstrated that the pulse transmission through the atomic array can be satisfactorily described by taking into account only relatively simple bound eigenstates states of the Bethe ansatz \cite{Yudson1984,Yudson1985} and the connection to solitons in the classical optics regime with large photon numbers has been made. More recently, the transition from the quantum to the classical nonlinear optics regime  has been analyzed  numerically for both non-chiral and chiral structures in Ref.~\cite{Calajo2022}. In Appendix~\ref{app:2scat-chiral}, we present an alternative equivalent derivation using the Green's function technique from Refs.~\cite{Baranger2013,Fang2014,Poshakinskiy2016}. In the case when $\gamma\gg \gamma^\rightarrow$, the photon-photon correlation function is well described by the approximate equation
\begin{equation}\label{eq:g2Nchiral:asymp}
g_N^{(2)}(0)\approx \left(1-\frac{\sqrt{2} \gamma^\rightarrow}{2 \gamma}\exp\left(\frac{4N\gamma^\rightarrow}{\gamma}\right)\right)^2\
\end{equation}
and its dependence on the number of atoms $N$ and on the decay rate $\gamma$ is shown in Fig.~\ref{fig:g2-chiral-N}. The result is significantly more interesting than in the non-chiral situation, compare Fig.~\ref{fig:g2-chiral-N} and Fig.~\ref{fig:g2N}. Specifically, the dependence of the correlation function on the number of atoms is nonmonotonous: increase of $N$ leads first to the antibunching and then to the bunching. The bunching threshold corresponds to $N^*\sim \gamma/\gamma^\rightarrow$, or, more precisely, one has $g_{N*}^{(2)}(0)=1$ at
\begin{equation}\label{eq:Nstar-chiral}
N^*(\gamma)\approx \frac{\gamma}{8\gamma^\rightarrow}\left(3\ln 2+2\ln\frac{\gamma}{\gamma^\rightarrow}\right)
\end{equation}
(black curve in Fig.~\ref{fig:g2-chiral-N}).  Qualitatively, the nonmonotonous behavior of the correlation function is caused by the interference of two contributions to the photon pair transmission coefficient (first and second term in Eq.~\eqref{eq:g2Nchiral:asymp}). The two terms correspond to an independent transmission of two photons and to the transmission of the correlated photon pair, are of the opposite sign. The correlated contribution becomes dominant at larger $N$, when the single photon transmission is suppressed, leading to the photon bunching. The transition from independent photon propagation to antibunching to bunching with increase of the atom number has been first observed in Ref.~\cite{Prasad2020}. The experimental dependence $g_N^{(2)}(0)$ on $N$ from Ref.~\cite{Prasad2020} is presented in Fig.~\ref{fig:prasad}. The measured value of the correlation function has been tuned from $g^{(2)}(0)\approx 0.37\pm 0.12$ for $N\approx 160$ atoms to $g^{(2)}(0)\approx 24\pm 7$ for $N\approx 200$ atoms. In the experiment, the antibunching was not perfect due to the fluctuations of the optical density resulting from the uncertainty in the preparation of an atomic ensemble and from the photon shot noise. This can be seen by comparing green and orange curves in Fig.~\ref{fig:prasad}. While the dash-dotted green curve has been calculated in Ref.~\cite{Prasad2020} for an idealized situation and corresponds to Eq.~\eqref{eq:g2Nchiral:asymp}, the solid orange curve takes the uncertainties into account and describes the experiment quantitatively. 

The correlated photon transport has been also studied for arrays of three-level atoms coupled to a symmetric \cite{Roy2011b,Roy2014,Song2017} and a chiral waveguide~\cite{Pohl2021}. For the chiral waveguide, it has been predicted that in  conditions  of electromagnetically induced transparency (EIT) of the three-level medium, a high degree of antibunching and photon transmission can be maintained in the presence of moderate losses~ \cite{Pohl2021}. While in this section we focused on a fully chiral setup, the behavior of quantum photon-photon correlations in the  case of general asymmetric coupling is a standing problem despite some recent progress in this field~\cite{jen2021bound}.

\subsection{Two-dimensional atomic arrays}\label{sec:2d}
As has been already mentioned in Sec.~\ref{sec:platforms}, an ordered two-dimensional atomic array with a small lattice period scatters incident light in the far field only in a certain direction, determined by in-plane light wave vector, because diffraction is not possible. Thus, the problem of photons interaction with the array is quasi-one-dimensional and  similar to a typical WQED problem with the whole array playing the role of an effective atom~\cite{Rui2020}. In this section, we discuss light reflection from such an effective atom in more detail.

Reflection of light from the planar array of scatterers is a well-known problem in classical optics~\cite{lagendijk_review,Khitrova2007nat,Gippius2009}. More recently, it has been extensively studied in the field of metasurfaces, artificial two-dimensional arrays of resonant scatterers with the spacing smaller than the light wavelength. Such scatterers can be metallic nanoparticles with plasmonic resonances~\cite{Decker2011}, dielectric and semiconductor particles, e.g. made of silicon, that have the Mie optical resonances ~\cite{Kuznetsov2016,Kivshar2018}.  The two-dimensional atomic arrays could be viewed as quantum metasurfaces with strong optical nonlinearities at the single-quantum level~\cite{Bettles2020,Bekenstein2020,Moreno2021,Solomons2021,Zhang2022photonphoton}.

\begin{figure}[!t]
 \centering\includegraphics[width=0.48\textwidth]{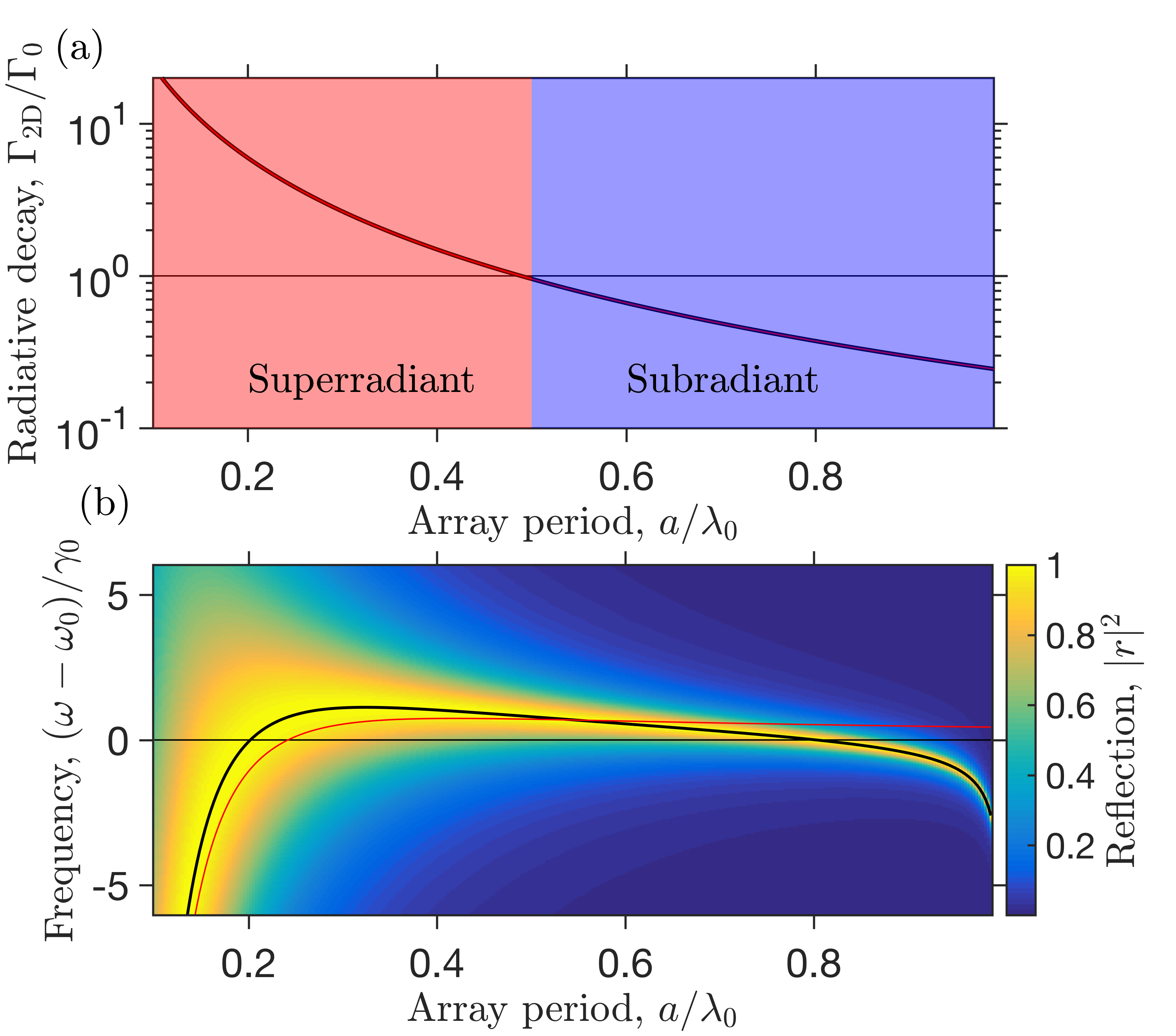}
 \caption{(a) Collective radiative decay rate of an atomic array depending on the ratio of the array period $a$ to the light wavelength at the atomic resonance $\lambda_0$.
(b) Light reflection spectra depending on the ratio $a/\lambda_0$.
Thick black line shows the collective resonance frequency $\widetilde\omega_0$   Eq.~\eqref{eq:w0g0meta}, 
solid red line shows the analytical answer found using Eq.~\eqref{eq:Cappr}. Calculated neglecting the nonradiative decay rate $\Gamma$ for the state with zero in-plane wave vector.
   }\label{fig:Lukin}
\end{figure}

Reflection and transmission coefficients for a single photon, normally incident upon the   array, can  presented in the form
\begin{equation}\label{eq:rt2d}
r=\frac{\rmi \gamma_{\rm 2D}}{\widetilde\omega_0-\omega-\rmi (\gamma + \gamma_{\rm 2D})},\quad 
t=\frac{\widetilde\omega_0-\omega-\rmi \gamma}{\widetilde\omega_0-\omega-\rmi (\gamma + \gamma_{\rm 2D})}
\end{equation}
that reminds reflection and transmission coefficients Eq.~\eqref{eq:rtN} of just one atom, coupled to a waveguide. Here, $ \gamma_{\rm 2D}$ is the collective radiative decay rate of the atomic array and $\widetilde \omega_0$ is the resonance frequency modified by the collective coupling with light (that can be viewed as cooperative Lamb shift), given by
\begin{equation}\label{eq:w0g0meta}
\widetilde\omega_0=\omega_0-\frac{3\gamma_0 \lambda_0^3}{16\pi^3} \Re C,\quad 
\gamma_{\rm 2D}=\gamma_0+\frac{3\gamma_0 \lambda_0^3}{16\pi^3}\Im C\:,
\end{equation}
where $\Gamma_0 = 2\gamma_0$ is the spontaneous decay rate of a single atom in free space and $C$ is the so-called interaction constant~ \cite{Simovski1999, belov2005}. The explicit expression for $C$ and the derivation details are presented in Appendix~\ref{app:lattice-sum}, see in particular Eq.~\eqref{eq:C}. In the limit when the spacing between the atoms is much smaller than the light wavelength, one can show that  \begin{equation}\label{eq:Cappr}
 C\approx \frac{2\pi\rmi \omega}{c a^2}+\frac{S+(\omega/c)^2S'}{2},
\end{equation}
where $S\approx 9.03/a^3$ and $S'\approx -3.90/a$. Substituting Eq.~\eqref{eq:Cappr} into Eq.~\eqref{eq:w0g0meta} we find that \cite{Kavokin1991,Ivchenko1992b} 
\begin{equation}\label{eq:Gamma0appr-2D}
\gamma_{\rm 2D}=\gamma_{0}\frac{3\lambda^2}{4\pi a^2}\:.
\end{equation}
Hence, similarly to the classical Dicke formula for the dense three-dimensional array, where the collective decay rate scales as $\gamma_{\rm 3D}\sim \gamma_{0}(\lambda/a)^3$, the decay rate of the two-dimensional array exhibits cooperative enhancement with the factor of the order of number of atoms per wavelength square. 
Figure~\ref{fig:Lukin}(a) shows the dependence of the collective radiative decay rate $\gamma_{\rm 2D}$ on the ratio of the array period to the light wavelength $a/\lambda_0$. For a small period, the array exhibits a superradiant behavior, $ \gamma_{\rm 2D}>\gamma_{0}$, but the radiative linewidth quickly decays with the growth of lattice spacing and for $a/\lambda_0>\sqrt{3/(4\pi)}\approx 0.5$ the structure becomes a subradiant one, $\gamma_{\rm 2D}<\gamma_{0}$.
This enhancement has been observed experimentally for the
quantum wire arrays \cite{Ivchenko1992b} and 
quantum dot arrays \cite{Khitrova2007nat}, but detailed studies of collective light-matter coupling were prevented by the strong inhomogeneous broadening. Much more experimental progress has been made for metamaterials and, recently, optical lattices~\cite{Rui2020}.

Figure~\ref{fig:Lukin}(b) presents the reflection spectrum depending on the array period (we have recalculated it after Ref.~\cite{Yelin2017}). While the linewidth of the reflection resonance decreases monotonously for larger spacing, following Fig.~\ref{fig:Lukin}(a), the behavior of the resonance frequency $\widetilde\omega_0$ is more subtle and non-monotonous. For small spacings, the cooperative Lamb shift diverges for $a\to 0$ as $\widetilde\omega_0-\omega_0\propto -\gamma_{\rm 1D}(\lambda_0/a)^3$ as follows from Eq.~\eqref{eq:Cappr}, red line in Fig.~\ref{fig:Lukin}(b). Qualitatively, the Lamb shift is determined by the near-field dipole-dipole interactions between atoms. However, the value of the Lamb shift crosses zero for $a\approx 0.2 \lambda_0$ due to the destructive interference of the two terms $S$ and $S'(\omega/c)^2$ in Eq.~\eqref{eq:Cappr}. This point corresponds to the perfect reflection when $|r(\omega_0)|=1$. Another special point when the Lamb shift vanishes and $|r(\omega_0)|=1$ is $a\approx 0.8\lambda_0$ ~\cite{Bettles2016,Yelin2017}.

We now discuss in a bit more detail a recent theoretical proposal of  atom-made metasurfaces for generation of highly entangled photon states~\cite{Bekenstein2020}, see also Ref.~\cite{Srakaew2022} for preliminary experimental results.
Generation and manipulation of Schr\"{o}dinger cat states has been recently demonstrated experimentally in Rydberg atom arrays \cite{Omran2019} and superconducting qubit arrays \cite{Song574}. Atom-made metasurfaces present a natural further step in this direction. The proposal of Ref.~\cite{Bekenstein2020} is based on placing a single ancillary atom near the metasurface. Next, by changing the quantum state of the ancillary atom one can control via the Rydberg interactions whether photons will be fully reflected or fully transmitted by the metasurface. This is because even  a single atomic layer with subwavelength spacing can realize perfect reflection, thus enhancing the coupling of the ancillary atom with photons. Specifically, it is proposed to coherently drive a 2D array of three-level atoms in the electromagnetically (EIT) induced transparency regime~\cite{FleischhauerRMP}. 
The EIT condition is also modified by the Rydberg interactions. If the ancillary atom is in the ground state $|g'\rangle$, the metasurface is in the $|U\rangle$ state, uncoupled from the incident light.  When the atom is in its Rydberg state $|r'\rangle$, the  metasurface is detuned from the EIT condition to the state $|C\rangle$ and fully reflects light.
 One starts by preparing the ancillary atom in the state $\psi=\tfrac{1}{\sqrt 2}(|g'\rangle+|r'\rangle)$,
so that the fully system atom+metasurface is in the state
$\tfrac{1}{\sqrt 2}(|g'\rangle|U\rangle+|r'\rangle|C\rangle)$. Next, one sends $N$ initially unentangled photons to different points in the array plane, and performs the projective measurement of the ancillary  atom in the basis $\tfrac{1}{\sqrt 2}(|g'\rangle\pm |r'\rangle)$ . As a result, the scattered light is in the Greenberger--Horne--Zeilinger (GHZ) photonic state $\tfrac{1}{\sqrt 2}( |0\rangle^{\otimes N}+|1\rangle^{\otimes N})$, where the states $|0\rangle$ and 
$|1\rangle$ correspond to transmitted and reflected photons. Such quantum scattering  corresponds to a  controlled-NOT (CNOT) gate for photons that processes photons in parallel due to the planar array geometry. The process efficiency is limited by the finite range of the Rydberg interactions, controlling the EIT condition. The GHZ state fidelity also depends on the number of atoms.  According to the calculation, the fidelity of over $90\%$ requires arrays with more than $20\times 20$~atoms. This protocol can be further developed to realizing more complex quantum states \cite{Bekenstein2020}. For example,  changing the ancillary atom state between the photon scattering processes should lead to generation photon cluster states. Highly entangled free-space photon states could be realized by coupling several ancillary atoms to the metasurface.


\section{Experimental demonstrations}\label{sec:applications}
We will now discuss experimental demonstrations, highlighting potential applications of the waveguide quantum electrodynamics platform. 
Since this field is rapidly evolving, covering all relevant works does not seem feasible. Instead, we chose to consider in detail  several experiments representing major research directions. We start in Sec.~\ref{sec:superradiance} by discussing superradiance in the waveguide-coupled atomic arrays measured in Ref.~\cite{Solano2017} and proceed to the generation of collective atomic excitations and quantum light in this setup (Sec.~\ref{sec:generation}), focusing on the experiment of Ref.~\cite{Corzo2019}. Next, we consider the slow light effect under conditions of the electromagnetically induced transparency reported in ~\cite{Gouraud2015,Sayrin2015} and potential applications on the quantum memory in Sec.~\ref{sec:memory}.
Section~\ref{sec:ustinov} is devoted to the demonstration of the subradiant modes in the transmon qubit array reported in \cite{brehm2020waveguide}.  
Section~\ref{sec:Bragg} presents experimental results \cite{Corzo2016} for the Bragg-spaced atomic arrays with both unidirectional and chiral interactions.

{An important comment should be made regarding disorder in the atomic Bragg arrays which is inevitably present at the current level of experimental technologies. The achievable filling factor is around $\sim 0.5 \div 0.8$ with random occupation of the lattice sites by the atoms \cite{Goban2012, Prasad2020, Corzo2016}. At the same time, the disorder in the atomic system may harm the quantum states such as subradiant states leading to non-homogeneous broadening as it is shown in Fig.~\ref{fig:Rui} for 2D atomic array systems. In 1D systems, the effects of disorder has been extensively studied theoretically in the case of achiral~\cite{Vahid2016, Kornovan2019} and chiral \cite{Mirza2017, Mirza2018, Jen2020a, fedorovich2020disorder} interactions. Alternatively, the influence of disorder on spectral properties of  semiconductor polaritonic lattices has been also studied previously \cite{Malpuech1999,Kosobukin2003,Kosobukin2007}. }


\subsection{Superradiance and subradiance in waveguide-coupled atomic arrays }\label{sec:superradiance}

\begin{figure}[!t]
 \centering\includegraphics[width=0.4\textwidth]{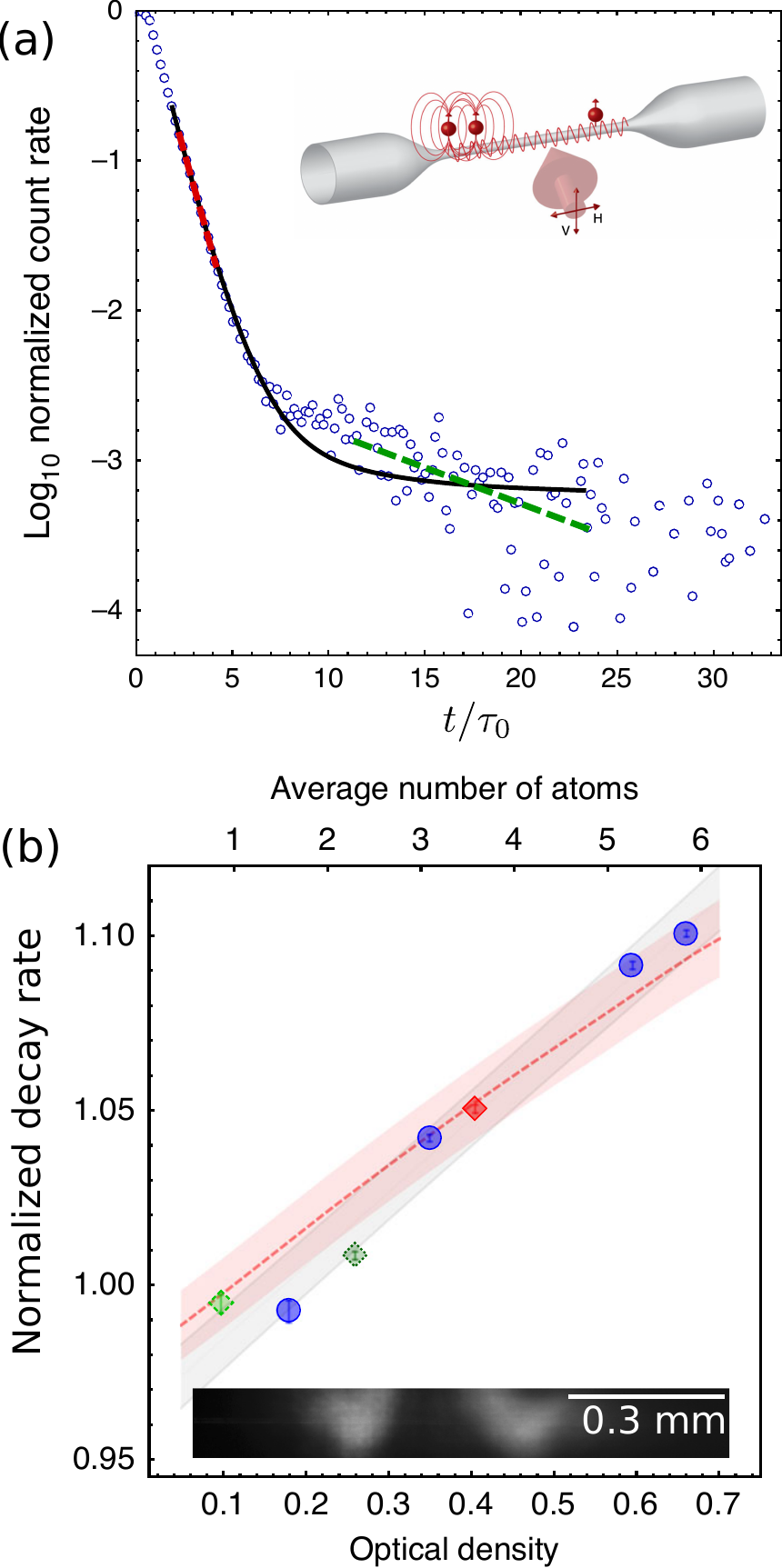}
 \caption{Reproduced from Ref.~\cite{Solano2017}. (a)
Experimental spontaneous emission kinetics of  photons in a nanofiber. Time is normalized to natural atom lifetime ($\tau_0=1/\Gamma_{0} = 26.24~$ns). 
 Inset schematically illustrates the setup and the waveguide-mediated coupling between distant atoms.
 (b) Dependence of the spontaneous decay rate on the average number of atoms and corresponding optical density (OD) for two distant atomic clouds, shown by the fluorescence image in the inset. The blue circles correspond to the signals from a single cloud of atoms. Blue circles, dotted dark green diamonds, and the solid red square correspond to the right atomic cloud, left atomic cloud, and the combination of both clouds, respectively. The red dashed line is the theoretical prediction. 
   }\label{fig:solano}
\end{figure}
While collective superradiant coupling can be observed for atoms in the free space~\cite{Devoe1996,Kaiser2016}, waveguides offer new opportunities to control the interactions between atoms. It is potentially possible to place atoms far enough so that the dipole-dipole interaction becomes irrelevant, and realize collective  superradiant or subradiant states mediated by long-ranged waveguide-mediated coupling~\cite{Tudela2011,Shahmoon2013}. An important milestone has been reported in Ref.~\cite{goban2014atom,goban2015superradiance} for cesium atoms near an alligator photonic crystal waveguide (see Sec.~\ref{sec:platforms} for more details of this structure).  The total spontaneous emission rate has been found to scale linearly with the number of trapped atoms. However, the number of the atoms was relatively small, $N\sim 1\div 3$, and their distance along the waveguide was on the order of $10~\mu m$ and not controllable. Thus, it was an important milestone to observe the controllable superradiant coupling between distant clouds of atoms, separated by about $0.3~$mm along the waveguide ~\cite{Solano2017}. The main experimental results from this work are reproduced in Fig.~\ref{fig:solano}. Instead of the alligator photonic crystal in Ref.~\cite{goban2014atom,goban2015superradiance}, Solano et al. have used an optical nanofiber overlap magneto-optical trap that contained $^{87}$Rb atoms. Fig.~\ref{fig:solano}(a) shows an example of measured spontaneous emission kinetics for an atomic cloud after initial excitation probe has been turned off. The average number of atoms for this realization was about $N=6$ with the optical density $\approx 0.7$. The initial faster component of the signal has been fitted as an exponential decay (red dashed line). Dependence of this decay constant on the array geometry is presented in  Fig.~\ref{fig:solano}(b) depending on the number of atoms in the system. 
All the measured decay rates scale linearly with the number of atoms which is a manifestation of a superradiant collective behavior. Blue circles correspond to a single atomic cloud and solid red square correspond to  atoms split into two separated clouds  separated by about 400 wavelengths (see the fluorescence image in the inset). This measurement satisfies the same linear scaling law and provides an unambiguous proof of long-range waveguide-mediated interactions for distant clouds. We note however, that the overall modification of the radiative decay rate as compared to that of a single atom in vacuum is not large, on the order of $10\%$. It has been limited by the coupling efficiency of atomic emission into the waveguide mode that has been estimated as  $\beta\approx 13\%$. Interestingly, Fig.~\ref{fig:solano}(a) also reveals a slower decaying tail, attributed to subradiant modes of the cloud. 

Importantly, here we have focused only on single-photon superradiance, resolved in the weak excitation regime. There also exists an opposite regime, considered in the original Dicke proposal when all the atoms of the array are initially in the excited state and then rapidly emit light. We refer the reader to the recent theoretical work \cite{Masson2020} studying many-body signatures of collective decay in atomic arrays, and references therein for more details of this regime. The phenomenon of superradiance can also be studied theoretically in more complicated setups. For example, Ref.~\cite{Wang2020supercorrelated} considered theoretically the quantum emitters coupled to a waveguides formed by an array of coupled cavities. There also exists an  interesting proposal of superradiant lattices in reciprocal momentum space, realized for an array of 3-level atoms coupled to an external wave~\cite{Scully2015}. Such a setup might be useful as a simulator of solid-state physics in a quantum optical setup.

\subsection{Generation of collective  excitations of atomic array}\label{sec:generation}
Generation of collective quantum states of atoms coupled to light presents one of the main potential applications of the WQED platform. Previously, in Sec.~\ref{sec:Prasad} we described  how arrays of atoms weakly chirally coupled to the waveguide have been used to experimentally demonstrate both photon bunching and antibunching depending on the photon number ~\cite{Prasad2020}.
There exist also preliminary reports on the two-photon entanglement and squeezing from the same setup ~\cite{hinney2020unraveling}. Schr\"odinger cat states have been recently realized in  an array of atoms coupled to an optical waveguide ~\cite{Leong2020}.

\begin{figure}[!t]
 \centering\includegraphics[width=0.45\textwidth]{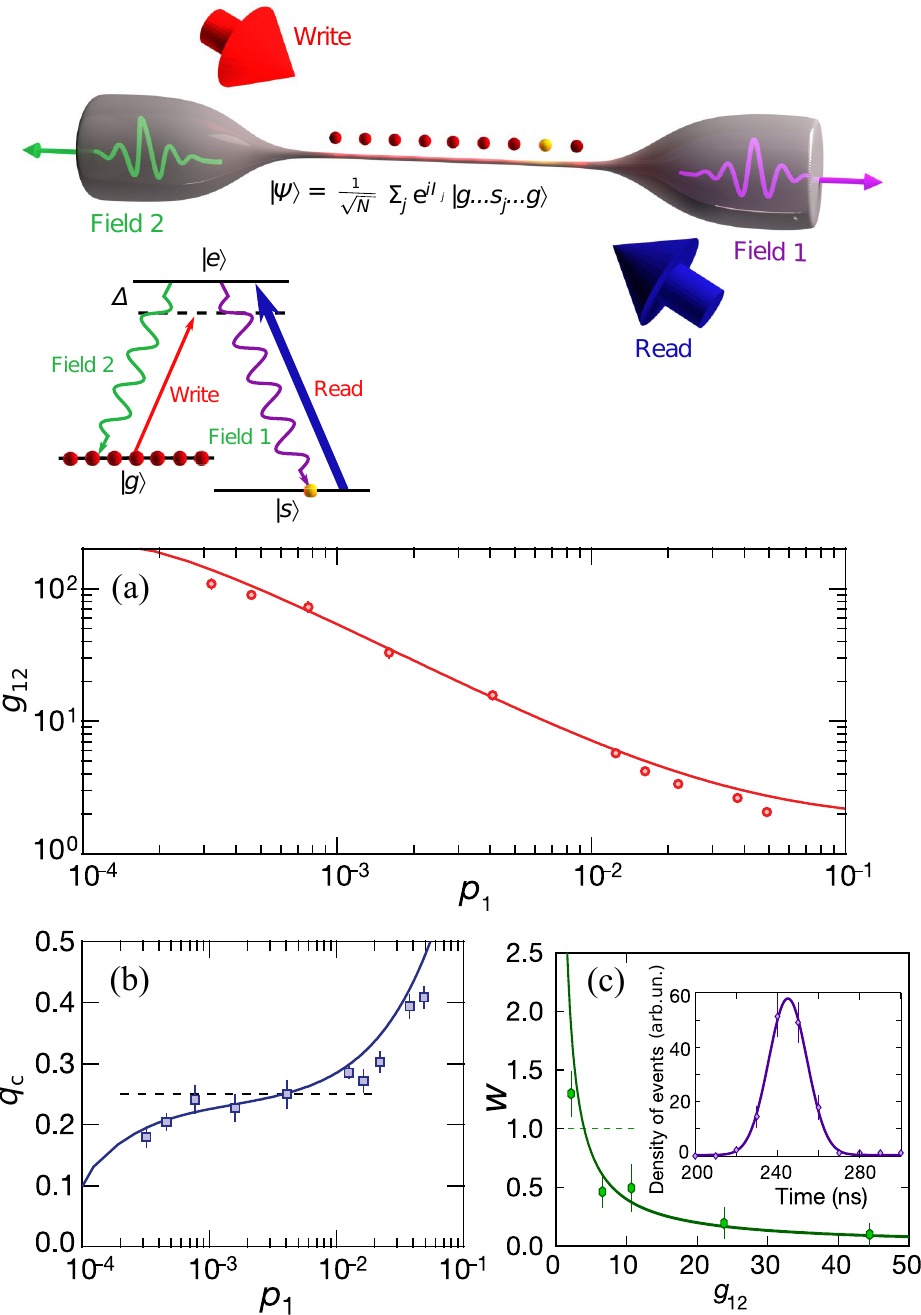}
 \caption{Reproduced from Ref.~\cite{Corzo2019}. Waveguide-coupled collective excitation of an atomic array. The top panel: DLCZ scheme \cite{DLCZ2001}. External to the waveguide write pulse, that is detuned from the atomic $|g\rangle \rightarrow |e\rangle$ transition, creates one single-flip excitation in a chain of $N$ atoms trapped in the vicinity of an optical nanofiber. This process is heralded by the detection of a photon in the guided Field-1 mode. Later, a read pulse resonant to $|s\rangle \rightarrow |e\rangle$ atomic transition converts the excitation into a single photon in the guided Field-2 mode. Lower panel: Characterization of the collective excitation. (a) Normalized cross-correlation function $g_{12}$ between the Field-1 and the Field-2. (b) Conditional retrieval efficiency $q_c$ as a function of the probability $p_1$ to detect a heralding photon in the Field-1. (c) Suppression of the two-photon component of the Field-2. }
 \label{fig:DLCZ}
\end{figure}

We will now discuss in more detail another important experimental demonstration~\cite{Corzo2019}. A  collective atomic excitation  in an atomic array has been prepared and then read out with an external laser pulse leading to a single photon emission into the guided mode
 by using Duan-Lukin-Cirac-Zoller (DLCZ) protocol \cite{DLCZ2001}.
To implement this protocol, 2000 atoms of $^{133}$Cs were initially prepared in the ground state $|g\rangle = |6S_{1/2}, F = 4\rangle$ and trapped  along the optical nanofiber, see the top panel of Fig.~\ref{fig:DLCZ}.  
Specifically,  a weak pulse in $y$-polarization and detuned by $\Delta = -10$ MHz from the $|g\rangle \rightarrow |e\rangle$ atomic transition creates a single collective excitation in an array.
This excitation  was heralded by detecting a single photon a guided mode of the nanofiber, quasi-linearly polarized along $x$-axis (Field-1 mode). After a programmable delay, an external read pulse resonant with $|s\rangle \rightarrow |e\rangle$ atomic transition was sent to the ensemble. The read pulse maps the collective excitation into a Field-2 photon that escapes the atomic ensemble and propagates in the opposite direction of the Field-1. The Field-2 mode is also a guided one, but it has a quasi-linear polarization along the $y$-axis. The readout process  benefits from the enhanced atom interaction with the guided mode. Namely, the single-atom coupling has been around $\beta=\Gamma_{\rm 1D}/\Gamma \approx 10^{-2}$, where $\Gamma_{\rm 1D}$ and $\Gamma$ are the radiative decay rates into the guided mode and into free space.   As has been described in Sec.~\ref{sec:principles}, this has allowed  to achieve optical depth (OD) equal to $300$ for $2000$ atoms, which means that the  atom-waveguide interaction is enhanced by about two orders of magnitude as compared to the effective free-space interaction.  The one-dimensional geometry of the problem presents another advantage as compared to the 3D free-space setup. Namely, since the initial pulse is incident from the side of the array and is external to the nanofiber, it is possible  to collectively excite all the atoms with the same amplitude.

After the implementation of the DLCZ protocol, the non-classical correlations between the Field-1 and the Field-2 should be characterized. This can be done with the normalized cross-correlation function $g_{12} = p_{12}/(p_1p_2)$, where $p_{12}$ is the joint probability of detecting a pair of photons and $p_1$, $p_2$ are the probabilities of detecting a photon in fields 1 and 2. The dependence of $g_{12}$ on the probability $p_1$ is shown in the Fig.~\ref{fig:DLCZ} (a). One can see, that the value of $g_{12}$ increases when the excitation probability is reduced. For efficient retrieval of the stored collective excitation, the conditional retrieval efficiency is a crucial parameter. It can be found via a measurement of the conditional probability of detecting a guided photon in the Field-2 after retrieval $p_c = p_{12}/p_1$. Memory efficiency can be found as a ratio $q_c = p_c/\eta_2$, with $\eta_2$ being the overall detection efficiency. Fig.~\ref{fig:DLCZ}(b), displays the retrieval efficiency as a function of $p_1$. One can observe here three different regimes. In the first one, characterized by the large value of $p_1$, $q_c$ increases with $p_1$, and corresponds to a multi-excitation process in the write field. The second region with a plateau in the $q_c$ corresponds to a single-excitation regime. And the third region corresponds to low excitation probability, where the noise background creates false heralding events becoming predominant.

Finally, a single-photon character of the heralded excitation can be confirmed by measuring the degree of suppression $w$ of the two-photon component of the retrieved Field-2 compared to a coherent state. This value can be found from the ratio $w = (p_1p_{1,2a,2b})/(p_{1,2a}p_{1,2b})$, where $p_{1,2a,2b}$ indicates the probability for triple coincidences and $p_1,2a$ and $p_{1,2b}$ are probabilities for coincidences between detectors. In Fig.~\ref{fig:DLCZ}(c), one can see the antibunching value $w$ as a function of the cross-correlated parameter $g_{12}$. The temporal mode of the guided single photon is given in the inset of Fig.~\ref{fig:DLCZ}(c).

These experimental achievements demonstrate that the collective quantum state can be characterized by the subsequent on-demand emission of a guided single photon and that this non-classical state can be preferentially coupled to a waveguide.

\subsection{Slow guided light and quantum memory}\label{sec:memory} 
Interfacing guided light with an atomic array has been foreseen as a promising alternative, enabling longer interaction length and large optical depth which are crucial for the quantum memory applications~\cite{Gorshkov2007}. In an optical nanofiber, a propagating single-mode field experiences a dispersion due to the dispersive material contents of the core and clad. The group velocity of the envelope of such a propagating fiber mode is $v_g = d\omega/dk_{\text{1D}}$ with $k_{\text{1D}}$ being the propagation constant of the fiber mode. The group velocity of the fiber mode can be significantly reduced under conditions of electromagnetically induced transparency (EIT). The light delay propagating in an optical nanofiber was theoretically studied by Hakuta in Ref.~\cite{Patnaik2002} and Kwek in Ref.~\cite{Song2017} and the first experimental demonstrations were done by groups of J.~Laurat in Ref.~\cite{Gouraud2015} and A.~Rauschenbeutel in Ref.~\cite{Sayrin2015}. The main difference of these two experiments consists in prepared atomic systems. Thus, in \cite{Gouraud2015}, a cloud of laser-cooled atoms overlapped with a nanofiber, while in \cite{Sayrin2015} laser-cooled cesium atoms were confined in a one-dimensional optical lattice realized in the evanescent field surrounding an optical nanofiber.


\begin{figure}[!t]
 \centering\includegraphics[width=0.48\textwidth]{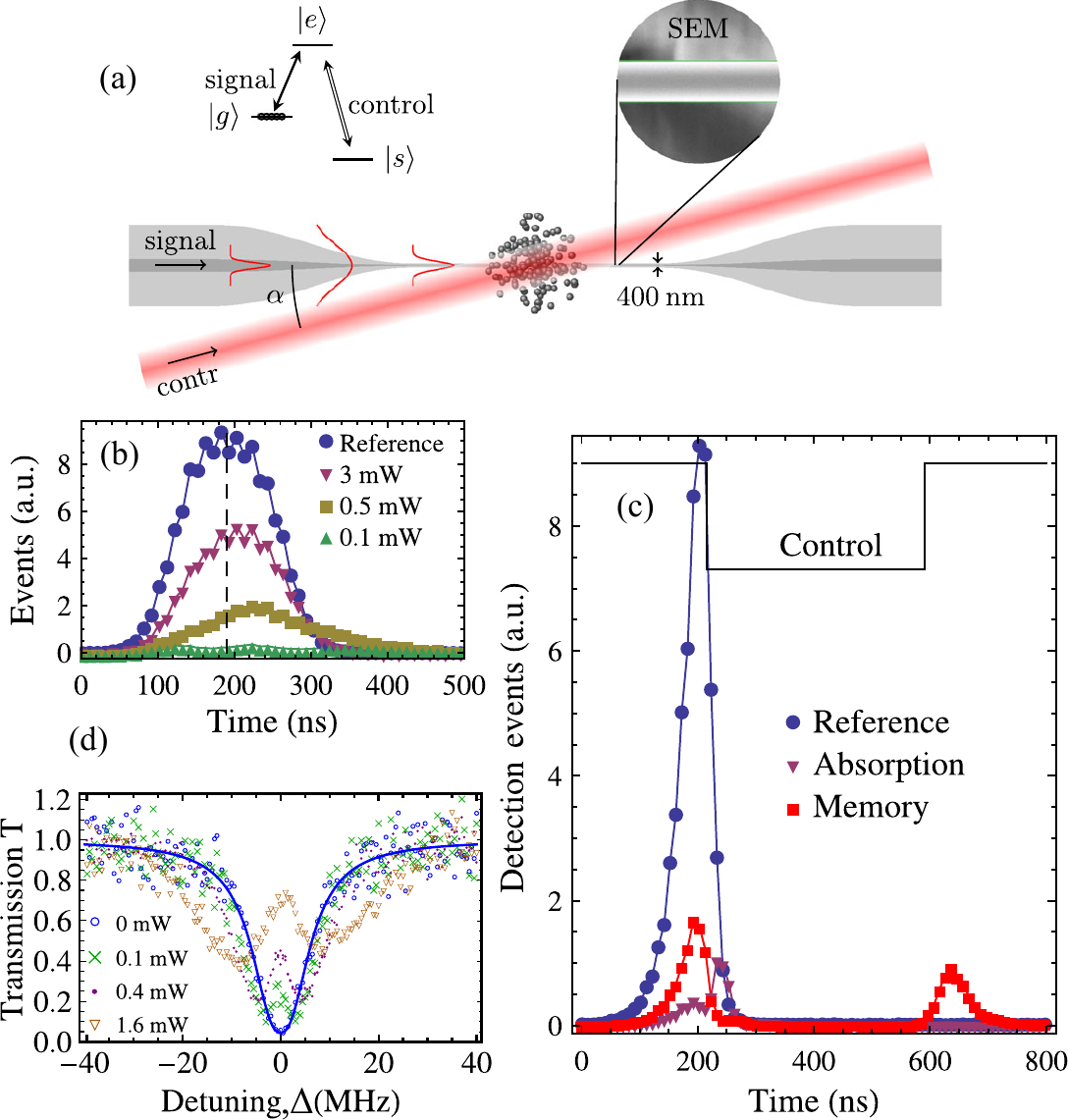}
 \caption{Reproduced from \cite{Gouraud2015}. (a) A nanofiber with a diameter $400$ nm is overlapped with an ensemble of cold atoms of $^{133}$Cs. The signal pulse is a guided mode field, the control pulse is external to the nanofiber. (b) Transmitted pulses for different control powers. The reference is measured without atoms. (c) Storage and retrieval of the guided light with an exponentially-rising profile with a full width at half-maximum of $60$ ns. In the absence of the control field, the blue and purple points give the transmitted pulse without and with atoms. The red data corresponds to the memory sequence, showing leakage and retrieval. The black line indicates the control timing. (d) EIT for the guided light. The control field is on resonance with the $|s\rangle \rightarrow |e\rangle$ transition while the signal is detuned by $\Delta$ from the $|g\rangle \rightarrow |e\rangle$ atomic resonance. }
 \label{fig:EIT}
\end{figure}

Guided light propagating through an optical nanofiber has a complex polarization pattern, including a significant non-transverse component. The transmission coefficient of weak intensity light propagating through a dilute atomic cloud with the optical depth $\text{OD}$ has an exponential dependence $\sim \text{exp}[-\text{OD}/(1+(2\Delta/\Gamma_{\text{tot}})^2)]$, where $\Delta$ and $\Gamma_{\text{tot}}$ are detuning from the atomic resonance and the total decay rate of the atomic excited state. The transmission coefficient of the guided light in a nanofiber-mediated atomic cloud is shown by blue in Fig.~\ref{fig:EIT}(d).
\begin{figure}[!t]
 \centering\includegraphics[width=0.48\textwidth]{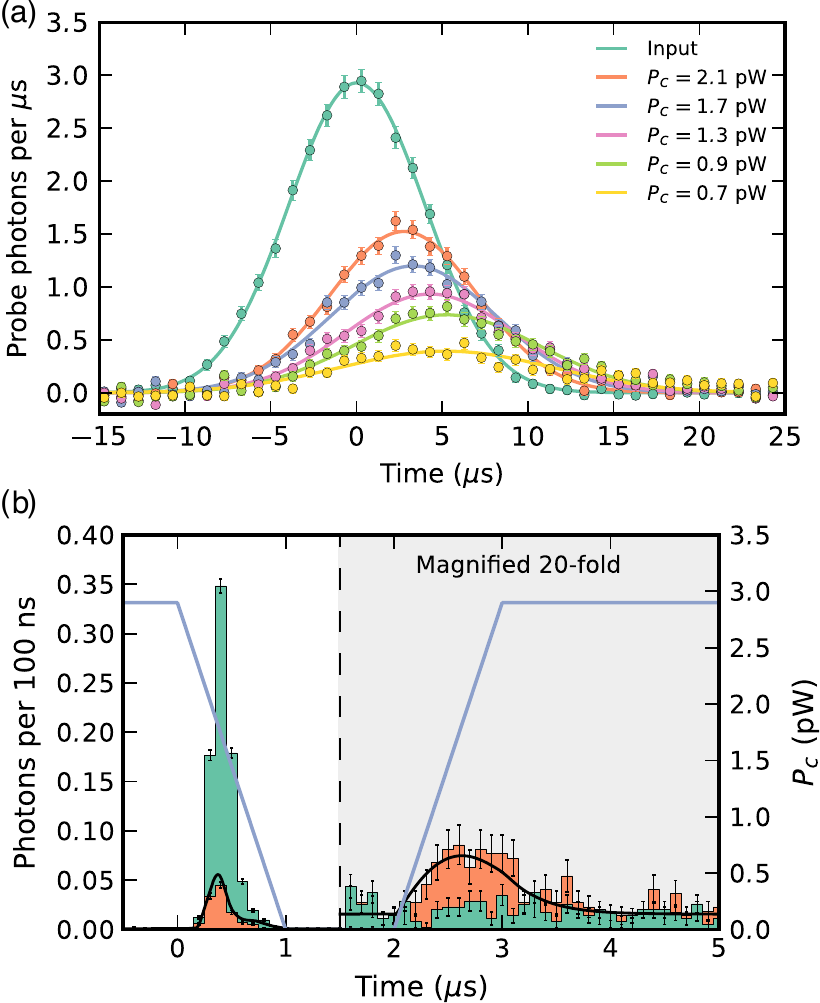}
 \caption{Reproduced from \cite{Sayrin2015}. (a) Slow guided light. A light delay with respect to a reference pulse (dark green) is clearly visible. The solid lines correspond to  Gaussian fits of the experimental data. (b) Storage of light in a nanofiber-trapped ensemble of cold atoms. A pulse duration $\tau = 0.2 \mu$s that contains $0.8$ photons on average.The storage time was chosen as $1 \mu$s. The green data is reference recorded without atoms, black line is simulated time trace. The homogeneous magnetic field is $B_{\text{off}} = 15$ G.}
 \label{fig:EIT_Sayrin}
\end{figure}

The fitting of the experimental data with the exponential profile yielded $\text{OD} = 3$ and $\Gamma_{\text{tot}}/2\pi = 6.8$ MHz. One can notice that this value is $30 \%$ larger than the natural linewidth in free space $\Gamma/2\pi = 5.2$~MHz, resulting from the finite temperature, surface interactions and modification of the spontaneous emission rate in the vicinity of the fiber. The authors of Ref.~\cite{Sayrin2015} obtained $\text{OD} = 6$ and $\Gamma_{\text{tot}}/2\pi = 6.4$ MHz with the similar fitting.

It is well known that a strong control field changes the transmission characteristics of the probe field. Figure~\ref{fig:EIT}(d) shows an example of the transmission profiles of the signal as a function of its spectral detuning $\delta$ from the resonance of the atomic transition $|g\rangle \rightarrow |e\rangle$ for different values of control field power, taken from Ref.~\cite{Gouraud2015}. When the control field is applied, a transparency window appears, providing a first signature of EIT in this evanescent-field configuration. Transparency close to $80\%$ was achieved in both experiments. After having EIT transparency, the slow-light effect resulting from the guided light propagation under EIT condition can be measured. As a signal pulse, a weak laser pulse at a single-photon level was used. Results of light delay are demonstrated in Fig.~\ref{fig:EIT}(b) for \cite{Gouraud2015} and in Fig.~\ref{fig:EIT_Sayrin}(a) for \cite{Sayrin2015}. One can see that larger delays are obtained when the control field is decreased due to the narrower transparency window. For a $0.5$~mW control field power, the light delay in $60$ ns was observed in \cite{Gouraud2015} and the light delay in $5\mu$s was achieved with the control field power $0.7$ pW in \cite{Sayrin2015}.

Also, the storage of the guided light can be demonstrated by switching off the control field. This corresponds to implementation of the dynamical EIT protocol. While the light is slowed down, the control is ramped down to zero and the signal pulse is converted into a collective atomic excitation. Later after a controllable delay, the control field can be switched on again and the light can be retrieved back in a well-defined spatio-temporal mode due to the collective enhancement provided by the atomic ensemble. Figure~\ref{fig:EIT}(c) and Figure~\ref{fig:EIT_Sayrin}(b) demonstrate the storage results for a signal with a mean photon number per pulse equal to $0.6$ and $0.8$, respectively. Due to the limited delay, the pulse cannot be contained entirely in the ensemble, and a leakage is observed before the control pulse is switched off. The crucial parameter characterizing the memory is its efficiency, which could be defined as a ratio of the photodetection events in the retrieved pulse to the ones in the reference. The efficiency $\eta = 10\%$ and $\eta = 3\%$ were obtained in these experiments. These efficiencies are compatible with the limited OD used in the experiments. 

These two experiments have demonstrated that the interaction of the evanescent field propagating through an optical nanofiber with the surrounding atoms provides an intrinsically-fibered memory that has potential applications.

\subsection{Subradiant excitations in the qubit array}\label{sec:ustinov}

So far we considered experiments probing symmetric superradiant modes of the atomic array. However, as has been discussed above in the context of Fig.~\ref{fig:rta-period}, there also exist subradiant modes with the radiative decay strongly suppressed as compared to a single atom. The lifetime of the darkest subradiant modes tends to increase with the number of atoms $N$, either as $N^3$ [Eq.~\eqref{eq:sub1},\cite{Zhang2020d}] or even faster, \cite{Kornovan2019}. 
\begin{figure}[t]
\includegraphics[width=\columnwidth]{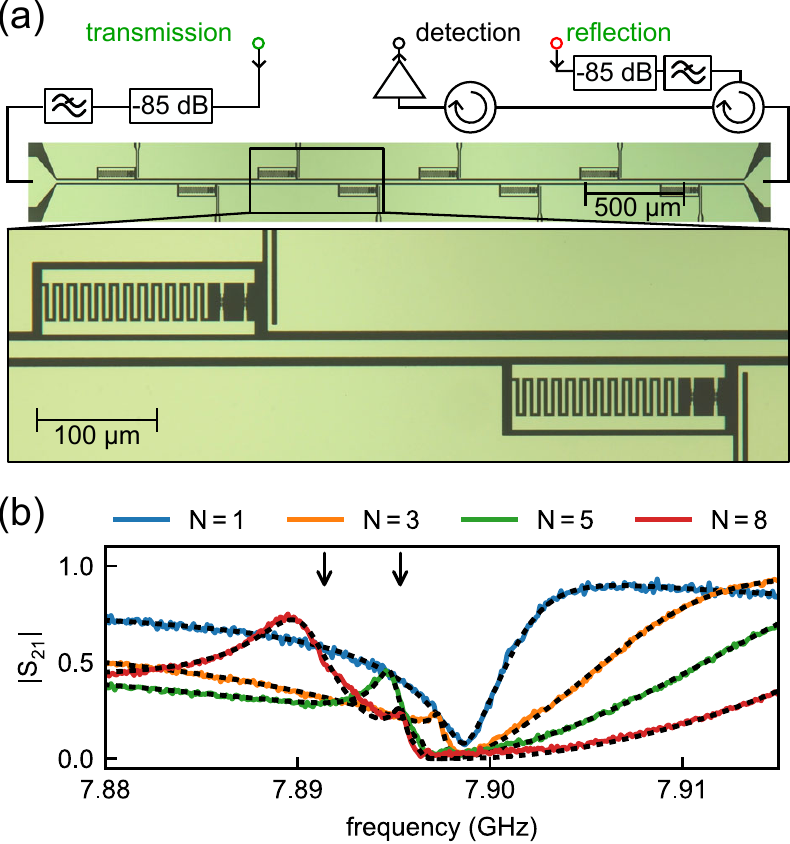}
\caption{Represented from Ref.~\cite{brehm2020waveguide}. (a) Optical micrograph of the $8$-qubit metamaterial composed of  superconducting transmon qubits capacitively coupled to a coplanar waveguide. Local flux-bias lines provide individual qubit frequency control in the range $3$--$8$~GHz. (b) Transmission spectra $|S_{21}|$ for different numbers $N$ of resonant qubits with the resonant frequency  $f= 7.898~$GHz and low drive powers. With increasing $N$, the emergence of subradiant states (visible as peaks in transmission) can be observed. Black dotted lines are fits to the expected transmission using a transfer matrix calculation. Black arrows mark calculated frequencies of the two brightest subradiant states for $N = 8$.	}
\label{fig:Jan1}
\end{figure}

However, the large radiative lifetime also means that it is hard to address subradiant states in experiments. We will illustrate this by discussing the  state-of-the-art experiment  Ref.~\cite{brehm2020waveguide} for an array of artificial atoms, superconducting qubits, coupled to the waveguide. The structure considered of $8$ transmon qubits as it is illustrated in Fig.~\ref{fig:Jan1}(a). The spacing of the qubits was approximately $40$ times smaller than the electromagnetic wavelength at the qubit resonance, so that the array can be viewed as a quantum metamaterial. In the experiment, the amplitude transmission coefficient of the electromagnetic wave through the metamaterial $S_{21}\equiv t(\omega)$ has been measured. The setup allows one to  tune the resonant frequencies of all the qubits independently by applying external voltage. Thus, by tuning a given number of the consecutive qubits $N\le 8$ to the resonance and detuning the remaining qubits, it has been possible to study the dependence of the transmission spectra on $N$. The experimental results are presented in Fig.~\ref{fig:Jan1}(b). The experiment has been performed for low excitation powers. For $N=1$, the transmission spectrum has a dip at the qubit resonance frequency. In contrast to the theoretically predicted Lorentzian, the experimental spectrum is slightly asymmetric. This  Fano-like asymmetry originates from  the interference between the resonant  scattering on the qubit and the multiple reflections from the edges of the waveguide. Figure~\ref{fig:Jan1}(b) demonstrates that increase of the number of qubits leads to suppression of the transmission around the qubit resonance. This reflects the formation of the polariton band gap for the coupled photon-qubit excitations. At the same time,  additional peaks appear with the increase of $N$ below the qubit resonance. These peaks correspond to subradiant excitations of the array, discussed above in Sec.~\ref{sec:single_photon}. Namely, they can be obtained by diagonalizing the effective Hamiltonian matrix Eq.~\eqref{eq:Hmn}.  Unfortunately, only up to two brightest subradiant modes have been resolved in the experiment. This is due to the quality of the sample, the ratio of nonradiative to radiative decay rates $\Gamma/\Gamma_{\rm 1D}$ was on the order of 10\%. Thus, the darkest subradiant modes decay mostly nonradiatively and are not resolved in transmission spectra. Still, the experiment of Ref.~\cite{brehm2020waveguide} reveals the potential of the superconducting qubit arrays. Natural extension of this work would be the demonstration of the slow light effect due to lower group velocity of polaritons near the resonance. It is also potentially possible to further increase both the quality and the number of qubits which will enable slowing and storing microwave pulses, propagating through the array~\cite{Leung2012}.

Another application of the subradiant states with superconducting qubits could be related to the quantum measurements. Quantum non-demolition  single-microwave photon detector remains a challenge~\cite{Blais2018,Grimsmo2021,blais2020circuit}.
The proposal of Ref.~\cite{Blais2018} is based on a signal  waveguide photon being absorbed by an array of superconducting  qubits. Next, the absorption of photon by the array leads to the  coherent state displacement in an additional harmonic mode which is detected using homodyne measurement. 
The essence of the proposal is the engineering of the system in such a way that a signal photon is absorbed into the bright state of the qubit array, transferred to a long-lived subradiant state, and after some time, returns to the bright state where it is re-emitted. The measurement efficiency is enhanced due to the long lifetime of the collective subradiant state. 
\subsection{Bragg-spaced  arrays}\label{sec:Bragg}

\begin{figure}[!t]
 \centering\includegraphics[width=0.48\textwidth]{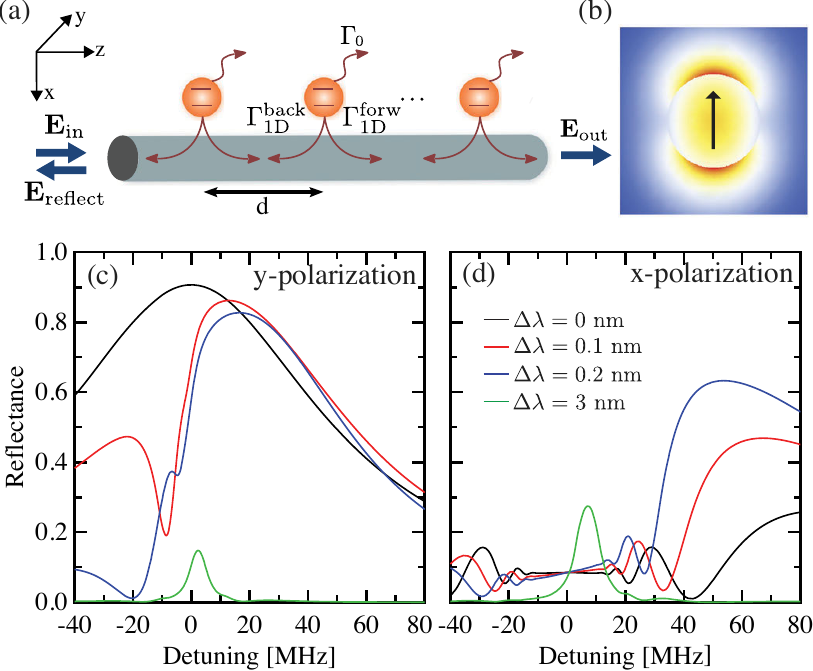}
 \caption{Bragg reflection from atoms coupled to a one-dimensional waveguide. (a) $N$ atoms are trapped near a waveguide and exhibit radiative decay rates $\Gamma^\rightarrow = \Gamma^\leftarrow$ into the right and left propagating modes, respectively, and $\Gamma' \sim \Gamma$ into all other modes. (b) Electric field distribution in the transverse plane of a nanofiber for a guided probe field with a quasilinear polarization (indicated by the arrow). (c) Theoretical reflection spectra for a probe quasilinearly polarized along the $y$ direction (symmetric decay rates) and (d) along $x$ direction (asymmetric decay rates). The spectra are given for different distances between the atoms, with values close to the commensurate case. $\Delta\lambda$ stands for the trap detuning from the resonance, with $d = \lambda_0/2 + \Delta\lambda/2$. The calculation parameters are $N = 2000$, $\Gamma_{\rm 1D}/\Gamma = 0.01$, $\Gamma^{\rightarrow} = 2.8\Gamma_{\rm 1D}$, $\Gamma^\rightarrow/\Gamma^\leftarrow = 12$. }
 \label{fig:Bragg_PRL1}
\end{figure}

In this section, we discuss in more detail the experimental  results of Ref.~\cite{Corzo2016} for a waveguide-coupled atomic array. Theoretical background for Bragg structures has been presented earlier in Sec.~\ref{sec:single_photon}.
 In this experiment, the array of $N = 2000$ trapped atoms of $^{133}$Cs was prepared in the evanescent field of a $400$~nm diameter nanofiber with a lattice constant $d$ close to $\lambda_0/2$, where $\lambda_0$ is the wavelength of the atomic transition. {For such an array of $N = 2000$ atoms, the reflection up to $75\%$ was achieved. For comparison, the $80\%$ reflection  was achieved in free space experiment with $10^7$ atoms \cite{Schilke2011}.}
 While previously in this section, we considered a symmetric non-chiral situation, the experiment of Ref.~\cite{Corzo2016} has also demonstrated the effect of the waveguide chirality arising from the complex polarization pattern. Namely, each atom exhibits a radiative decay rate {$\gamma_{\rm 1D}^\rightarrow$} and {$\gamma_{\rm 1D}^\leftarrow$} into the right and left propagating modes respectively, and $\Gamma' \sim \Gamma_0$ into all other modes, with $\Gamma_0$ being the free space radiative decay rate, see Fig.~\ref{fig:Bragg_PRL1} (a). For a guided probe field quasilinearly polarized along the $y$ direction, the two decay rates are equal, $\gamma^\rightarrow = \gamma^\leftarrow = \gamma_{\rm 1D}/2$. In contrary, for an orientation along the $x$ direction, the couplings to the waveguide become strongly asymmetric \cite{Kien2014}. In the case of asymmetric coupling, the forward decay rate is increased by sixfold while the backward decay rate is suppressed by about one order of magnitude.  In order to examine the effect of asymmetric coupling in experiment, one can compare the light reflection spectra in $x$- and $y$- polarizations.

Figure~\ref{fig:Bragg_PRL1}(c) and Fig.~\ref{fig:Bragg_PRL1}(d) provide theoretical reflection spectra calculated for different small detunings $\Delta\lambda$ of the trap wavelength to atomic resonance and for the two orthogonal polarizations. The calculations have been performed using the transfer matrix formalism discussed in Appendix~\ref{app:transfer}.
For atoms separated exactly by $\lambda_0/2$, the reflection spectrum has a broadened Lorentzian profile in the symmetric coupling case while the reflectance is strongly suppressed in the chiral case. Indeed, the amount of chirality and number of atoms result in a finite bandwidth around resonance where reflection is suppressed. One can see, that close to commensurate array, the Bragg condition is fulfilled out of the resonance. This leads to a maximum reflectance shifted to the blue but also results in an increased reflectance for the chiral case. Large reflectance values can be then obtained for both polarizations as the single-atom reflection coefficients are similar in the chosen configuration.

\begin{figure}[t]
 \centering\includegraphics[width=0.48\textwidth]{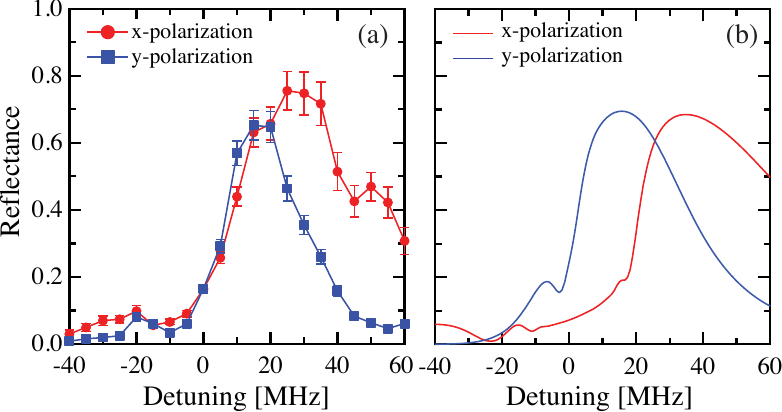}
 \caption{Effect of the chiral character of the waveguide on the Bragg reflection. (a) Measured reflection spectra for $x$ and $y$ quasilinear polarizations, with $\Delta\lambda = 0.2$ nm. (b) Theoretical simulations done with $N = 2000$, {$\Gamma_{\rm 1D}/\Gamma_0 = 0.007$}, {$\Gamma^\rightarrow/\Gamma^\leftarrow = 12$}, $f = 0.3$. }
 \label{fig:Bragg_PRL2}
\end{figure}

The measured reflection spectra for both $x-$ and $y$- polarization profiles are shown in Fig. \ref{fig:Bragg_PRL2} (a). {The achievable filling factor is around 0.5  with random occupation of the lattice sites by the atoms.  In Fig.~\ref{fig:Bragg_PRL2}, the estimated filling factor was around 0.3 and, nevertheless, one can observe the pronounced Bragg peak in the reflection  spectrum averaged over multiple realizations. One of the possible mechanisms of such tolerance is related to directional interaction with the waveguide mode and suppressed back-reflection.} The trap detuning here was fixed at $\Delta\lambda = 0.2$ nm {, that adds slight disorder in the atomic array. However,} one can see that the reflection spectrum is significantly shifted and broadened in the asymmetric case of $x$ polarization (red curves). These features are compelling signatures of the chiral character of the waveguide on the reflection, as confirmed by the associated simulations shown in Fig. \ref{fig:Bragg_PRL2} (b). One can notice that the maximal observed reflectance of $(0.75 \pm 0.06)$ was obtained in the asymmetric case at a probe detuning of $25$ MHz. Beyond their fundamental significance, the observation of the chiral character of the nanofiber demonstrates key ingredients for the exploration of a variety of emerging and potentially rich protocols based on 1D reservoirs coupled to atoms.



\section{Summary and outlook}\label{sec:summary}
In this review, we tried to provide an introduction in the emerging field of waveguide quantum electrodynamics (WQED). We considered various experimental platforms and described in detail several recent groundbreaking experiments devoted to light-matter interactions in the quantum regime, including among others tunable bunching and antibunching of emission due to formation of bound photon states, resonant light reflection from atom-made quantum metasurfaces, and topological quantum optics. We also provide a detailed introduction into various theoretical techniques useful in quantum optics of ordered one-dimensional atomic arrays in the waveguide. 

 Given the tremendous advances in the WQED field in the last 5--10 years, it is  hard to predict specific directions of future development but  substantial progress can certainly be expected. Not only the current  WQED platforms  will continue to develop but also new types of structures could become prominent, for example, those based on Rydberg superatoms, topologically nontrivial waveguides, or even M\"ossbauer nuclei \cite{Rhlsberger2021}, see the discussion in Sec.~\ref{sec:outlook}. We can also envisage arrival of hybrid quantum systems involving interacting excitations of different origin. For example, one could think of tripartite systems with interacting atomic excitations, photons and atomic vibrations in the quantum regime. There have already been a number of relevant experimental demonstrations~\cite{Lecocq2016,Bothner2020}
and theoretical predictions ~\cite{Chang2013b,Iorsh2020,sedov2020chiral} and the field of WQED might be soon complemented by waveguide quantum optomechanics.  
Another type of quasiparticles, that can be interfaced with superconducting circuits, it offered by collective spin excitations in ferromagnetic crystals, see the review~\cite{LachanceQuirion2019}.
The whole concept of the WQED might evolve, fuse with circuit QED, and include structures with two-dimensional arrays of waveguides~\cite{Marques2021} or even more complicated topologies.  
 
Second, the focus of research will probably  shift from pioneering demonstrations of cooperative light-matter interactions, such as generation of super- and sub-radiant states, to  the studies of  advanced high-excited and high-entangled quantum states of atoms and photons. Different protocols are currently being developed, that explore both the possibility of cascaded photon processing in one-dimensional structures~\cite{Guimond2020} and the potential possibility of two-dimensional arrays to parallelize photon processing~\cite{Bettles2020,Bekenstein2020} for generation of photon cluster states and cat states.
In addition to potential applications for quantum information processing, WQED systems can act as quantum simulators useful for fundamental problems from many-body and condensed matter physics. Some examples include topological quantum Hall phases induced by interactions \cite{Poshakinskiy2020} or external magnetic field~\cite{debernardis2020lightmatter}.  An advantage of the optical setup over the more conventional condensed matter ones would be the possibility to visualize interesting quantum states, such as the fractional quantum Hall phase~\cite{Perczel2020}. Very recently, self-ordering of photons in the WQED setup with three-level atoms with Laughlin-like photon states has been predicted in Ref.~\cite{iversen2021selfordering}.
The situation becomes especially interesting when the disorder is taken into account. Single-photon properties of disordered arrays have already been studied \cite{Vahid2016,Lodahl2018,fedorovich2020disorder,Song2021}, but the  many-body problem of photons interacting with atoms in the disordered structure seems especially interesting. For example, recent numerical calculations indicate, that the WQED system exhibits many-body localization instead of conventional thermalization \cite{fayard2021manybody}. The localization phase occurs provided that the excitation filling factor in the array of two-level atoms is less than 1/2. At larger filling factors, the states are delocalized which can be related to saturation of optical transitions. The filling factor of 1/2 seems to be a special value because subradiant states in subwavelength-array also disappear above this threshold~\cite{Poshakinskiy2021dimer}. The interplay of disorder and dissipation in this system certainly deserves future studies.
Another interesting type of many-body phases in the Dicke model is related to spin glasses~\cite{Sachdev2011,Goldbart2011,Rotondo2015b}. Specifically, cavity-mediated coupling between the atoms can be mapped to the effective  long-ranged spin-spin  interactions. One of the interesting features of the spin glass model is the similarity to the Hopfield associative memory. Thus, certain  elements of ``machine learning behavior" emerge in the strongly coupled multimode Dicke model~\cite{Fiorelli2020}.  Namely, the atomic array behaves as a   basic associative memory with stationary states corresponding to the retrieval phase of the Hopfield neural network~\cite{Hopfield1982}, characterized by the ability to recall previously stored information.
To the best of our knowledge, the spin glass behavior has  been so far theoretically studied for atomic arrays in a cavity, and it is not yet clear what happens in the strongly dissipative waveguide setup. The  progress  in artificial intelligence and machine learning will probably inspire a lot of studies also in the realm of WQED. One can not only seek for   analogies between processes in quantum systems and  neural networks, but also apply machine learning approaches to understand better the results of numerical calculations \cite{Che2020} and experiments \cite{Ahmed2020}. So far we have discussed stationary many-body phases.
It could be also interesting to examine how the so-called time-crystal phases, that break time-translation symmetry \cite{Carollo2020}, are manifested  in the WQED setup. 

To summarize, the possibility to tune energies and interactions of atoms and to probe the wavefunctions at individual atoms is unprecedented for conventional solid-state systems and will keep to inspire many beautiful experiments. We also refer the reader to recent reviews \cite{Noh2016,KimbleRMP2018}  and original theoretical works on quantum simulators based on the atomic arrays ~\cite{GonzlezTudela2015,Douglas2015,Hung2016}. We hope that all these intriguing predictions will soon drive new theoretical concepts, experimental demonstrations and eventually practical applications for quantum technologies.

\acknowledgments
{The authors are  grateful to J.~Brehm, N.V.~Corzo,  G.~Fedorovich, E.L.~Ivchenko,  Y.~Ke, Yu.~Kivshar, D.~Kornovan, E.~Shahmoon, U.~Smilansky, A.~Ustinov, E.~Vlasyuk , V.~Yudson and J.~Zhong for fruitful collaborations and useful discussions.}
ASS thanks the Russian Science Foundation (Grant No. 21-72-10107) and Russian Foundation Basic Research joint with CNR (project No. 20-52-7816), MIP thanks the Priority 2030 Federal Academic Leadership Program.

\appendix
\section{Input-output formalism}\label{sec:input_output}
Here we outline the  input-output formalism \cite{Gardiner1985,walls2007quantum} for an ensemble  of two-level atoms interacting with a single modes of photons with linear dispersion, propagating in a  waveguide and discuss its limitations. The derivation mostly follows Ref.~\cite{Caneva2015}. 

The system under consideration is described by the sum of a photon Hamiltonian 
$H_{\rm phot}$,  atoms Hamiltonian $H_{\rm atom}$ and an atoms-photon interaction Hamiltonian $H_{\rm atom-phot}$:
\begin{gather}\label{eq:H1atom1b}
H=H_{\rm phot}+H_{\rm atom}+H_{\rm atom-phot}\\\nonumber
H_{\rm phot}=\sum\limits_{k} \omega_k a_k^\dag a_k^{\phantom{}}\:,
H_{\rm atom}=\omega_0\sum\limits_{m=1}^N\sigma_m^\dag \sigma_m^{\vphantom{\dag}}\:,\\\label{eq:Hatomphot2}
H_{\rm atom-phot}=-\sum\limits_{m=1}^N\hat {\bm d}\cdot \hat {\bm E}(z=z_m)\:,
\end{gather}
Here, we use the units with  $\hbar=1$, $\omega_k=c|k|$ is the frequency of the photonic mode with the wave vector $k$, $c$ is the light speed in the waveguide, and  $\sum_k\equiv L\int_{-\infty}^{\infty} \rmd k/(2\pi)$, where $L$ is the normalization length. Bosonic operators $a_k^\dag$ and $b^\dag$ describe creation of photon in the waveguide and excitation of the atom  and $\sigma\equiv |e\rangle\langle g|$ where  $|g\rangle$ and $|e\rangle$ are the ground and excited atom states. The light-atom coupling is treated in the dipole approximation with the dipole momentum operator, 
$
\hat{\bm d}=\bm d b+\bm d^*b^\dag\:,
$
where 
 $\bm d$ is the matrix element of electric dipole momentum between ground and excited states of the (point-like) atom. 
The quantized electromagnetic field operator reads
\begin{equation}
\bm E(z,t)=\sum\limits_k\sqrt{\frac{2\pi \omega_k}{LA}}\e^{\rmi kz} \bm e_k a_k(t)+{\rm H.c.}\:,
\end{equation}
where $A$ is the normalization area (effective cross-section of the waveguide), 
$\bm e_k $ is the unit photon polarization wave vector and we use the Gaussian units system. In this section, we consider resonant rotating wave approximation when 
the Hamiltonian $H_{\rm atom-phot}$ can be reduced to
\begin{multline}\label{eq:RWA}
H_{\rm atom-phot}=\frac{1}{\sqrt{L}}\sum\limits_{m=1}^N\sum\limits_{k} (g_{k}\e^{-\rmi k z_m} a_k^\dag \sigma_m^{\vphantom{\sigma}}\\+
g_{k}^*\e^{\rmi k z_m}a_k^{\vphantom{\sigma}} \sigma_m^\dag)
\end{multline}
with $g_{k}=-\sqrt{2\pi \omega_k/A}\bm d\cdot \bm e_{k}^*$.  This rotating wave approximation holds provided that the array is excited  resonantly and the atom-photon coupling is reasonably weak, $|g_k|/\sqrt{L}\ll \omega_0$.  
An ultrastrong coupling regime, when $g/\sqrt{L}\sim \omega_0$ and the counter-rotating terms can not be ignored,   is considered in Sec.~\ref{sec:ultrastrong}.
In the same Markovian approximation, the dependence of $g$ on the light wave vector $k$ can be simplified to
$g_k=g_{\omega_0/c}\equiv g_+$ for $k>0$ and $g_k=g_{-\omega_0/c}\equiv g_-$ for $k>0$.
We also introduce annihilation operators corresponding to forward- and backward-propagating photons
\begin{equation}\label{eq:a_forward_backward}
a_\rightarrow(z,t)=\sum\limits_{k>0}a_k(t)\e^{\rmi k z},\quad 
a_\leftarrow(z,t)=\sum\limits_{k<0}a_k(t)\e^{\rmi k z}\:.
\end{equation}
The Heisenberg equations for these operators read \cite{Caneva2015}
\begin{multline}
a_\rightarrow(z,t)=a_{{\rm in},\rightarrow}(t-z/c)\\-\frac{\rmi g_+\sqrt{L}}{c}\sum\limits_{m=1}^N\theta(z-z_j) \sigma_m(t-|z-z_m|/c)\:,\label{eq:aleft}
\end{multline}
where the operator $a_{{\rm in},\rightarrow}(t-z/c)$ describes the field, incident from the left. For the backward-propagating operators  $\theta(z-z_m)$ in Eq.~\eqref{eq:aleft} is to be replaced with $\theta(z_m-z)$, $a_{{\rm in},\rightarrow}(t-z/c)$ with the backward-propagating input field $a_{{\rm in},\leftarrow}(t+z/c) $ and $g_+$ with $g_-$.
The equation of motion for atomic operators is \cite{Caneva2015}
\begin{multline}\label{eq:sigmam}
 \frac{\rmd  \sigma_m}{\rmd t}=-\rmi\omega_0\sigma_m-\frac{\rmi (1-2\sigma_m^\dag \sigma_m)}{\sqrt{L}}[g_+^*a_{{\rm in},\rightarrow}(t)+g_-^*a_{{\rm in},\leftarrow}(t)]
\\ -\gamma_{\rm 1D}\sum\limits_{n=1}^N\sigma_{n}\left(t-|z_m-z_n|/c\right)\:,
\end{multline}
where  we have introduced the rate of spontaneous emission into the waveguide $\gamma_{\rm 1D}=|g_+^2|/c=|g_-^2|/c$.
From now, we consider a situation when $|g_+|=|g_-|$, a more general case  with chiral coupling is discussed in Sec.~\ref{sec:chiral}. 

We now use the Markovian approximation 
\begin{equation}
\sigma_{n}\left(t-|z_m-z_n|/c\right)\approx \sigma_{n}(t)\e^{\rmi \omega_0 |z_m-z_n|/c}
\end{equation}
that is provided that the structure is not too long. More general, non-Markovian input-output approach can be found in Ref.~\cite{Ciccarello2018}.
Equation~\eqref{eq:sigmam} then reduces to
\begin{multline}\label{eq:sigmam2}
 \frac{\rmd  \sigma_m}{\rmd t}=\rmi [H_{\rm eff},\sigma_m]\\-\frac{\rmi(1-2\sigma_m^\dag \sigma_m)}{\sqrt{L}}
[g_+^*a_{{\rm in},\rightarrow}(t)+g_-^*a_{{\rm in},\leftarrow}(t)]\:.
\end{multline}
and
\begin{equation}\label{eq:Heff-io2}
H_{\rm eff}=\sum\limits_{m=1}^N\omega_0\sigma_m^\dag \sigma_n^{\vphantom{\dag}}-\rmi \gamma_{\rm 1D}\
\sum\limits_{m,n=1}^N \sigma_m^\dag \sigma_{n}\e^{\rmi \omega_0 |z_m-z_n|/c}
\end{equation}
is the effective atomic Hamiltonian with traced out photonic degrees of freedom. It is equivalent to Eq.~\eqref{eq:Heff-io} in the main text if the replacement $k_z\to \omega_0/c$ is made.

The treatment becomes simpler when the input field is in the coherent state. In this case, it is convenient to determine the system density matrix $\rho$ from the master equation 
$ \partial_t \rho=\rmi [\rho,H]+\mathcal L\rho$ and then find the scattered light. Specifically, we consider case
when the structure is driven from the left by a  coherent field  at the  frequency $\omega$. The Hamiltonian $H$, entering the master equation, then reads
\begin{equation}
H=\frac{\Omega}{2}\sum\limits_{m=1}^N(\e^{\rmi\omega( z_m/c- t)}\sigma_m^\dag+{\rm H.c.})\\+
\frac{ H_{\rm eff}+H_{\rm eff}^\dag}{2}
\end{equation}
where $\Omega$ is the Rabi frequency. This Hamiltonian includes the Hermitian part of the effective Hamiltonian Eq.~\eqref{eq:Heff-io}. The non-Hermitian part describing the decay processes is incorporated into the Lindblad operator
\begin{multline}
\mathcal L\rho=\sum\limits_{m,n=1}^{N} \left[ \gamma_{\rm 1D} \cos\frac{\omega_0 (z_m-z_{n})}{c} +\delta_{m,n}\gamma\right]\\\times
(2\sigma_{m}\rho \sigma_{n}^{\dag}-\sigma_{m}^{\dag}\sigma_{n}\rho-\rho \sigma_{m}^{\dag}\sigma_{n})\:.\label{eq:Lgen}
\end{multline}
We have also added nonradiative damping term $\gamma$ to the Lindblad operator.
Once the density matrix has been found, the coherent reflection coefficient is given by
\begin{equation}\label{eq:r-io}
r=\frac{2\rmi \gamma_{\rm 1D}}{\Omega}\sum\limits_{m=1}^N\e^{\rmi \omega_0z_m/c} \langle \sigma_m (t)\rangle \e^{\rmi \omega t}\:,
\end{equation}
and the coherent transmission coefficient reads 
\begin{equation}\label{eq:t-io}
t=1+\frac{2\rmi \gamma_{\rm 1D}}{\Omega}\sum\limits_{m=1}^N\e^{-\rmi \omega_0z_m/c} \langle \sigma_m(t) \rangle \e^{\rmi \omega t}\:.
\end{equation}
 More details on the input-output formalism in the WQED setup can be also found in Refs.~\cite{Blais2013,Sorensen2018}.

\section{Transfer matrix method}\label{app:transfer}
Probably, the easiest way to calculate single-photon reflection and transmission coefficients for an arbitrarily spaced atomic array coupled to the waveguide, is offered by the transfer matrix method~\cite{Corzo2016}. 
Electric field to the left and right from an atom located at the point $z=0$ is presented as
\begin{equation}
E(z)=\begin{cases}
E_L^\rightarrow\e^{\rmi \omega z/c}+
E_L^\leftarrow\e^{-\rmi \omega z/c}, & (z<0)\:,\\
E_R^\rightarrow\e^{\rmi \omega z/c}+
E_R^\leftarrow\e^{-\rmi \omega z/c}, & (z>0)
\end{cases}
\end{equation}
(we assume the $\e^{-\rmi \omega t}$ time dependence).
The fields to the left and right of the atom are linked  
\begin{equation}
\begin{pmatrix}
E_R^\rightarrow\\
E_R^\leftarrow
\end{pmatrix}=M_{\rm atom}
\begin{pmatrix}
E_L^\rightarrow\\
E_L^\leftarrow
\end{pmatrix}
\end{equation}
by the transfer matrix 
\begin{equation}\label{eq:Matom:chiral}
M_{\rm atom}=\frac{1}{t_\leftarrow}\begin{pmatrix}
t_{\rightarrow}t_{\leftarrow}-r^2&r\\-r&1
\end{pmatrix}
\end{equation}
that is expressed via the reflection and transmission coefficients $r$ and $t_{\rightarrow/\leftarrow}$, given by Eqs.~\eqref{eq:rt1oneway}.
Equation~\eqref{eq:Matom:chiral} is written for the case of general chiral coupling, when 
forward $t_{\rightarrow}$ and backward $t_{\leftarrow}$ transmission coefficients can differ, see Eqs.~\eqref{eq:rt1oneway}.
The transfer matrix through the free part of the waveguide with the length $d$ can be expressed as
\begin{equation}
M_{d}=\begin{pmatrix}
\e^{\rmi \omega d/c}&0\\0&\e^{-\rmi \omega d/c}
\end{pmatrix} \:. 
\label{eq:Md}
\end{equation}
By multiplying these matrices, we obtain the total transfer matrix through an array of $N$ atoms with the period $d$ in the following form:
\begin{equation}
M_N=(M_{d}M_{\rm atom})^N,
\end{equation}
that allows to find reflection and transmission coefficients as
\begin{equation}\label{eq:rN0}
r_N^{\leftarrow}=-\frac{[M_N]_{2,1}}{[M_N]_{2,2}},\quad 
t_N^{\rightarrow}=\frac{\det M_N}{[M_N]_{2,2}}\:.
\end{equation}
In case of symmetric coupling, when $t_{\rightarrow}=t_{\leftarrow}=t$, 
 it is also possible to obtain an analytical expressions for Eqs.~\eqref{eq:rN0} that read \cite{Ivchenko2005}
\begin{equation}
\label{eq:rN}
r_N^\leftarrow = \frac{\tilde r \sin NKd}{\sin NKd-\tilde t\sin (N-1)Kd }\:,\:
t_N^\rightarrow=\frac{\tilde t \sin Kd}{ \tilde r \sin NKd}r^\leftarrow_N\:,
\end{equation}
where $\tilde t=t\e^{\rmi \omega d/c}$, $\tilde r=r\e^{\rmi \omega d/c}$
are the transmission and reflection coefficients through one period of the array.

\section{Photon pair scattering from a single atom}\label{app:2scat-kspace}
Here, we  solve the problem of a photon pair scattering on a single  atom.
General diagrammatic Green's function approach to solve a more general problem for $N$ atoms is discussed in 
Appendix~\ref{app:2scat-exciton}. The goal of the current section is to present a more straightforward technique that does not require prior knowledge of the Green's function theory. Instead of solving the real-space differential equations for the two-photon wave function, as has been done in \cite{Fan2007,Shen2007}, we solve the Schr\"odinger equation directly in double-excited subspace of the Hilbert space. The procedure  can be viewed as a ``poor man" version of the Bethe ansatz technique, that will be discussed in Appendix~\ref{app:2scat-Bethe}.
We start from the Hamiltonian 
\begin{multline}\label{eq:H-NatomsU1}
H=H_{\rm phot}+H_{\rm atom}+H_{\rm atom-phot}\\=
\sum\limits_{k} \omega_k a_k^\dag a_k^{\phantom{}}+\omega_0b^\dag b+\frac{U}{2}b^\dag b(b^\dag b-1)
+\frac{g}{\sqrt{L}}\sum\limits_{k}(a_k^\dag b^{\vphantom{\dag}}+a_k b^\dag)\:.
\end{multline}

 The wavefunction is sought in the form
\begin{equation}\label{eq:psi2atom1}
|\psi\rangle=\sum\limits_{k}\sum\limits_{k'}E_{kk'}\frac{a_{k}^{\dag}a_{k'}^{\dag}}{2}|0\rangle +\sum\limits_{k}P_{k}a_{k}^{\dag}  b^{\dag}|0\rangle+Q\frac{b^{\dag,2}}{2}|0\rangle\:.
\end{equation}
Here,  $|0\rangle$ is a state with zero photons and the atom in its ground state.
The state Eq.~\eqref{eq:psi2atom1} contains all possible combinations of the double excited states, namely the states with two photons, the states with one photon absorbed and the atom in the double excited state and the state with a double excited atom. The Schr\"odinger equation for the double-excited states reads
\begin{equation}\label{eq:E-2-phot0}
E_{kk'}\omega_{k}+\omega_{k'}+\frac{g}{\sqrt{L}}(P_{k}+P_{k'})=2\eps E_{kk'}\:,
\end{equation}
where $2\eps$ is the total energy. However,  we need to take into account  that the structure is excited from the left by the two photons with the energy $\eps$ and the wave vector $k=\eps/c$. To describe this we add an inhomogeneous term to Eq.~\eqref{eq:E-2-phot0}, corresponding to the excitation, so that  in the absence of atoms one has $E_{kk'}=\delta_{k,\eps/c}\delta_{k',\eps/c}$. The result is
\begin{equation}\label{eq:E-2-phot1}
E_{kk'}+\frac{g/\sqrt{L}}{\omega_{k}+\omega_{k'}-2\eps-\rmi 0}(P_{k}+P_{k'})=\delta_{k,\eps/c}\delta_{k',\eps/c}\:.
\end{equation}
Here the term $-\rmi 0$ in denominator means an infinitely small imaginary part that has been added  for regularization purposes.
The Schr\"odinger equation for the states, where one photon has been absorbed reads
\begin{equation}\label{eq:P-2-phot1}
(\omega_{k}+\omega_{0}-2\eps)P_{k}+\frac{g}{\sqrt{L}}\sum\limits_{k'}(E_{kk'}+E_{k'k})+\frac{g}{\sqrt{L}}\sqrt{2}Q=0\:,
\end{equation}
Expressing electric field from Eq.~\eqref{eq:E-2-phot1} and substituting into Eq.~\eqref{eq:P-2-phot2}, we find
\begin{multline}\label{eq:P-2-phot2}
(\omega_{k}+\omega_{0}-2\eps)P_{k}=\frac{g^{2}}{L}P_{k}\sum\limits_{k'}\frac1{\omega_{k}+\omega_{k'}-2\omega-\rmi 0}
\\+\frac{g^{2}}{L}\sum\limits_{k'}\frac1{\omega_{k}+\omega_{k'}-2\omega-\rmi 0}P_{k'}
-g\sqrt{\tfrac{2}{L}}Q\\+\tfrac{2g}{\sqrt{L}}\delta_{k,\eps/c}\:.
\end{multline}
This summation can be carried out exactly
in the rotating wave approximation, 
\begin{equation}\label{eq:sum-rwa}
\int\limits_{-\infty}^{\infty} \frac{\rmd k}{2\pi}\frac1{\omega_k-\omega}\approx 
2\lim_{\delta\to 0}\int \limits_{-\infty}^\infty \frac{\rmd k}{2\pi}\frac1{ck-\rmi \delta-\omega}=\frac{\rmi}{c}\:.
\end{equation}
Here, we have split the integration into two parts, $\int_{-\infty}^{\infty}\rmd k=\int_{0}^{\infty}\rmd k+\int_{-\infty}^{0}\rmd k$ and then  extended each of two resulting integrals back to the full axis, which results in the   prefactor of $2$. 
The result reads
\begin{multline}\label{eq:P-2-phot3}
(\omega_{k}+\omega_{0}-2\eps-\rmi\gamma_{\rm 1D})P_{k}=\frac{g^{2}}{L}\sum\limits_{k'}\frac1{\omega_{k}+\omega_{k'}-\rmi 0-2\eps}P_{k'}\\-g\sqrt{\tfrac{2}{L}}Q+\tfrac{2g}{\sqrt{L}} \delta_{k,\eps/c}.
\end{multline}
The term $-\rmi\gamma_{\rm 1D}P_k$ in the left hand side of Eq.~\eqref{eq:P-2-phot3} describes the spontaneous decay of the state ``propagating photon+excited atom" into the state with two propagating photons. We now divide both parts of Eq.~\eqref{eq:P-2-phot3} by $\omega_{k}+\omega_{0}-2\eps$
and sum over $k$ to find
\begin{equation}\label{eq:P-2-phot4}
\sum\limits_{k}P_{k}=\frac{2g/\sqrt{L}}{\omega_{0}-\eps-\rmi\gamma_{\rm 1D}}-\frac{\rmi g}{c} \sqrt{2L}Q\:,
\end{equation}
where we have again used Eq.~\eqref{eq:sum-rwa}. Importantly, the sum stemming from the first term in the right-hand-side of Eq.~\eqref{eq:E-2-phot1} is zero,
\begin{equation} \label{eq:sum2-zero}
\sum\limits_k\frac1{(\omega_{k}+\omega_{0}-2\eps-\rmi 0 )(\omega_{k}+\omega_{k'}-2\omega-\rmi 0)}=0\:.
\end{equation}
The Schr\"odinger equation for the double-excited state is
\begin{equation}\label{eq:Q2phot}
(2\omega_{0}+U-2\eps)Q+g\sqrt{\tfrac{2}{L}}\sum\limits_{k}P_{k}=0\:.
\end{equation}
Combining Eq.~\eqref{eq:Q2phot} and Eq.~\eqref{eq:P-2-phot3}, we find the amplitude of the double-excited state
\begin{equation}
Q=-\frac{2\sqrt{2}g^{2}/L}{(\omega_{0}+U/2-\eps-\rmi \gamma_{\rm 1D})(\omega_0-\eps-\rmi\gamma_{\rm 1D})}\:.
\end{equation}
We now proceed to solve Eq.~\eqref{eq:P-2-phot3} for $P_k$. This can be done iteratively, by treating the first term in the right-hand side as a perturbation:
\begin{equation}\label{eq:P-2-phot5}
P_{k}=P_k^{(0)}+P_k^{(1)}+\widetilde  P_k^{(1)}\:,
\end{equation}
where 
\begin{equation}
P_{k}^{(0)}=\frac{2g}{\sqrt{L}}\frac{\delta_{k,\eps/c}}{\omega_{0}-\eps-\rmi\gamma_{\rm 1D}}\:,
\end{equation}
\begin{multline}
 P_{k}^{(1)}=\frac{g^{2}}{L}\sum\limits_{k'}\frac1{\omega_{k}+\omega_{k'}-2\eps}P^{(0)}_{k'}
=\frac{2g^{3}}{L^{3/2}}\\\times\frac1{(\omega_{k}+\omega_{0}-2\eps-\rmi\gamma_{\rm 1D})(\omega_k-\eps-\rmi 0)(\omega_{0}-\eps-\rmi\gamma_{\rm 1D})}\:,
\end{multline}
and
\begin{multline}
\tilde P_{k}^{(1)}=-g\sqrt{\tfrac{2}{L}}\frac{Q}{\omega_{k}+\omega_{0}-2\eps-\rmi\gamma_{\rm 1D}}\\=
P_{k}^{(1)}\frac{\omega_k-\eps}{\omega_0+U/2-\eps-\rmi \gamma_{\rm 1D}}\:.
\end{multline}
The key observation is that the solution Eq.~\eqref{eq:P-2-phot5} is actually exact and all the higher order terms are zero. Mathematically, the reason of the cancellation is the same as for the sum Eq.~\eqref{eq:sum2-zero}.

We are now in the position to find the amplitude of the incoherent two-photon scattering process.
To this end, we substitute the solution Eq.~\eqref{eq:P-2-phot5} into Eq.~\eqref{eq:E-2-phot1}.
The terms $\propto P_k^{(0)}$ describe the incoherent scattering. The amplitude of the scattering matrix for the incoherent scattering process is given by
\begin{equation}
S(\omega_k,\omega_{k'}\leftarrow \eps,\eps)=2\pi\rmi\delta (\omega_k+\omega_{k'}-2\eps)M(\omega_k,\omega_{k'}\leftarrow \eps,\eps)
\end{equation}
where 
\begin{equation}\label{eq:M-2phot-4terms}
M(\omega_k,\omega_{k'}\leftarrow \eps,\eps)=\frac{L^2}{c^2}\frac{g}{\sqrt{L}}(P_{k}^{(1)}+\widetilde P_{k}^{(1)}+P_{k'}^{(1)}+\widetilde P_{k'}^{(1)})
\end{equation}
with  $\omega_{k'}+\omega_k=2\eps$.
Performing the summation of four terms in Eq.~\eqref{eq:M-2phot-4terms}, we obtain 
\begin{equation}
M(\omega_k,\omega_{k'}\leftarrow \eps,\eps)\\
=-\frac{4U\gamma_{\rm 1D}^2 s(\omega_k)s(\eps)s(\omega_k')}{(2\omega_0+U-2\eps-2\rmi \gamma_{\rm 1D})}\:,
\end{equation}
where $s(\omega)=1/(\omega_0-\omega-\rmi\gamma_{\rm 1D})$\:. In the limit of two-level atom, $U\to \infty$, this expression exactly matches the general result Eq.~\eqref{eq:M-2phot-1} for $N=1$ atom.

The derivation above becomes especially simple for a two-level atom, where $U\to \infty$ so that $Q=0$ and the terms $\widetilde P^{(1)}$ can be neglected. It also explains how the incoherent scattering vanishes for a harmonic atom, $U=0$. In this case, the terms $P_{k}^{(1)}+P_{k'}^{(1)}$ and 
$\widetilde P_{k'}+ \widetilde P_{k'}^{(1)}$, resulting from the single-excited and double-excited states, cancel each other exactly.


\section{Photon pair scattering: the Bethe ansatz}\label{app:2scat-Bethe}
In this section, we consider the Dicke problem of photons interacting with an array of $N$ identical two-level atoms. The derivation mostly follows the Bethe ansatz approach from \cite{Yudson1984,Yudson2008}
and \cite{Fan2007,Roy2013}\:.

We start by rewriting the problem Hamiltonian Eq.~\eqref{eq:H-NatomsU1} in the real space
\begin{multline}\label{eq:H-chiral1}
H=-\rmi c\int\rmd x [a_\rightarrow^\dag (x)\partial_xa_\rightarrow(x)-a_\leftarrow^\dag (x)\partial_xa_\leftarrow(x)] \\+\sum\limits_{\nu=\rightarrow,\leftarrow}\bigl[ g [a_\nu^\dag(0) \sigma+a_\nu(0) \sigma^\dag]\bigr]\:.
\end{multline}
Here, we assume the two-level atoms with the excitations characterized by the destruction operators $\sigma_j$,  $\sigma=\sum_{j=1}^N \sigma_j$, and  $a_\nu^\dag$ are the creation operators for right- ($\nu=\to$) and left- ($\nu = \leftarrow$) going photons; $[a_\nu^\dag(x),a_\nu(x')]=\delta(x-x')$. We assume the rotating wave approximation measuring the energies from the atomic resonance $\omega_0$ and also set the normalization length $L$ to unity in this section.

Due to the mirror reflection symmetry $x\to -x$, the problem described by the Hamiltonian Eq.~\eqref{eq:H-chiral1} can be solved separately in even- and odd- scattering channels. Namely, if the new operators $a(x)$ and $a_0(x)$
\begin{equation}\label{eq:aa0}
a(x)=\frac{a_\to(x)+a_\leftarrow(-x)}{\sqrt{2}},\quad 
a_0(x)=\frac{a_\to(x)-a_\leftarrow(-x)}{\sqrt{2}}
\end{equation}
are introduced, the Hamiltonian Eq.~\eqref{eq:H-chiral1} is separated as $H=H_{\rm even}+H_{\rm odd}$, where
\begin{equation}\label{eq:H-chiral0}
H_{\rm even}=-\rmi c\int\rmd x a^\dag (x)\partial_x a(x) +\tilde g [a^\dag(0) \sigma+a(0) \sigma^\dag]
\end{equation}
with $\tilde g=\sqrt{2g}$. In the odd scattering channel, the photons do not interact with atoms at all,
$H_{\rm odd}=-\rmi c\int\rmd x a_0^\dag (x)\partial_x a_0(x)$, and the problem is trivial. We will now focus on the scattering problem in the even channel.
\begin{figure}[t!]
\centering\includegraphics[width=0.3\textwidth]{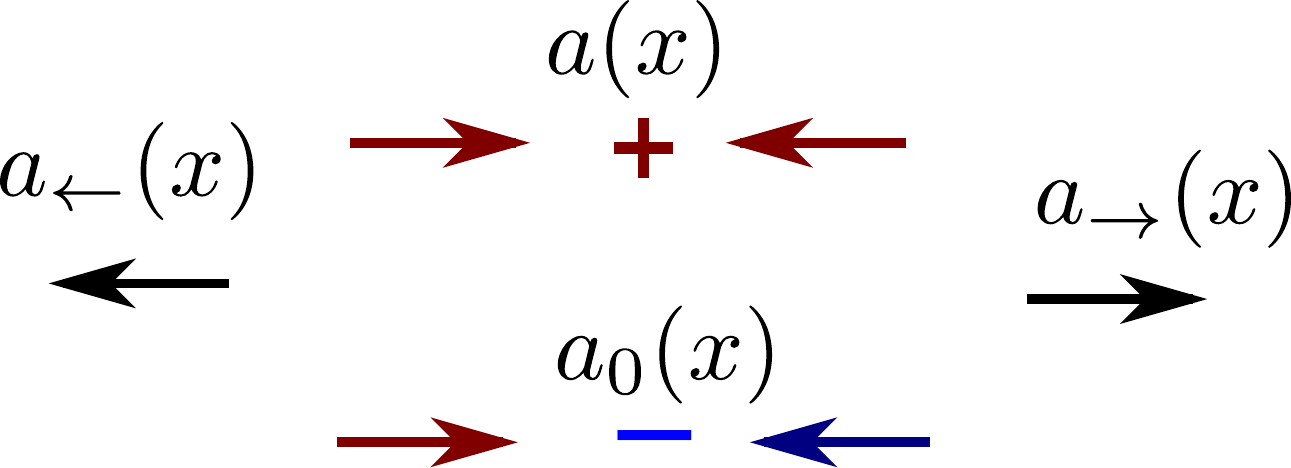}
\caption{Schematics of the separation  Eq.~\eqref{eq:aa0} of the problem with left- and right- propagating photons into two problems with even and odd symmetry. }\label{fig:a-chiral}
\end{figure}

Single-excited eigenstates of Eq.~\eqref{eq:H-chiral1} can be written as a superposition of the states with one photon and the states with zero photons and atoms excited to a symmetric Dicke state:
\begin{equation}\label{eq:Bethe-k-1}
|k\rangle=\int\rmd xE(x) a^\dag(x) |0\rangle+P \sigma^\dag |0\rangle\:.
\end{equation}
The Schr\"odinger equation reads
\begin{align}\label{eq:Bethe1photon-real}
-\rmi c\partial_xE+\tilde g\delta(x)P=\eps E\:,\\
N\tilde gE(0)=\eps P\nonumber
\end{align}
and has eigenstates with the energy
$\eps=ck$ and 
\begin{equation}
E_k(x)=\theta(-x)\e^{\rmi kx}+\theta(x) t^{\rm even}_k \e^{\rmi kx},\:
\end{equation}
where 
$
t^{\rm even}_k=(ck-\rmi N\gamma_{\rm 1D})/(ck+\rmi N\gamma_{\rm 1D})
$ is the transmission coefficient in the even channel and 
$\gamma_{\rm 1D}=\tilde g^2/(2c)\equiv g^2/c$ is the radiative decay rate.
 We note, that the reflection and transmission coefficients for one atom can be found as $r=(1-t_k^{\rm even})/2$, $t=(1+t_k^{\rm even})/2$ taking into account that $\eps=\omega-\omega_0$. The eigenstate \eqref{eq:Bethe-k-1} can be also written in a compact way as
\begin{equation}\label{eq:Bethe-k-2}
|k\rangle =\int\rmd x r_k^\dag(x)|0\rangle,\:
r_k^\dag(x)\equiv E_k(x) a^\dag(x)+P\delta(x)\sigma^\dag\:.
\end{equation}

\begin{figure}
\centering\includegraphics[width=0.4\textwidth]{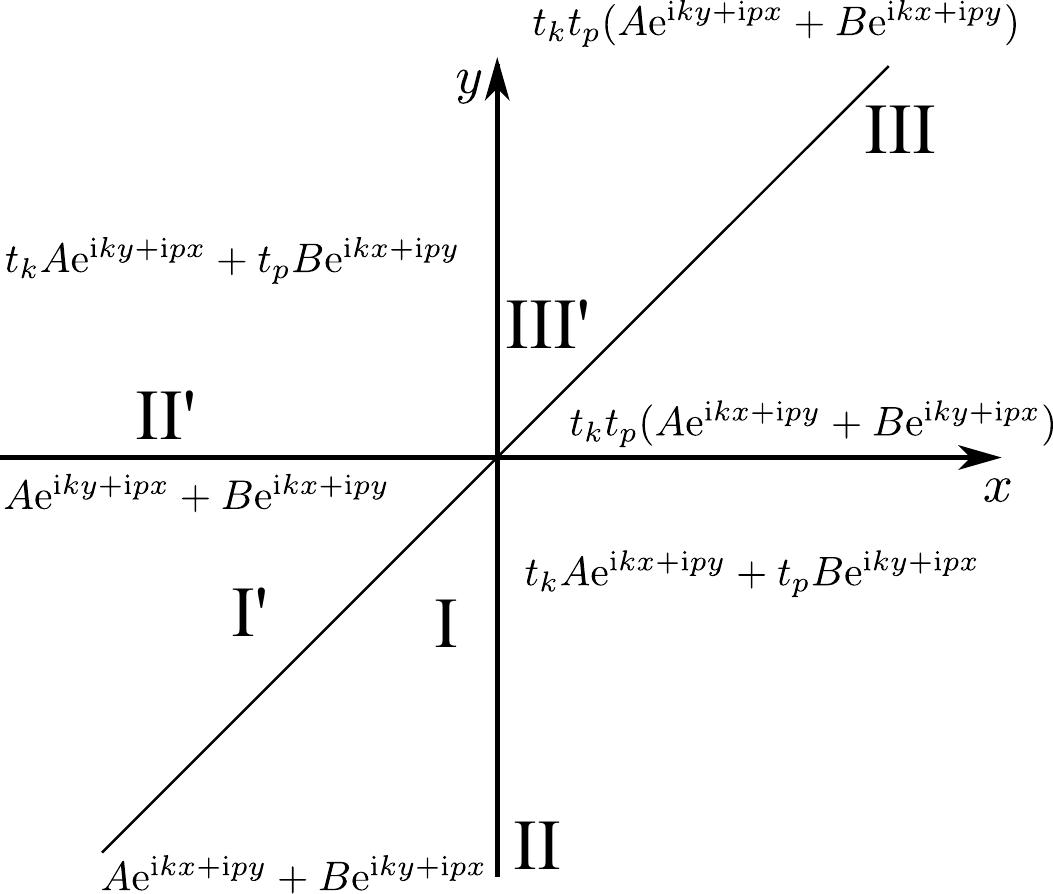}
\caption{Illustration of the Bethe ansatz Eq.~\eqref{eq:Bethe2photons-real2} for the two-photon states depending on the first and second coordinates $x$ and $y$. Region I corresponds to the incident state. In regions II and II$'$ either first or second photons is scattered  on the atom, in regions III and III$'$ both photons have scattered.}\label{fig:Bethe}
\end{figure}

\begin{figure}[t]
\centering\includegraphics[width=0.3\textwidth]{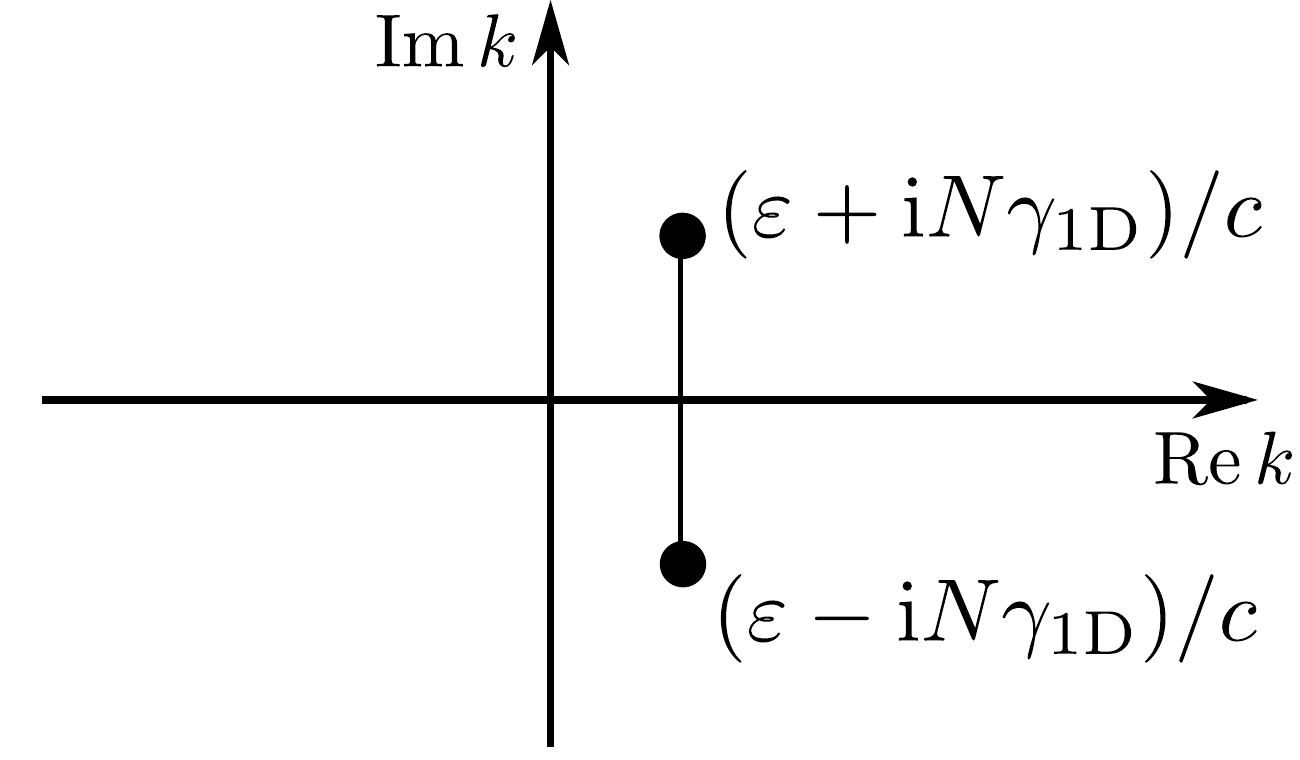}
\caption{Illustration of the so-called string between two rapidities, describing the bound two-photon state Eq.~\eqref{eq:Bethe-bound} in the Bethe ansatz.} \label{fig:string-Yudson}
\end{figure}

We now proceed to the double-excited states  having the energy $2\eps$ and described by the ansatz
\begin{multline}\label{eq:Bethe-k-3}
\iint\rmd x\rmd y  E(x,y) a^\dag(x)a^\dag(y) |0\rangle+ \int\rmd x  P(x) a^\dag(x) \sigma^\dag |0\rangle\\+\frac{Q}{2} \sum\limits_{j\ne j'}\sigma_j^\dag \sigma_{j'}^\dag|0\rangle\:,
\end{multline}
equivalent to Eq.~\eqref{eq:psi2atom1}. The last term accounts for the double-excited atomic array and is present only for $N>1$. Instead of Eqs.~\eqref{eq:Bethe1photon-real}, we obtain
\begin{align}\label{eq:Bethe2photons-real}
&-\rmi (\partial_x+\partial _y)E+\frac{\tilde g}{2}[\delta(x)P(y)+\delta(y)P(x)]=2\eps E\:,\\
&-\rmi \partial_x P+N\tilde g[E(x,0)+E(0,x)]+(N-1)\tilde g \delta(x)Q=2\eps P\:,\nonumber\\\nonumber
&2\tilde g P=2\eps Q\:.
\end{align}
Here, we assume the bosonic symmetry $E(x,y)=E(y,x)$ and also define the electric field at the singular lines $x=0$ or $y=0$ as
\begin{equation*}
\frac{E(x,0^+)+E(x,0^-)}{2}\equiv \lim_{\delta\to 0}\frac{E(x,-\delta)+E(x,+\delta)}{2}\:.
\end{equation*}
The essence of the Bethe ansatz approach is the representation of the amplitude $E(x,y)$ as a sum of free-space plane wave solutions outside the singular lines where $x=0$ or $y=0$,
\begin{equation}\label{eq:Bethe2photons-real2}
E(x,y)=\begin{cases}
A\e^{\rmi kx+\rmi py}+B\e^{\rmi ky+\rmi px},&\text{region I}\:,\\
At_k^{\rm even}\e^{\rmi kx+\rmi py}+Bt_p^{\rm even}\e^{\rmi ky+\rmi px},&\text{region II}\:,\\
t_k^{\rm even}t_p^{\rm even} (A\e^{\rmi kx+\rmi py}+B\e^{\rmi ky+\rmi px}),&\text{region III}\:,
\end{cases}
\end{equation}
where the regions I,II,III are indicated in Fig.~\ref{fig:Bethe}.
The states Eq.~\eqref{eq:Bethe2photons-real} have the energy $2\eps=c(k+p)$. The amplitude $P(x)$ is found 
from the first of Eqs.~\eqref{eq:Bethe2photons-real} 
as
\begin{equation}\label{eq:Bethe-jump}
P(x)=\frac{2\rmi}{\tilde g}\begin{cases}
E(x,0^+)-E(x,0^-),\quad x<0\:,\\
E(0^+,x)-E(0^-,x),\quad x>0\:.
\end{cases}
\end{equation}
Second and third Eqs.~\eqref{eq:Bethe2photons-real}  yield 
 the continuity condition 
\begin{equation}\label{eq:Bethe-continuity}
P(0^+)-P(0^-)=-\frac{\rmi (N-1)\tilde g^2}{2\eps}[P(0^+)+P(0^-)]\:.
\end{equation}
From Eqs.~\eqref{eq:Bethe-continuity} and \eqref{eq:Bethe-jump} we find  the relationship between the amplitudes $A$ and $B$:
\begin{equation}\label{eq:ABBethe}
\frac{A}{B}=-\frac{\rmi \tilde g^2+ck-cp}{\rmi \tilde g^2-ck+cp}\:.
\end{equation}
Taking into account that $E(x,y)=E(y,x)$, we rewrite the amplitude in the region where 
$x,y<0$ as
\begin{multline}
E(x,y)=\theta(x-y)[A\e^{\rmi kx+\rmi py}+B\e^{\rmi ky+\rmi px}]\\+
\theta(y-x)[A\e^{\rmi ky+\rmi px}+B\e^{\rmi kx+\rmi py}]\\=
\frac{A+B}{2}\e^{\rmi kx+\rmi py}\left(1+\sign(x-y)\frac{A-B}{A+B}\right)
+(k\leftrightarrow p)\:.
\end{multline}
Since 
$(A-B)/(A+B)=\rmi \tilde g^2/(ck-cp)$,
the eigenstate \eqref{eq:Bethe-k-3} can be presented in the following form:
\begin{multline}\label{eq:Bethe-k-4}
|k,p\rangle =C\iint\rmd x\rmd y  \left(1+\frac{\rmi (\tilde g^2/c)\sign(x-y)}{k-p}\right)
\\\times  r_k^\dag(x)r_p^\dag(y)|0\rangle\:,
\end{multline}
where $C$ is the normalization factor.
Such the factorization of \eqref{eq:Bethe-k-4} in products of single-excited eigenstates $r_k^\dag(x)r_p^\dag(x)|0\rangle$ is the central result of this section and means that the problem is solvable by the Bethe ansatz. In the Bethe ansatz formalism, the wavevectors $k$ and $p$ are termed as rapidities~\cite{Tsvelick1983}. For two-photon states with real-valued energies $2\eps=c(k+p)$, the rapidities can be either both real or can correspond to complex conjugated pairs  $k=p^*$ that are termed as ``strings", see Fig.~\ref{fig:string-Yudson}.
Of special interest are the ``strings" 
\begin{equation}\label{eq:string}
k=\frac{\eps+\rmi  N \gamma_{\rm 1D}}{c},\quad p=\frac{\eps-\rmi N\gamma_{\rm 1D}}{c},
\end{equation}
that correspond to the bound two-photon states.  
Specifically, if $k$ and $p$ are given by Eq.~\eqref{eq:string}, we find from Eq.~\eqref{eq:ABBethe} that 
$B=0$ and the two-photon state \eqref{eq:Bethe2photons-real2} is then simplified to
\begin{equation}\label{eq:Bethe-bound}
E(x,y)=\e^{\rmi \eps(x+y)/c}\e^{-N|x-y|\gamma_{\rm 1D}/c}\times\begin{cases}
1,& x,y < 0\\
t_k,& x,y < 0\\
t_k t_p,& x,y > 0\:.
\end{cases}
\end{equation}
The wavefunction amplitude for the bound state decays exponentially with increase of the  distance between the two photons $|x-y|$.

It has been proved in Ref.~\cite{Yudson1984} that the Bethe ansatz holds for an arbitrary number of excitations $M=1,2,3\ldots $, i.e. the quantum Dicke problem is integrable. The eigenstate characterized by the so-called rapidities $k_1\ldots k_M$ has the form
\begin{multline}\label{eq:Bethe-k-general}
|\bm k\rangle =C
\iint\rmd^M x\prod\limits_{m<n}  \left(1+\frac{\rmi  \tilde g^2\sign(x_m-x_n)}{c(k_m-k_n)}\right)\\\times\prod\limits_{n=1}^M r_{k_n}^\dag(x_n)|0\rangle
\end{multline}
with the energy $\eps=c\sum_{m=1}^Mk_m$.

In order to solve a scattering problem, when $M$ photons are incident at the atoms from the left, $x\to -\infty$,
one more step is required. Namely, the incident photons wavefunction has to be expanded over the Bethe eigenstates. This can be done in two ways. One is the ``brute-force" approach, when the input eigenstate is presented as a superposition of the states  Eq.~\eqref{eq:Bethe-k-4} with real rapidities $k$ and $p$ and a bound state  Eq.~\eqref{eq:Bethe-bound}. It has been proved in Ref.~\cite{Fan2007} that this set is complete and allows one to find the full scattering matrix of the problem. In order to perform such expansion, the eigenstates Eq.~\eqref{eq:Bethe-k-4},  Eq.~\eqref{eq:Bethe-bound} have to be properly normalized.

\begin{figure}[t!]
\centering\includegraphics[width=0.35\textwidth]{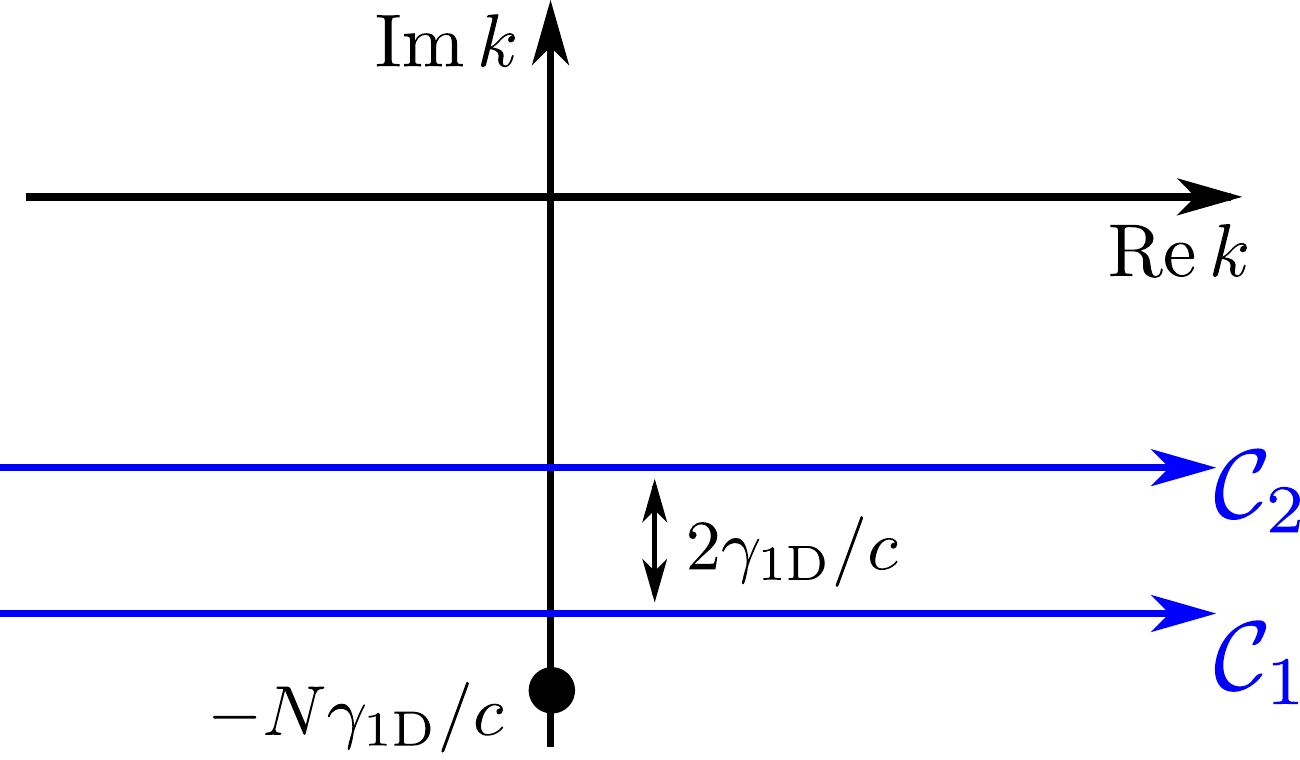}
\caption{Integration contours for the rapidities $k_1$ and $k_2$ used to calculate the two-photon scattering problem in Eq.~\eqref{eq:E-2-phot-scat-Bethe}.  }\label{fig:contour-Yudson}
\end{figure}

There exists an alternative approach that  circumvents the expansion of  the input state over the Bethe eigenstates and directly provides the scattered eigenstate ~\cite{Yudson1985}. Its particular application to the two-photon scattering problem is discussed in detail in \cite{Yudson2008}. The two-photon amplitude is expressed as
\begin{multline}\label{eq:E-2-phot-scat-Bethe}
E(x_1,x_2,t)=\int\limits_{\mathcal {C}_1}\frac{\rmd k_1}{2\pi}
\int\limits_{\mathcal {C}_2}\frac{\rmd k_2}{2\pi}
\left(1-\frac{2\rmi (\tilde g^2/c)\theta(x_2-x_1)}{k_1-k_2+\rmi \tilde g^2/c}\right)\\\times\e^{\rmi k_1(x_1-x_1^{(0)}-ct)}\e^{\rmi k_2(x_2-x_2^{(0)}-ct)}E_{k_1}(x_1)E_{k_2}(x_2),
\end{multline}
where the integration over the rapidities $k_1$ and $k_2$ is performed along the contours $\mathcal C_1$ 
and $\mathcal C_2$ in the complex plane shown in Fig.~\ref{fig:contour-Yudson}. Here, $x_2^{(0)}<x_1^{(0)}<0$ are the coordinates of two incident photons at $t=0$ when the  input state  is $a^\dag(x_1^{(0)})a^\dag(x_2^{(0)})|0\rangle $ . The advantage of the approach Eq.~\eqref{eq:E-2-phot-scat-Bethe} is that it can be generalized for an arbitrary number of incident photons. To this end, the integration contours should satisfy the relation \cite{Yudson2008}
\begin{equation}
\Im k_{n+1}-\Im k_{n}>\frac{2\gamma_{\rm 1D}}{c},\quad \Im k_{1}>-\frac{N\gamma_{\rm 1D}}{c}.
\end{equation}
Performing the integration in Eq.~\eqref{eq:E-2-phot-scat-Bethe}, we obtain
\begin{multline}\label{eq:Bethe-solution}
E(x_1,x_2,t)=\Phi(x_1,x_1^{(0)}+ct)\Phi(x_2,x_2^{(0)}+ct)\\+\Phi_c(x_1,x_2,x_1^{(0)}+ct,x_2^{(0)}+ct)\:.
\end{multline}
Here, 
\begin{multline}
\Phi(x,x^{(0)}+ct)=\delta(x-x^{(0)}-ct)\\-2\gamma_{\rm 1D} N\theta(0<x<x^{(0)}+ct)\e^{\gamma_{\rm 1D} N(x-x^{(0)}-ct)}
\end{multline}
is the solution of the scattering problem for input state $a^\dag(x^{(0)})|0\rangle $ with one photon at $t=0$.
Hence, the first term in Eq.~\eqref{eq:Bethe-solution} describes the independent scattering of two photons on the array of atoms. The second term describes the interaction between the photons and reads
\begin{multline}\label{eq:Bethe-solution2}
\Phi_c(x_1,x_2,x_1^{(0)}+ct,x_2^{(0)}+ct)=\\-\frac{2N\tilde g^4}{c^2}
\theta(0<x_1<x_2<x_2^{(0)}+ct<x_1^{(0)}+ct)\\\times\e^{N\gamma_{\rm 1D}(x_1+x_2-x_1^{(0)}-x_2^{(0)}-2ct)} \\\times[N-(N-1)\e^{2\gamma_{\rm 1D} (x_2^{(0)}+ct-x_2)}]\:.
\end{multline}
Here, the notation $\theta(0<x_1<x_2<\ldots )$ means the product of the corresponding Heaviside step-functions
that is unity if $0<x_1<x_2$ and zero otherwise. The second term in square brackets in Eq.~\eqref{eq:Bethe-solution} describes the contribution of photon scattering by an atomic system already excited to the Dicke state $b^\dag |0\rangle$. This term is absent for the single-atom case when $N=1$.

The scattering of two photons with certain incident energies $ck_1$, $ck_2$
($ck_1+ck_2=2\eps$) can be considered by performing the Fourier transform of  Eq.~\eqref{eq:Bethe-solution2}. We write the input state as
\begin{multline*}
|{\rm in}\rangle =\iint\rmd x_1^{(0)}\rmd x_2^{(0)}\theta(x_2^{(0)}<x_1^{(0)})\e^{\rmi k_1x_1^{(0)}+\rmi k_2x^{(0)}_2}\\\times
a^\dag_\to(x_1^{(0)})a^\dag_\to(x_2^{(0)})|0\rangle +(k_1\leftrightarrow k_2)\:.
\end{multline*}
The scattered state is obtained by separating the odd and even scattering channels
and applying Eq.~\eqref{eq:Bethe-solution2} in the even channel
(see Fig.~\ref{fig:a-chiral} and  Eq.~\eqref{eq:aa0}).
The scattered state for $x_2,x_1>0$ (transmission channel) can be presented as 
\begin{equation}
|{\rm out}\rangle= \iint\rmd x_1\rmd x_2 t(x_1,x_2)a^\dag(x_1^{(0)})a^\dag(x_2^{(0)})|0\rangle,
\end{equation}
where 
\begin{multline}\label{eq:t-Bethe}
 t(x_1,x_2)= t(x_2,x_1)\\=\frac1{2}[\e^{\rmi k_1x_1+k_2x_2}+(k_1\leftrightarrow k_2)]t_N(ck_1)t_N(ck_2)\\+
 \frac{1}{8}\int\limits^\infty_{x_2}\rmd x_2^{(0)} \int\limits^\infty_{x_2^{(0)}}\rmd x_1^{(0)} \Phi_c(x_1,x_2,x_1^{(0)},x_2^{(0)})\\
 \times [\e^{\rmi k_1x_1^{(0)}+\rmi k_2x^{(0)}_2}+(k_1\leftrightarrow k_2)] 
\end{multline}
and
\[
t_N(ck)=\frac{1+t_k^{ \rm even}}{2}=\frac{-ck}{ck+\rmi N\gamma_{\rm 1D}}
\]
is the transmission coefficient through $N$ atoms (equivalent to Eqs.~\eqref{eq:rtN}).
The first part of Eq.~\eqref{eq:t-Bethe} describes the independent transmission of two photons and the second part results from their interaction with each other. The prefactor $1/8$ comes from the conversion from the symmetric to the chiral problem and back ($\propto 1/2^2$) and also from the symmetrization of the transmission amplitude with respect to the permutations of $x_1$ and $x_2$. Performing the integration we obtain \cite{Yudson2008}
\begin{multline}\label{eq:t-Bethe2}
 t(x_1,x_2)=\frac1{2}[\e^{\rmi k_1x_1+k_2x_2}+(k_1\leftrightarrow k_2)]t_N(ck_1)t_N(ck_2)\\+
 \e^{\rmi \eps(x_1+x_2)}\frac{N\eps \gamma_{\rm 1D}^2\e^{(\rmi \eps-N\gamma_{\rm 1D})|x_1-x_2|/c}}{[\eps+\rmi (N-1)\gamma_{\rm 1D}](ck_1+\rmi N\gamma_{\rm 1D})(ck_2+\rmi N\gamma_{\rm 1D})}\:.
\end{multline}
Finally, the transmitted state  for two incident photons having
 incident energies $\omega_{1,2}=ck_{1,2}$ is given by
\[
|{\rm out}\rangle =\frac1{2}\int\frac{\rmd \omega_1'\rmd \omega_2' }{(2\pi)^2}
S(\omega_1',\omega_2'\leftarrow \omega_1,\omega_2) a_{\omega_1'/c}^{\dag}a_{\omega_2'/c}^{\dag}|0\rangle
\]
where 
\begin{multline}\label{eq:t-Bethe2b}
S(\omega_1',\omega_2'\leftarrow \omega_1,\omega_2)=(2\pi)^2[\delta(\omega_1-\omega_1')\delta(\omega_2-\omega_2')\\+(\omega_1\leftrightarrow \omega_2)]t_N(\omega_1)t_N(\omega_2)\\+2\pi\rmi M(\omega_1',\omega_2'\leftarrow \omega_1,\omega_2) \delta(\omega_1+\omega_2-\omega_1'-\omega_2')
\end{multline}
\begin{multline}
M(\omega_1',\omega_2'\leftarrow \omega_1,\omega_2)=4 \gamma_{\rm 1D}^2s(\omega_1)s(\omega_2)s(\omega_1')s(\omega_2')\\\times
\frac{N(\eps-\omega_0)(\eps-\omega_0+\rmi N\gamma_{\rm 1D})  }{\eps-\omega_0+\rmi (N-1)\gamma_{\rm 1D}} \:.
\end{multline}
and $s(\omega)=1/(\omega-\omega_0+\rmi N\gamma_{\rm 1D})$\:.
Here, we have restored the atomic resonance frequency $\omega_0$ to underline the resonant character of the scattering and introduced the frequencies of the scattered photons $\omega_{1,2}'$.

\section{Functional integral approach}\label{app:2scat-func}
Here, we show how the photonic degree of freedom can be integrated out to obtain the effective non-Hermitian Hamiltonian for the atomic system~\cite{Shi2009,Xu2015}. We start from the full Lagrangian of the system $L = L_{\rm atoms} + L_{\rm phot} + L_{\rm atom-phot}$, where
\begin{align}
&L_{\rm phot} = \sum_k  \left(\rmi a_k^\dag \frac{da_k}{dt}- \omega_k a_k^\dag a_k^{\phantom{}}\right) ,\\\nonumber
&L_{\rm atom-phot} = -\frac{g}{\sqrt{L}}\sum\limits_{k}\sum\limits_{j=1}^N(a_k b_j^\dag \e^{\rmi k z_j}+a_k^\dag b_j^{\vphantom{\dag}}\e^{-\rmi k z_j})\:.
\end{align}
and $L_{\rm atoms} =\rmi  \sum_j b_j^\dag (db_j/dt) - H_{\rm atoms}$ is the Lagrangian of the atoms that depends on $b_j$ and  $b_j^\dag$ only. The Green's functions of the atomic subsystem can be described by the generating functional, that is readily given by the functional integral
\begin{align}\label{FI:Z}
&Z [\zeta_1^*(t), ... , \zeta_N^*(t);  \zeta_1(t), ... ,\zeta_N(t) ]\\\nonumber
&= \int \e^{\rmi \int [L+ \sum_j(\zeta_j^*b_j +\zeta_j b_j^\dag)  ] \, dt}  \prod_j D[b_j]D[b_j^\dag] \prod_k D[a_k]D[a_k^\dag] .
\end{align}
In particular, the single-excitation Green's function is given by the functional derivative
\begin{align}
G_{jl}(t,t') = -\left.\frac{\delta^2\,\ln Z}{\delta \zeta_j^*(t)\, \delta \zeta_l(t') }\right|_{\zeta,\zeta^* = 0} \,.
\end{align}

We now perform integration over $D[a_k]$ and $D[a_k^\dag]$ in Eq.~\eqref{FI:Z}. To this end, we separate the part that depends on the photonic operators, and switch from temporal representation to the frequency domain, i.e., from $a_k(t)$ and $a_k^\dag(t)$ to their Fourier transforms $a_k(\omega)$ and $a_k^\dag(\omega)$: 
\begin{align}
&\int (L_{\rm phot} + L_{\rm atom-phot}) dt = \sum_k \int \big\{
(\omega - \omega_k) a_k^\dag(\omega) a_k(\omega) \nonumber\\
&- g \sum_j [a_k(\omega) b_j^\dag(\omega) \e^{\rmi k z_j}+a_k^\dag(\omega) b_j^{\vphantom{\dag}}(\omega)\e^{-\rmi k z_j}]
 \big\} d\omega \nonumber .
\end{align}
The above expression is quadratic in $a_k(\omega)$ and $a_k^\dag(\omega)$. Therefore, the corresponding functional integral is Gaussian an can be easily evaluated,
\begin{align}
\int \e^{\rmi \int (L_{\rm phot} + L_{\rm atom-phot}) dt}  \prod_k D[a_k]D[a_k^\dag] = C \e^{\rmi \delta S} ,
\end{align}
where $C$ is a constant and
\begin{align}
\delta S = -\int d\omega \sum_k \frac{g^2}{\omega- \omega_k + \rmi 0}  \sum_{j,l} b_j^\dag b_l \e^{\rmi k (z_j-z_l)}  \,.
\end{align}
Calculating the sum over $k$ separately for $k>0$ and $k<0$, we obtain
\begin{align}
\delta S = \rmi \gamma_{\rm 1D} \int d\omega   \sum_{j,l} b_j^\dag(\omega) b_l(\omega) \e^{\rmi \omega |z_j-z_l|/c}  \,,
\end{align}
where $\gamma_{\rm 1D} = g^2/c\equiv \Gamma_{\rm 1D}/2$. 

Coming back to the generating functional Eq.~\eqref{FI:Z}, we can now present it in the following form:
\begin{align}
Z = \int \e^{\rmi \int [L_{\rm atoms}^{\rm eff}+ \sum_j(\zeta_j^*b_j +\zeta_j b_j^\dag)  ] \, dt} \prod_j D[b_j]D[b_j^\dag] ,
\end{align}
where $\int L_{\rm atoms}^{\rm eff} dt = \int L_{\rm atoms} dt + \delta S$ is the effective action of the atomic system that accounts for photon-mediated interatomic interactions. Equivalently, we can introduce the effective Hamiltonian, $\int L_{\rm atoms}^{\rm eff}dt = \rmi  \sum_j b_j^\dag (db_j/dt) - H_{\rm atoms}^{\rm eff}(t)$, where 
\begin{align}
H_{\rm atoms}^{\rm eff}(t)  = H_{\rm atoms}(t) - \rmi \gamma_{\rm 1D} \sum_{j,l}  b_j^\dag (t) b_l (t-|z_j-z_l|/c) 
\end{align} 
and $H_{\rm atoms}(t) =\omega_0  \sum_j b_j^\dag (t) b_j (t)$. We note the two key features of the effective atomic Hamiltonian: it is non-Hermitian and non-Markovian. In the case of excitation by monochromatic light at frequency $\omega$,  we can use $b_l (t-\tau) = b_l \e^{-\rmi \omega \tau}$. Then, the effective Hamiltonian assumes the form 
\begin{align}
H_{\rm atoms}^{\rm eff} = \sum_{j,l} H_{jl}(\omega) b_j^\dag  b_l^{\vphantom{l}} \,,
\end{align} 
where the matrix
\begin{align}
H_{jl}(\omega) = \omega_0 \delta_{jl} - \rmi\gamma_{\rm 1D} \e^{\rmi \omega |z_j-z_l|/c} \,
\end{align}
agrees with Eq.~\eqref{eq:Hmn}.

\section{Photon pair scattering: the Green's function solution in an electron representation}\label{app:2scat-electron}

Here, we show how the two-photon scattering matrix for  a single two-level atom (qubit) can be calculated using  the Green function technique. First, we introduce the bare (disregarding the light-qubit interaction) Green function of the qubit in its ground and excited states,
\begin{align}
G_{0,g(x)}(\varepsilon) = \frac{1}{\varepsilon - \varepsilon_{g(e)} + \rmi 0} \,,
\end{align}
where $\varepsilon_{g(e)}$ is energy of the qubit's ground (excited) state, $\varepsilon_{e}- \varepsilon_{g} = \omega_0$. 
The Green function of the waveguide photon with the wave vector $k$ reads
\begin{align}
D_k(\omega) = \frac{1}{\omega - \omega_k + \rmi 0} \,
\end{align}
where $\omega_k = c|k|$ is the photon dispersion. 

\begin{figure}[t]
\includegraphics[width=0.45\textwidth]{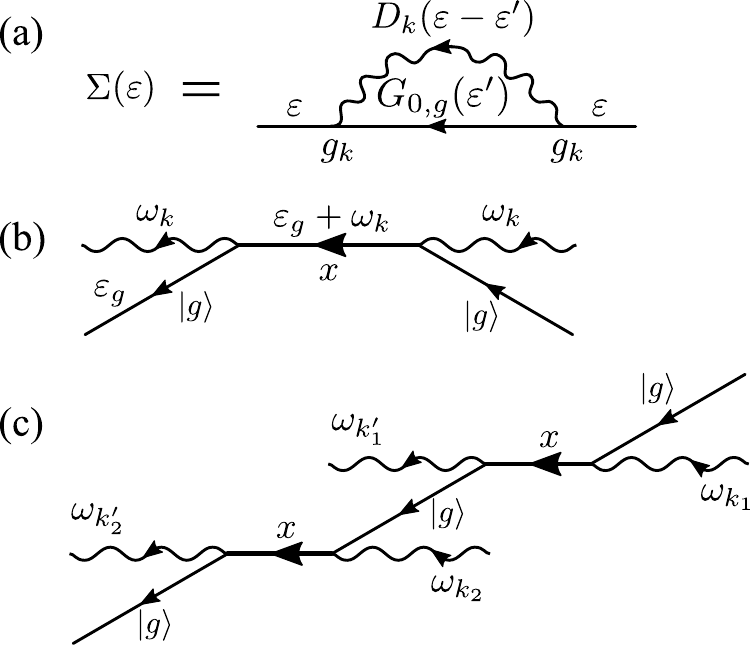}
\caption{(a) The diagrammatic representation of the self-energy correction for the excited atomic state. Straight line is the bare Green's function of the atom, wavy line is the photon Green's function. 
(b) The diagram representing the single photon reflection from an atom. Bold straight line represents the dressed Green's function of the atom in the excited state. 
(c) The diagrammatic representation of the two-photon scattering by an atom.}
\label{fig:S12}
\end{figure}

The interaction with light does not affect the ground state of the qubit, since it cannot emit a photon, while the excited state gets dressed. The dressing occurs due to the processes when the qubit in the excited state emits a photon and then reabsorbs it, as depicted in Fig.~\ref{fig:S12}(a). There, the solid lines denote the Green's functions of the qubit in its ground or excited state, and wavy lines stand for the photons Green's function. The vertex represents the process of photons absorption or emission by the atom and corresponds to the amplitude $g_k/\sqrt{L}$, where $g_k$ is the interaction constant and $L$ is the normalization length. The excited state self-energy corresponding to diagram in Fig.~\ref{fig:S12}(a) reads
\begin{align}
\Sigma(\varepsilon) &= -\rmi \sum_k \int \frac{d\varepsilon'}{2\pi}\left(\frac{\rmi g_k}{\sqrt L}\right)^2 G_{0,g}(\varepsilon') D_k(\varepsilon-\varepsilon') = \\
&= -\rmi  \int \frac{dk}{2\pi}\, \frac{g_k^2}{\varepsilon-\varepsilon_g-\omega_k+\rmi 0} \,.
\end{align}
We separate the real and the imaginary parts of $\Sigma$ using the Sochocki formula, $\Sigma = \Sigma' + \rmi \Sigma''$,
\begin{align}
&\Sigma'(\varepsilon)  =  \mathcal{P}\int \frac{dk}{2\pi}\, \frac{g_k^2}{\varepsilon-\varepsilon_g-\omega_k},\,\\
&\Sigma''(\varepsilon)  = -\int \frac{dk}{2\pi}\, g_k^2\,   \pi \delta(\varepsilon-\varepsilon_g-\omega_k) = -\frac{g_{k_0}^2}{v_{k_0}} = -\gamma_{\rm 1D}\,.\end{align}
Here, the real part $\Sigma'$ describes the radiative correction to the energy of the excited state (the Lamb shift), while the imaginary part $\Sigma''$ corresponds to the lifetime of the excited state, as calculated previously. Note, that if we ignore dependence of $g_k$ on $k$ and linearize the photon dispersion near $k_0 = \omega_0/c$, $\omega_k \approx \omega_{k_0} + v (k-k_0)$, the $\Sigma'$ vanishes while $\Sigma''$ is constant. 

The dressed Green's function of the excited state is readily expressed via the self-energy $\Sigma$,
\begin{align}
{G}_e(\varepsilon) = \frac{1}{G_{0,e}^{-1}(\varepsilon)-\Sigma(\varepsilon)} = \frac{1}{\varepsilon-\varepsilon_0+ \rmi\gamma_{\rm 1D}}\,.
\end{align}
The single-photon reflection coefficient is given by the diagram in Fig.~\ref{fig:S12}(b), where bold line denotes dressed Green's function of the excited state, which yields
\begin{align}
r(\omega_k) = -\rmi\gamma_{\rm 1D} {G}_e(\varepsilon_g+\omega_k)  = -\frac{\rmi\gamma_{\rm 1D}}{\omega_k -\omega_0 + \rmi\gamma_{\rm 1D}}\,.
\end{align}

A diagram describing simultaneous reflection of two photons is shown in Fig.~\ref{fig:S12}(c).
The scattering occurs in the 4-th order in interaction constant. The higher-order contributions are easily taken into account by using the dressed Green's function of the atom in the excited state (bold solid line). The scattering matrix element corresponding to the diagram reads
\begin{align}
S_1 &=   - \frac{\rmi g^4}{L^2}  {G}_e(\varepsilon_g+\omega_{k_1}) G_{0,g}(\varepsilon_g+\omega_{k_1}-\omega_{k_1'}) \, \rmi {G}_e(\varepsilon_g+\omega_{k_2'}) \nonumber\\ &\times 2\pi\delta(\omega_{k_1}+\omega_{k_2}-\omega_{k_1'}-\omega_{k_2'})  \nonumber\\
&=  \frac{-\rmi g^4/L^2}{(\omega_{k_1}-\omega_0+\rmi\GO)(\omega_{k_1}-\omega_{k_1'}+\rmi 0)(\omega_{k_2'}-\omega_0+\rmi\GO)}
\nonumber\\ & \times 2\pi\delta(\omega_{k_1}+\omega_{k_2}-\omega_{k_1'}-\omega_{k_2'})
\,,
\end{align}
where we suppose that $g_k$ is $k$-independent, $g_k = g$. Using the Sochocki formula we decompose the result in two terms,
\begin{align}
&S_1 = S_1^\text{(coh)} + S_1^\text{(incoh)}\:, \\
&S_1^\text{(coh)} = \frac{- g^4/L^2}{(\omega_{k_1}-\omega_0+\rmi\GO)(\omega_{k_2'}-\omega_0+\rmi\GO)}
 \, \pi \delta(\omega_{k_1}-\omega_{k_1'}) \:,
 \nonumber \\ & \times 2\pi\delta(\omega_{k_1}+\omega_{k_2}-\omega_{k_1'}-\omega_{k_2'}) 
 \nonumber \\ & = \frac12 \, r(\omega_{k_1}) r(\omega_{k_2}) \, \delta_{|k_1|,|k_1'|} \delta_{|k_2|,|k_2'|} \:,
  \\
 &S_1^\text{(incoh)} =  \frac{-\rmi g^4/L^2}{(\omega_{k_1}-\omega_0+\rmi\GO)(\omega_{k_2'}-\omega_0+\rmi\GO)(\omega_{k_1}-\omega_{k_1'})} 
 \nonumber \\ & \times
  2\pi\delta(\omega_{k_1}+\omega_{k_2}-\omega_{k_1'}-\omega_{k_2'}) \,.
\end{align}
$S_1^\text{(coh)}$ describes the process where photons are reflected independently, each of them conserving its frequency. The amplitude of such process is given by the product of the single-photon reflection coefficients $r(\omega)$ for the two photons. In contrast, $S_1^\text{(incoh)}$ describes the process where the photons interact and the energy is redistributed between them. 

Apart from the contribution $S_1$, the scattering matrix features three more,
\begin{align}
S = S_1 +S_2 +S_3 +S_4\:,
\end{align}
that are obtained by permutations: $S_2$ by $k_1 \leftrightarrow k_2$, $S_3$ by $k_1' \leftrightarrow k_2'$, and $S_4$ by both $k_1 \leftrightarrow k_2$ and $k_1' \leftrightarrow k_2'$. The calculation yields for the coherent term
\begin{align}
S^\text{(coh)} =  r(\omega_{k_1}) r(\omega_{k_2}) \, \left[ \delta_{|k_1|,|k_1'|} \delta_{|k_2|,|k_2'|} + \delta_{|k_1|,|k_2'|} \delta_{|k_2|,|k_1'|} \right]\:.
\end{align}
For the incoherent term, after somewhat cumbersome algebra we obtain
\begin{align}
&S^\text{(incoh)} = \frac{2\rmi g^4}{L^2}\, \frac{\omega_{k_1} + \omega_{k_2} -2\omega_0 + 2\rmi\GO}{(\omega_{k_1}-\omega_0+\rmi\GO)(\omega_{k_2}-\omega_0+\rmi\GO)}  \nonumber\\ 
&\times
\frac{2\pi\delta(w_1+w_2-w_1'-w_2')}{(\omega_{k_1'}-\omega_0+\rmi\GO)(\omega_{k_2'}-\omega_0+\rmi\GO)}
 \,.
\end{align}
The above results for two-photon scattering matrix for a single atom coincide with those obtained by other methods. However, the above approach does not allow the direct generalization to the case of many atoms~\cite{Kocabas2016}.   

\section{Photon pair scattering: the Green's function solution in an exciton representation}\label{app:2scat-exciton}
\begin{figure}[t]
 \centering\includegraphics[width=0.45\textwidth]{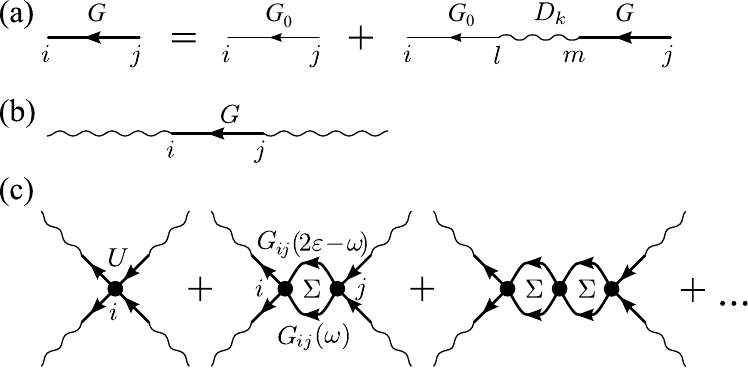}
 \caption{(a) Dyson-like equation for the dressed exciton Green's function (thick straight line). Thin straight and wavy line represent bare exciton and bare photon Green's functions. (b) The diagram representing single photon reflection. (c) The series corresponding to the two-photon scattering. Bold dots indicate exciton-exciton interaction $U$. }\label{fig:diaU}
\end{figure}

Here, we describe the Green's function technique for calculation of scattering matrix of an atomic array that is based on excitonic representation of the atomic Hamiltonian in Eq.~\eqref{eq:H1atom1b}. We start from the bare Green's functions of the atomic excitations and waveguide photons, 
\begin{align}
G_0(\omega)  = \frac{1}{\omega-\omega_0+ \rmi 0 } \,,\\
D_k(\omega) = \frac{1}{\omega - \omega_k + \rmi 0} \,\,.
\end{align}
Similarly to the approach of Appendix~\ref{app:2scat-electron}, we start by dressing the atomic excitations by photons. The dressed exciton Green's function ${G}_{ij}(\omega)$ can be calculated from the Dyson-like equation depicted in Fig.~\ref{fig:diaU}(a), that yields
\begin{align}\label{eq:G11}
{G}_{ij}(\omega) &= {G}_{0,ij}(\omega)\\
\nonumber &+ \frac{1}{L}\sum_k g_k^2 \sum_{l,m} {G}_{0,il}(\omega)\,  \e^{\rmi k z_l}\, D_k(\omega)\,  \e^{-\rmi k z_m} \,{G}_{mj}(\omega) \,.
\end{align}
Summation over $k$ can be easily performed assuming linear dispersion $\omega_k = c |k|$ and constant $g_k = g$\:,
\begin{align}
\frac{1}{L}\sum_k \frac{g_k^2 \e^{\rmi k (z_l-z_m)}}{\omega-\omega_k + \rmi 0} = -\rmi\, \frac{g^2}{c}\, \e^{\rmi\omega|z_l-z_m|/c} \,,
\end{align}
see also  Eq.~\eqref{eq:sum-rwa}\:.
Then, Eq.~\eqref{eq:G11} assumes the form
\begin{equation}  \label{eq:defG}
(\omega-\omega_0) {G}_{ij}(\omega)+\rmi\gamma_{\rm 1D}\sum_m \e^{\rmi (\omega/c)|z_i-z_m|} {G}_{mj}(\omega)= \delta_{ij}\:,
\end{equation}
where $\gamma_{\rm 1D} = g^2/c$. In other words, the exciton Green's function can be found by inverting the matrix as
\begin{align}
[{G}^{-1}(\omega)]_{ij} = (\omega-\omega_0)\delta_{ij} + \rmi\gamma_{\rm 1D} \, \e^{\rmi (\omega/c)|z_i-z_j|} \,.
\end{align}

The amplitude of the photon reflection from the left is given by the diagram of Fig.~\ref{fig:diaU}(b).
It reads
\begin{align}
r  = -\rmi\gamma_{\rm 1D} \sum_{ij} \, {G}_{ij}\, \e^{\rmi (\omega/c) (z_j+z_i)},
\end{align}
where we replace the $2\pi \delta(0)$  term with the time $T=L/c$. The result can be simplified by the use of relations, which follow from Eq.~\eqref{eq:defG},
\begin{align} \label{eq:spdef}
&s_i^+(\omega) \equiv \sum_{j} \, {G}_{ij}\, \e^{\rmi (\omega/c) z_j}  
\\\nonumber &= \frac{\e^{\rmi (\omega/c) z_{\text{min}}}}{\rmi\gamma_{\rm 1D}} \left[ \delta_{i,i_\text{min}} - (\omega-\omega_0) {G}_{i,i_\text{min}} \right] \,,\\ \label{eq:smdef}
&s_i^-(\omega) \equiv  \sum_{j} \, {G}_{ij}\, \e^{-\rmi (\omega/c) z_j}  
\\\nonumber &= \frac{\e^{-\rmi (\omega/c) z_{\text{max}}}}{\rmi\gamma_{\rm 1D}} \left[ \delta_{i,i_\text{max}} - (\omega-\omega_0) {G}_{i,i_\text{max}} \right] \,,
\end{align}
where $i_\text{max(min)}$ are the indices of the atoms with the maximal (minimal) $z$-coordinate value $z_\text{max}$($z_\text{min}$). We then obtain
\begin{align}
r = \frac{\e^{2\rmi(\omega/c)z_\text{min}}}{\rmi\gamma_{\rm 1D}} \left[ (\omega-\omega_0-\rmi\gamma_{\rm 1D}) - (\omega-\omega_0)^2 {G}_{i_\text{min},i_\text{min}} \right] \,.
\end{align}
A similar calculation gives the transmission coefficient
\begin{align}
t &=1 -\rmi\gamma_{\rm 1D} \sum_{ij} \, {G}_{ij}\, \e^{\rmi (\omega/c) (z_j-z_i)}
\\\nonumber &= \frac{(\omega-\omega_0)}{\rmi\gamma_{\rm 1D}} \left[ \delta_{i_\text{max},i_\text{min}} - (\omega-\omega_0) {G}_{i_\text{max},i_\text{min}}\right] \,.
\end{align}

The diagrams corresponding to the two-photon scattering are shown in Fig.~\ref{fig:diaU}(c). There, bold solid lines represent the exciton Green's function and the dots stand for the exciton-exciton interaction with the amplitude $U$. 
The scattering matrix element is given by the geometric series~[\onlinecite{Poshakinskiy2016}]
\begin{multline}\label{S}
S^{\text{incoh}}(\omega_1',\omega_2';\omega_1,\omega_2)\nonumber \\
= \frac{2 g^4}{L^2}\sum_{i,j=1}^{N} s_i^-(\omega_1')s_i^-(\omega_2')\bigl[ -\rmi U \delta_{ij}   + U^2 \Sigma_{ij}  + \ldots\bigr] \\\times s_j^+(\omega_1)s_j^+(\omega_2)  2\pi\delta(\omega_1+\omega_2-\omega_1'-\omega_2') \nonumber \\
=2\pi\rmi M\left(\frac{c}{L}\right)^2\delta(\omega_1+\omega_2-\omega_1'-\omega_2'),
 \end{multline}
where $s^{\pm}_i(\omega)$ is defined by Eqs~\eqref{eq:spdef},\eqref{eq:smdef},
\begin{equation}\label{eq:M-2phot-gen}
M= -2\rmi \frac{g^4}{c^2}\sum_{i,j=1}^{N} s_i^-(\omega_1')s_i^-(\omega_2')Q_{ij}s_j^+(\omega_1)s_j^+(\omega_2) \,
\end{equation}
is the two-exciton self-energy, and the matrix $Q$ is given by $ Q=-\rmi U(1-\rmi U \Sigma)^{-1}$,
and the matrix $\Sigma$ has the elements
\begin{align}\label{eq:gdef}
\Sigma_{ij}(\varepsilon) = \int  G_{ij}(\omega)G_{ij}(2\varepsilon-\omega) \frac{\rmd\omega}{2\pi} \,,
\end{align}
with $\varepsilon = (\omega_1+\omega_2)/2$. In the case of two-level atom, $U \to \infty$, we get $Q = \Sigma^{-1}$. 

It is instructive to compare the resonances of the scattering matrix $\Sigma$ with the eigenstates of the two-photon Schr\"odinger equation. To this end, we substitute the double excited states in the form $\sum_{m,n=1}^N \psi_{mn} b_m^\dag b_n^\dag |0\rangle$ into the effective Hamiltonian 
 \begin{align}\label{eq:H-atoms-gen-mark}
H_{\rm atoms}=\sum\limits_{mn}H_{mn}b_{m}^{\dag}b_{n}+\frac{U}{2}
\sum\limits_{m=1}^Nb_{m}^{\dag}b_{m}^{\vphantom{\dag}}(b_{m}^{\dag}b_{m}^{\vphantom{\dag}}-1)\:,
\end{align}
with $H_{mn}$ given by Eq.~\eqref{eq:Hmn} evaluated at $\omega=\omega_0$ 
 and obtain \cite{Ke2019}
\begin{equation}\label{eq:Sh2}
\sum\limits_{m'n'=1}^N(\mathcal H+\mathcal U)_{mn,m'n'}\psi_{m'n'}=2\eps \psi_{mn}
\end{equation}
with
$\mathcal H_{mn;m'n'}=\delta_{mm'}H_{nn'}+\delta_{nn'}H_{mm'}$
and 
$\mathcal U_{mn,m'n'}=\delta_{mn}\delta_{mm'}\delta_{nn'} U\:.$
We now note that the integral in Eq.~\eqref{eq:gdef} can be presented as
\begin{multline}
\int\frac{\rmd \omega}{2\pi} G_{ij}(\omega)G_{kl}(2\eps-\omega)\\=
\int\frac{\rmd \omega}{2\pi} \left[\frac1{H-\omega}\right]_{ij}\left[\frac1{H+\omega-2\eps}\right]_{kl}=
\\=\left[\frac{\rmi}{H\otimes 1+1\otimes H-2\eps}\right]_{ik,jl}\:.
\end{multline}
Hence,
\begin{equation}\label{eq:Sigma1}
\Sigma_{mn}=\left[\frac{\rmi}{\mathcal H-2\eps}\right]_{mm,nn}\!,\:
Q_{mn}=\rmi U\left[\frac{2\eps-\mathcal H}{\mathcal H+\mathcal U-2\eps}\right]_{mm,nn}\!\:.
\end{equation}
So the two-photon scattering matrix indeed  has resonances when the sum of the energies 
of two incident photons $2\eps$ matches the energy of the double-excited state Eq.~\eqref{eq:Sh2}.
In the limit of $U\to \infty$ Eq.~\eqref{eq:Sigma1} can be further simplified to
\begin{equation}
Q_{mn}=2(\rmi \eps-\gamma_{\rm 1D})\delta_{mn}+\sum\limits_{\nu=1}^{N(N-1)/2}\frac{2\rmi d^\nu_{m}d^{\nu}_{n}}{\eps_\nu-\eps}\label{eq:Q3}\:,
\end{equation}
where $\eps_\nu$ are the two-photon state energies found  from Eq.~\eqref{eq:Sh2},
and $d_m^\nu=\mathcal H_{mm;m'n'}\psi^\nu_{m'n'}$ with the normalization condition for two-photon states being $\sum_{m'n'}(\psi^\nu_{m'n'})^2=1$.
\section{Generation of GHZ state and quantum state transfer}\label{app:Guimond}
In this section, we provide some more details on the protocols of the generation of  the Greenberger--Horne--Zeilinger (GHZ) state and the quantum state transfer, proposed in Ref.~\cite{Guimond2020}, and illustrated in Fig.~\ref{fig:Guimond} in the main text.

The proposed quantum protocol works in the regime when the interaction of a waveguide photon through a given dimer is described by the operator
\begin{equation}
\label{eq:scatq}
|{\rm phot}\rangle\to \left[|1_q\rangle\langle 1_q|\rangle-|0_q\rangle\langle 0_q|\right] |{\rm phot}\rangle\equiv -\sigma_{z,q}|{\rm phot}\rangle\:,
\end{equation}
where $|0_q/1_q\rangle$ are the ground and excited states of the stationary qubit coupled to the dimer. This can be realized by choosing the energy of the incoming photon in between the bare dimer qubit resonance and the  dimer qubit resonance, shifted by the excitation with the stationary qubit~\cite{Guimond2020}. We also introduce the Hadamard states of the stationary qubits and photons
$$
|\pm\rangle=(|1\rangle\pm |0\rangle)/\sqrt{2}.
$$
In the Hadamard basis, the photon transmission is described by the Pauli matrix $\sigma_x$, which corresponds to transitions $|+\rangle\to |-\rangle$ and $|-\rangle \to |+\rangle$.

The Hadamard gate acting on the qubit $n$ is defined as 
$\hat{\mathcal H}_n=|+\rangle_n \langle 0|_n+|-\rangle_n \langle 1|_n\:.
$
The photons in upper and lower waveguides shown in Fig.~\ref{fig:Guimond} play the role of "flying qubits". The waveguides can be linked by beam splitters that act as the  Hadamard gates for photons. The only difference is that the basis of the states $|0\rangle$ and $|1\rangle$ is replaced by $|d\rangle$ and $|u\rangle$, corresponding to the photons in upper and lower waveguides. 
Specifically, photon entering the lower arm is transformed into 
$(|d\rangle+|u\rangle)/\sqrt{2}$ and 
photon entering the lower arm is transformed into 
$(|d\rangle-|u\rangle)/\sqrt{2}$.

We will now illustrate the protocol to generate the  GHZ state  for the simplest case of just $N=2$ qubits. 
The system is  initialized as
$
\psi_{\rm in}=|+\rangle_1|+\rangle_2 |d\rangle\:,
$
i.e. the stationary qubits are in the product state and 
one photon is incident in the lower waveguide.  
The cascaded photon processing in the array is detailed below:
\begin{align}
\text{1st beamsplitter:~~} &|+\rangle_1|+\rangle_2 \frac{|d\rangle+|u\rangle}{\sqrt{2}}\:,\\
\text{after qubit 1:~~} & \frac{1}{\sqrt{2}} |-\rangle_1|+\rangle_2|u\rangle+ \frac{1}{\sqrt{2}} |+\rangle_1|+\rangle_2|d\rangle\:,\nonumber\\
\text{after qubit 2:~~} &  \frac{1}{\sqrt{2}} |-\rangle_1|-\rangle_2|u\rangle+ \frac{1}{\sqrt{2}} |+\rangle_1|+\rangle_2|d\rangle \nonumber\:,\\\nonumber 
\text{2nd beamsplitter:~~} & \frac{1}{2} |d\rangle \left(|+\rangle_1|+\rangle_2-|-\rangle_1|-\rangle_2\right)\\&+\frac{1}{2} |u\rangle (|+\rangle_1|+\rangle_2+|-\rangle_1|-\rangle_2)\nonumber .
\end{align}
Hence, after projecting the output on the state $|d\rangle$ or $|u\rangle$ (performing the measuring) one arrives to the qubit array in the GHZ state.

Similar approach can be used for the quantum state transfer, see Fig.~\ref{fig:Guimond}(b). We start with the first qubit in an arbitrary quantum state and one photon in the lower waveguide
\begin{equation}
\psi_{\rm in}=(c_+|+\rangle_1+c_-|-\rangle_1)|+\rangle_2 |d\rangle\:,
\end{equation}
with $|c_+|^2+|c_-|^2=1$.
Using the same logic as above, we find that the output state after the third beamsplitter is
\begin{multline}
|\psi_1\rangle=-\frac{\sqrt{2}}{4}\bigl\{[(c_++c_-)|-\rangle_2+(c_+-c_-)|+\rangle_2]|-\rangle_1\\+[(c_++c_-)|-\rangle_2-(c_+-c_-)|+\rangle_2]|+\rangle_1\bigr\}|d\rangle+(\ldots )|u\rangle \nonumber\:.
\end{multline}
Now, we start performing the quantum measurements. We project the state $|\psi_1\rangle$ onto the state $|d\rangle$ , apply $\sigma_{z,1}$ matrix and project the first qubit in the $|-\rangle_1$ state:
\begin{equation} 
\psi_2=\langle d|\langle -|_1\sigma_{z,1} |\psi_1\rangle=-\tfrac{\sqrt{2}}{4}[(c_-+c_+)|-\rangle_2+
(c_+-c_-)|+\rangle_2]\:.\nonumber
\end{equation}
Now we apply the Hadamard transformation followed by  $\sigma_{z,2}$ operation to the second qubit:
\[
2\sigma_z
\mathcal{\hat{H}}\psi_2=c_+|+\rangle_2+c_-|-\rangle_2\:.
\]
Thus, the quantum state has been transferred to the second qubit.

The  setup from Fig.~\ref{fig:Guimond} can be easily tailored to construct   arbitrary matrix product states Eq.~\eqref{eq:MPS} of the stationary qubits, with the rank of the $A$ matrix being equal to two. Even more complicated  states could be generated by adding waveguides to the system~\cite{Guimond2020}.
\section{Photon pair scattering from a chiral atomic array}\label{app:2scat-chiral}
Here, we generalize the technique from Appendix~\ref{app:2scat-exciton} to the case of photon pair scattering from a chiral array of two-level atoms.  The problem is still described by the Hamiltonians \eqref{eq:H1atom1b}-\eqref{eq:RWA}, but the summation over $k$ is carried out only over the right-going photon states, where $k>0$.
The effective Hamiltonian describing light-mediated interaction between single-excited atomic states is
\begin{equation}\label{eq:Hnm-chiral}
H_{nm}=\begin{cases}
\omega_0-\rmi (\gamma^\rightarrow+\gamma),& n=m\\
-2\rmi\gamma^\rightarrow,& n>m\\\:
0,&n<m\:,
\end{cases}
\end{equation}
where $\gamma^\rightarrow=g^2/(2c)$ is the radiative decay rate of a single atom and $\gamma$ is the nonradiative decay rate. This Hamiltonian is very similar to the Hamiltonian~\eqref{eq:Hmn} for a non-chiral waveguide, but now the atoms with $n<m$ are not coupled to each other by light.

The Green's function for single-excited states is given by $G=(\omega-H)^{-1}$. Since the atoms with $n<m$ are decoupled, it can be found recursively, as 
\begin{equation}
G_{11}=\frac{1}{\omega-\widetilde\omega_0},
\end{equation}
($\widetilde\omega_0=\omega_0-\rmi \gamma^\rightarrow-\rmi\gamma$) and 
\begin{equation}
G_{nm}=G_{11}+2\rmi\gamma^
\rightarrow\sum\limits_{m'=0}^{n-1}G_{nm'},
\quad G_{nm}=0\text{ for }m>n\:. 
\end{equation}
Using \eqref{eq:gdef}, we now find the two-photon scattering kernel
\begin{equation}
\Sigma_{nm}(\eps)\equiv  \int  G_{nm}(\omega)G_{nm}(2\varepsilon-\omega) \frac{\rmd\omega}{2\pi}= \frac{\rmi \gamma^\rightarrow\delta_{nm}}{\widetilde\omega_0-\eps}\:.
\end{equation}
The scattering matrix is given by \eqref{eq:M-2phot-gen} and reads
\begin{multline}
M(\omega,2\eps-\omega\leftarrow \eps,\eps)\\= 
\frac{4g^2}{c} \frac{1}{(\eps-\widetilde\omega_0)(\omega-\widetilde\omega_0)(2\eps-\omega-\widetilde\omega_0)}\frac{t_\eps^{2N}t_\omega^Nt_{2\eps-\omega}-1}{t_\eps^{2}t_\omega t_{2\eps-\omega}-1},
  \end{multline}
where 
\begin{equation} 
t_\omega=\frac{\omega_0-\omega+\rmi \gamma^\rightarrow-\rmi\gamma}{\omega_0-\omega-\rmi \gamma^\rightarrow-\rmi\gamma}
\end{equation}
is the single-photon transmission coefficient of one atom. 
The photon-photon correlation function is found from Eq.~\eqref{eq:g2gen} and is determined by the residue of $N$-th order at $\omega=\widetilde\omega_0$ that can be calculated as
\begin{equation}\label{eq:g2Nchiral}
g_N^{(2)}(0)=\left(1+\frac{\rmd^{N-1}}{\rmd x^{N-1}} \mathcal M\bigr|_{x=0}\right)^2
\end{equation}
with
\begin{equation}\label{eq:Mchiral}
\mathcal M=\frac {(1+\frac{\gamma^\rightarrow}{\gamma}) \left( {\frac { ( 1-x)( x-\gamma/\gamma^\rightarrow) }{\gamma/\gamma^\rightarrow+1-x}} \right) ^{N}
}{\frac{1}{2}- \left( \frac{1}{2}+\frac{\gamma}{2\gamma^\rightarrow}-x \right) ^{2}+\left(\frac{\gamma}{\gamma_{\rm 1D}^{\rightarrow}}\right)^2 }\:.\end{equation}
For relatively small values of $N\lesssim 100$, the residue can be readily calculated by expanding the products in
Eq.~\eqref{eq:Mchiral} in the binomial series~\cite{Lodahl2018}. For large values of $\gamma/
\gamma^\rightarrow$, it is possible to use a simple asymptotic expression Eq.~\eqref{eq:g2Nchiral:asymp}
that is valid for an arbitrarily large $N$.
In order to derive this expression, we rewrite the residue in Eq.~\eqref{eq:g2Nchiral} back as a contour integral
\begin{equation}\label{eq:Rres}
\frac{\rmd^{N-1}}{\rmd x^{N-1}} \mathcal M\bigr|_{x=0}=
\frac1{2\pi\rmi}\oint\limits_{x=R}\mathcal M\rmd x\:.
\end{equation}
For $\gamma\gg \gamma^\rightarrow$, the value of the integral is mostly determined by a simple pole at $x=x^*$, where  
\begin{equation}
x^*=\frac{\gamma}{\gamma^\rightarrow} + \frac{1}{2} - \sqrt{2\left(\frac{\gamma}{\gamma^\rightarrow}\right)^2 + \frac{1}{2}}
\end{equation}
is a zero of the denominator which results in Eq.~\eqref{eq:g2Nchiral:asymp}.

\section{Photon reflection from a planar atomic array}\label{app:lattice-sum}
In this section, we outline the calculation of the light reflection from a two-dimensional atomic array, considered in Sec,~\ref{sec:2d}.
The light-array interactioin can be treated  in the discrete dipole approximation ~\cite{Draine1994}:
\begin{equation}\label{eq:dda}
\bm p_j=\alpha\bigl[\bm E_0(\bm r_j)+\sum\limits_{j'\ne j} G(\bm r_{j'}-\bm r_j)\bm p_{j'}\bigr],
\end{equation}
where $\bm p_j$ is the electric dipole moment of the atom number $j$,
\begin{equation}
\alpha=\frac{3\rmi c^3}{2\omega_0^3}\frac{\gamma_0}{\omega_0-\omega-\rmi (\gamma_0 + \gamma_\text{nonrad})}
\end{equation}
is the single atom polarizability, characterized by the resonance frequency $\omega_0$, radiative decay rate $\gamma_\text{0}\equiv \Gamma_\text{0}/2$ and nonradiative decay rate $\gamma_\text{nonrad}\equiv \Gamma_\text{nonrad}/2$. 
Here, $\bm E_0$ is the electric field of the incident wave and 
\begin{equation}
G_{\mu\nu}(\bm r,\omega)=\left(\delta_{\mu\nu}+\left(\frac{c}{\omega}\right)^2\frac{\partial^2}{\partial x_\mu\partial x_\nu}\right)\frac{\e^{\rmi \omega r/c}}{4\pi r}
\end{equation}
is the electromagnetic tensor of the Green's function at the frequency $\omega$ satisfying the equation
$\rot\rot G=(\omega/c)^2G+\delta(\bm r)$. The first term in Eq.~\eqref{eq:dda}  describes the polarizability of the atom by the incident wave and the second term accounts for the sum of electric fields emitted by all other atoms. We consider for simplicity the case when light is incident in the normal direction, so that $\bm E_0(\bm r_j)\equiv \bm E_0$ and $\bm p_j=\bm p$. In this case Eq.~\eqref{eq:dda} is readily solved yielding 
\begin{equation}\label{eq:pdda}
\bm p=\widetilde\alpha\bm E_0,\quad \widetilde \alpha=\frac{\alpha }{1-C\alpha}
\end{equation}
where 
\begin{equation}\label{eq:C}
C=4\pi\left(\frac{\omega}{c}\right)^2\sum_{\bm r\ne 0}G_{xx}(\bm r)
\end{equation}
is the so-called interaction constant describing light-induced coupling between the given atom and all the other atoms in the array. Hence, the coefficient $\widetilde \alpha$ in Eq.~\eqref{eq:pdda} is the polarizability, renormalized by collective coupling between the atoms. The amplitude light reflection and transmission coefficients from the array $r$ and $t$ 
can be found by summing the field emitted from all the dipoles in the normal direction. The results read~\cite{Ivchenko2000,Yugova2009}
 \begin{equation} 
r=\frac{2\pi\rmi \omega }{c a^2}\tilde \alpha,\quad t=1+r\:.\label{eq:rtarray}
\end{equation}
We consider the array with $a<\lambda$, when all the diffracted waves are evanescent. In this case, the reflection and transmission coefficients reduce to Eqs.~\eqref{eq:rt2d} from the main text.

The sum in Eq.~\eqref{eq:C} for arrays with tens of atoms, occurring in practice ~\cite{Rui2020}, can be readily evaluated directly. In the theoretical limit of infinite array the convergence of the sum is quite slow due to the far-field interactions. The commonly used approach to evaluate the lattice sum is the Ewald summation that is based on splitting the sum in two parts. The first part corresponding to smaller values of $r$ in the near field zone, is evaluated in the real space, and the second part corresponding to the far field zone $r\gtrsim c/\omega$ is Fourier transformed into the reciprocal spaced using the identity 
\begin{equation}\label{eq:G0b}
\frac{\e^{\rmi \omega r/c}}{r}=\sum\limits_{\bm b}\frac{2\pi \rmi }{k_b a^2}\e^{\rmi k_b|z|+\rmi \bm b\bm \rho},
\end{equation}
where $\bm r=(\bm r,z)$ and the reciprocal lattice vectors $\bm b$ form a square lattice with the spacing $2\pi/a$, $k_b\equiv \sqrt{(\omega/c)^2-b^2}$. More details on the Ewald summation can be found e.g. in Ref.~\cite{Kambe1967}. Another efficient summation technique, that is in our experience even more efficient is the Floquet summation technique developed in \cite{belov2005}. Specifically, the sum is given by Eq.~(A37) of Ref.~\cite{belov2005} that has to be complex conjugated and also multiplied by $4\pi$ to take into account the time dependence convention $\e^{j\omega t}$ and a different definition of the Green's function used in Ref.~\cite{belov2005}. 

We now briefly discuss how to obtain the approximate expression for the lattice constant, given by Eq.~\eqref{eq:Cappr} in the main text. The first term $2\pi\rmi \omega/(ca^2)$ is given by the term with $b=0$ in Eq.~\eqref{eq:G0b} multiplied by $q^2$. It describes the radiative decay due to the emission of the waves propagating normally to the array, where $b=0$, or, in another words, results from far-field radiative coupling between the atoms. The last term $S/2$ is given by the near field and can be obtained by setting $\omega$ in Eq.~\eqref{eq:C} to zero:
\begin{multline}
\frac{S}{2}=\sum_{\bm r\ne 0} \frac{\partial^2}{\partial x^2}\frac{1}{r}\equiv 
\sum_{\bm r\ne 0}\frac{3x^2-r^2}{r^5}\\=\frac{1}{2}\sum\sum_{\bm r\ne 0}\frac{1}{r^3}\approx\frac{9.03}{2a^3}\:.
\end{multline}
The sum for the square lattice converges rapidly enough and can be calculated directly.
The term $S'(\omega/c)^2/2$ results from the field in Eq.~\eqref{eq:C} in the intermediate zone between the far field and the near field reads as
\begin{equation}
\frac{S'(\omega/c)^2}{2}=\sum\limits_{\bm r\ne 0}
\frac{1-(x/r)^2}{r}\frac{\e^{\rmi\bm b\cdot\bm r+\rmi \omega r/c}}{r}\:.
\end{equation}
The term $1-(x/r)^2$ for the square lattice can be replaced by 1/2, and we can write it as
\begin{equation}\label{eq:S11}
S'=\lim_{z\to 0}\lim_{\omega \to 0}\left(\Re \sum\limits_{\bm r }\frac{\e^{\rmi  \omega\sqrt{r^2+ z^2}/c}}{\sqrt{r^2+z^2}}-\frac{\e^{\rmi \omega |z|/c}}{|z|}\right)\:.
\end{equation}
Taking the Fourier transformation of the first term with the help of Eq.~\eqref{eq:G0b}, we find 
\begin{multline}\label{eq:S12}
S'=\lim_{z\to 0}\lim_{\omega\to 0}\left(
\frac{2\pi \rmi}{a}\sum\limits_{\bm b}\frac{\e^{\rmi \sqrt{q^2-b^2}|z|}}{\sqrt{(\omega/c)^2-b^2}}
-\frac{\e^{\rmi \omega|z|/c}}{|z|}\right)\\=\lim_{z\to 0}\left(
\frac{2\pi }{a}\sum\limits_{\bm b}\frac{\e^{-b|z|}}{b}-\frac{1}{|z|}
\right)\:.
\end{multline}
Let us first check the cancellation of the singular diverging terms $\propto 1/z$ in \eqref{eq:S12}.
To do this, we can replace the summation by integration:
\begin{equation}
 \frac{2\pi }{a}\sum\limits_b\frac{\e^{-bz}}{b}\approx 
\int_0^\infty bdb \frac{\e^{-b|z|}}{b}= \frac1{|z|}.
\end{equation}
Thus, the terms $1/z$ cancel each other and  Eq.~\eqref{eq:S12} has a finite limit of the order $1/a$.
Numerical calculation for a square lattice yields to
\begin{equation}
S'=\lim_{z\to 0}\left(
\frac{2\pi }{a}\sum\limits_{\bm b}\frac{\e^{-b|z|}}{b}-\frac{1}{|z|}
\right)=-\frac{3.90}{a}.
\end{equation}


%

\end{document}